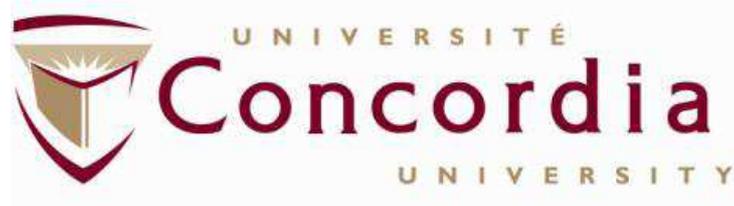

*Department of Computer Science and Software Engineering*

# Comparative Studies of 10 Programming Languages within 10 Diverse Criteria

Jiang Li
Concordia University
Montreal, Quebec, Concordia
j_lxi43@cse.concordia.ca

Sleiman Rabah
Concordia University
Montreal, Quebec, Concordia
s_rabah@cse.concordia.ca

Mingzhi Liu
Concordia University
Montreal, Quebec, Concordia
l_mingz@cse.concordia.ca

Yuanwei Lai
Concordia University
Montreal, Quebec, Concordia
la_yua@cse.concordia.ca

This page was intentionally left blank

**Abstract**

There are many programming languages in the world today.Each language has their advantage and disavantage. In this paper, we will discuss ten programming languages: C++, C#, Java, Groovy, JavaScript, PHP, Schalar, Scheme, Haskell and AspectJ. We summarize and compare these ten languages on ten different criterion. For example, Default more secure programming practices, Web applications development, OO-based abstraction and etc. At the end, we will give our conclusion that which languages are suitable and which are not for using in some cases. We will also provide evidence and our analysis on why some language are better than other or have advantages over the other on some criterion.

# 1 Introduction

Since there are hundreds of programming languages existing nowadays, it is impossible and inefficient to put effort on analyzing each languages. But we can classify the some representative categories of languages and make deep research on them according to some certain criteria. Thus our research problem is aiming to compare and contrast 10 languages according to 10 specified criteria with the purpose of determining the suitability and applicability of the languages for each criterion, distinguish them their pros and cons, evaluate and explore the related features on those languages, illustrate the points either with code examples or related work. In our project, we will evaluate our languages based on following criteria:

1. Default more secure programming practices
2. Web applications development
3. Web services design and composition
4. OO-based abstraction
5. Reflection
6. Aspect-orientation
7. Functional programming
8. Declarative programming
9. Batch scripting
10. UI prototype design

Depends on your choice of languages, some of them may have something in common on certain aspect while some part may totally different.

## 1.1    Related work

In order to complete our comparison work, we do some relevant search among conference papers, text books, Wikipedia, official websites and discuss with classmate and teacher in courses.

## 1.2    Overview

The rest of this paper is organized as follows. First we introduce the formatting basics in Sec-tion 1.3 and Section 1.4. We then briey introduce the languages being compared in Section 1.4, 1.5, 1.6, 1.7, 1.8, 1.9, 1.10, 1.11, 1.12, and Section 1.13. Next, we present our analysis of the criteria in Section 2 for pair-wise comparison of the assigned languages. We then move on to macro analysis and synthesis of our results in a consolidated form in Section 3. We conclude and outline our future work plans in Section 4 and Section 4.1 respectively.

## 1.3    Programming Language

## 1.4    C++

C++ is a <u>statically typed</u>, <u>free-form</u>, <u>multi-paradigm</u>, <u>compiled</u>, general-purpose <u>programming language</u>. Some people say that C++ is a middle language because it has the features of <u>high-level</u> and <u>low-level</u> language. As one of the most popular programming languages in the world, C++ is widely used in the software industry.[1] C++ is also used for <u>hardware design</u>to analyze structure. Some of its application domains include systems software, application software, device drivers, embedded software, high-performance server and client applications, and entertainment software such as <u>video games</u>.

## 1.5    AspectJ

AspectJ is a general-purpose Aspect-Oriented extension to java programming language [2]. It was created at Palo Alto Research Center Incorporated (PARC), now it is an open source project and part of the Eclipse Foundation. AspectJ has everything that Java has and more which means every valid Java program is also a valid AspectJ program [2]. The main goal of AspectJ development is modularizing crosscutting concerns such as logging, error checking and handling, synchronization, context-sensitive behavior, performance optimizations, monitoring and logging, debugging support, and multi-object protocols [4].

Aspect-oriented programming (AOP) is a programming paradigm built on top of the object-oriented paradigm and aims to modularize crosscutting concerns [2] by isolating secondary functions from the program's business logic [3]. AOP enhances code readability and reuse.

AspectJ compiler produces java bytecode, an AspectJ program can run on any Java compatible virtual machine. The runtime library "aspectjrt.jar" is required to run any AspectJ program. AspectJ development tool (AJDT) is a plug-in for the Eclipse IDE which can be used to compile and run AspectJ programs.

## 1.6    Haskell

**Haskell** is an advanced, standardized, general-purpose purely functional programming language incorporating many recent innovations in programming language design. Haskell provides higher-order functions, non-strict semantics, static polymorphic typing, user-defined algebraic datatypes, pattern-matching, list comprehensions, a module system, a monadic I/O system, and a rich set of primitive datatypes, including lists, arrays, arbitrary and fixed precision integers, and floating-point numbers. In Haskell, a function is a primary control construct of the programming language. It allows rapid development of robust, concise, correct software. Haskell is easier to produce flexible, maintainable high-quality software due to its strong support for integration with other languages, built-in concurrency and parallelism, debuggers, profilers, rich libraries and an active community. [5]

## 1.7    PHP

PHP is a wide use general purpose scripting language which used to make dynamic interact web pages. It can embedded with HTML source document used in server side. Influenced by C, Perl, Java, C++, thus it support multiple paradigm in programming, such as object-oriented (OO) and imperative. In the meantime, its type system is loosing typing and dynamic type checking. As the updates of version, it supports more new features to make the PHP more functional and diversify. Now, the latest version is PHP 5.33, and will be used in the following.

## 1.8    Scheme

Scheme is a general-purpose, functional and multi-paradigm programming language. Scheme derives some of its dialects and features from LSIP. Scheme is primarily intended to be a functional programming language; it supports lambda calculus, lexical scope and recursion. Today, Scheme is almost every where: it is used in many software development projects such as text editors, compilers optimization, expert systems, etc [9].

There are many implementations of Scheme providing different features [6] based on IEEE Scheme standards. The Gambit project provides the Gambit-C compiler as part of Gambit programming system. Gambit-C generates portable C code and executable [7]. PLT Scheme (known as Racket project) is one of Scheme implementations which provides a massive set of libraries for supporting many    features    such    as    GUI,    macros,    classes    and    objects    and    many    more    [8].

## 1.9    Groovy

Groovy is an agile and dynamic language for the Java Virtual Machine. It has some features similar to those of Python, Ruby, Perl, and Smalltalk. It can be used as a scripting language for the Java Platform.[10] Groovy uses a Java-like bracket syntax. Groovy compiles into Java bytecode and extends the Java API and libraries. It runs on Java 1.4 or newer. Most Java code is also syntactically valid Groovy.Groovy support XML and HTML.

## 1.10    Java

**Java** is a powerful, platform independent, object-oriented, strongly-type, interpreted and compiled, general-purpose programming language with build-in automatic memory management. It is a programming language originally developed by James Gosling at Sun Microsystems (which is now a subsidiary of Oracle Corporation) and released in 1995 as a core component of Sun Microsystems' Java platform. The language derives much of its syntax from C and C++ but has a simpler object model and fewer low-level facilities. It is intended to let application developers "write once, run anywhere". Java is currently one of the most popular programming languages in use, and is widely used from application software to web applications. [11]

## 1.11    JavaScript

**JavaScript** is a small, lightweight, prototype-based object-oriented, interpreted, cross-platform scripting language. Today, it is the most popular http://en.wikipedia.org/wiki/Scripting_language scripting language on the internet, and works in all major browsers, such as IE, Firefox, Chrome, Opera, and Safari. JavaScript was designed to add interactivity to HTML pages and usually is embedded directly into HTML pages. But it may also be used at outside webpage, such as server side.

## 1.12    Scala

Scala is a general purpose programming languages which support multiple paradigms. It extends the object-oriented characteristic with functional extension. It integrated many other languages features to itself. It is designed to express common programming patterns in a concise, elegant, and type-safe way [12]. The Scala run its code on the JVM, which is byte code compatible with Java. That is to say you can utilize all the libraries or existing resource in Java. This not only can benefit the java programmer to make productive and efficient product using Scala, it gives a higher start point for Scala completive with other programming languages. Although it seems Scala integrate with Java seamlessly, Scala is not a subset of java, it has much more features rather than Java. Now the latest version of Scala is 2.80. And will be used in the following.

## 1.13   C#

**C#** is modern, general-purpose and multi-paradigm programming language enclosing object-oriented, imperative, functional, generic, event-driven and component-oriented programming styles (DLLs and Assemblies). It was designed by Andres Hejlsberg (the creator of Turbo Pascal), developed by Microsoft and first released in July 2000. C# was developed specially for the .NET platform with a main goal to provide a simple, powerful and a strongly-typed programming language allowing programmers to quickly build a wide range of applications for the *.NET platform* [13].

The .NET *platform* is composed of a Common Language Runtime (CLR) and a large rich class library known as the .NET *Framework* (DLLs files). The .NET Framework provides a wide powerful range of features, among them: multi-threading, user interface prototyping, database connectivity, web application and service-oriented application development.

The CLR is the core component of the .NET platform and the execution environment in which all the managed code runs. The CLR is the Microsoft's JVM equivalent that supports several programming languages and performs services such as memory management, exception handling, garbage collection, security and interaction with the operating-system services. The CLR provides a common development environment enabling developers to build applications using different languages such as C#, VB, C++, F# and Python, etc.

C# code source is compiled to an intermediate language presentation called Microsoft Intermediate Language (MSIL) and it is the Java byte-code's equivalent. MSIL is translated into machine code by the CLR at run-time. C# has an advantage over Java which is high interoperability with other languages such as C/C++, Python, VB.NET, etc [14].

C# can be used to develop the following types of applications [14]:

- **C**ommand line application, aka console applications with a text-based interface (text user interface), that can be run using a Command Line Interface e.g MS DOS.
- **W**indows applications for developing GUI using Windows Forms or the new Windows Presentation Foundation (WPF) which is first released with the .NET Framework 3.0 that enables rendering user interfaces.
- **W**eb applications such as Websites using *ASP.NET* technology, Web Services and service-oriented application using Windows Communication Foundation (*WCF)* framework.

The .NET Framework needs to be installed in order to run application written in C#.

Microsoft's Visual C# Integrated Development Environment (IDE) is used to create such applications. It provides a set of built in tools such as a C# compiler, a user interface designer for web and GUI development and also a powerful debugger. C# has been approved as a standard by the *European Computer Manufacturers Association* (ECMA) (ECMA-334) [15] [13].

*Mono* is an open source project which aims to create a cross-platform implementation of the .NET platform based on the EMCA standard.

# 2 Analysis

## 2.1 AspectJ vs C#

### 2.1.1 Source code size

The following examples consist of the Hello World program; they also show AsepctJ and C# syntax. We conclude that both languages have nearly the same syntax. Thus, the source code size is little bigger in AspectJ in this examples but also the have nearly the same compilation time.

```java
/**
 * Example illustrating a simple aspect
 *
 * @author Sleiman Rabah
 */
public aspect HelloFromAspectJ {

        // Intercepts the main method execution
        pointcut mainMethod() : execution(public static void main(String[]));

        // Will be executed at the end of the main method
        after() returning : mainMethod() {
                System.out.println("Hello from AspectJ");
        }
}

/**
 * A Java class.
 *
 * @author Sleiman Rabah
 */
public class HelloWorld {

        public static void main(String[] args) {
                System.out.println("Hello world!");
        }
}
```

*Hello word program in C# showing its syntax. Hello word program in C# showing its syntax.*

```csharp
using System;
// A "Hello World" program in C#
namespace HelloWorld
{
    /// <summary>
    /// <autor>Sleiman Rabah</autor>
    /// </summary>
    class Hello
    {
        static void Main()
        {
            System.Console.WriteLine("Hello World!");
        }
    }
}
```

## 2.1.2  Default more secure programming practices

*"Our civilization runs on software"*
*__ Bjarne Stroustrup*

Our reliance on software to automate things has resulted in including software in every industry from healthcare, education, aviation to defense.  But the question remains how often programs don't correctly work due to software bugs? And, can we develop better more robust software? Most of software bugs come from data mishandling (data conversion), unnecessary code and unsecure input and output handling.

AspectJ and C# are both strongly-typed languages and were designed to be secure; hence their type systems play a very important role in developing secure programs by ensuring the type-safety. AspectJ is statically typed language, but it does support some kind of dynamic typing such as down casting. Since .NET 4.0, C# supports dynamic-type checking after the introduction of the *dynamic* keyword. Dynamic type-checking can be disabled in C# by using an unsafe code block marked as unsafe (using the keyword *unsafe*)[16].

AspectJ and C# also derive a set of features from C and C++ except pointers which has eliminated a major problem: manual memory management.
The JVM and the CLR have similar run-time services. As part of their specification, they manage code execution, automatic memory management (Garbage Collection, memory allocation and de-allocation) and exception handling. Garbage collection is an automatic memory management mechanism; it eliminates some bugs related to manual memory management such as dangling pointers and double free bugs (freeing a pointer twice) [17].

Both AspectJ and C# provide an exception handling mechanism allowing preventing application from crashing at run-time. Contrarily to AspectJ/Java primitive data type, C# has value types (used to store values) which are objects found in the *System* namespace [18].
Run-time bounds checking are provided by both Aspect and C# [19]. Bounds checking are used to control data structure manipulation such as arrays.
To ensure total safety in AspectJ applications, programmers have to be well experimented when expressing crosscutting. A minimal logical mistake can result in total system failure [20].

| Feature | AspectJ | C# |
|---|---|---|
| Memory management | Yes, provided by the JVM | Yes, provided by the CLR |
| Bounds checking | Yes, for arrays and data structure. | Yes, also for array bounds (raise *IndexOutOfRange* exception), buffer overflow. It can be disabled in C# [25] |
| Static Type checking | Yes, at compile time. | Yes at compile time. |
| Dynamic Type checking | Partial dynamic checking, i.e down casting. Also, when using reflection. | Yes, since C# 4.0 (.NET System.Dynamic namespace) |
| Type safety | Yes, Based on Java's access rules. | Yes, C# is strongly typed. The *unsafe* keyword can be used for allowing pointers use. |
| Exception handling | Yes, exceptions can be thrown in pointcuts. Handlers are used to catch exceptions in AspectJ. Within try-catch/finally block in Java code. | Yes, Within try-catch/finally block. Multiple catch blocks is supported in C#. |
| Compiled/Interpreted | Compiled, and the Java-byte code interpreted by a JVM compliant. | Compiled, Interpreted by the CLR |
| Conditional compilation | Yes (using SCoPE compiler, conditional pointcut evaluator) [23] | Yes (using preprocessor directives)[21] |
| Assertions | Yes, temporal assertions at run-time[24] | Yes, Managed code assertions [22] |

### 2.1.3 Web applications development

ASP.NET/C#: As part of the .NET Framework, ASP.NET is a new web development model and the successor of Aspect Server Pages (ASP) technology. ASP.NET enables developers to build a wide variety of secure server-side and browser-based applications such as e-commerce, dynamic websites, and e-learning applications. ASP.NET is compiled and run on the .NET platform where it takes advantage of the CLR secure and powerful environment and its multi-programming languages support. ASP.NET is more and more widely used in software industry thanks to the Visual Studio .NET IDE which is excellent for rapidly building variety of applications in a Rapid Application Development (RAD) environment.

C# language is mainly and heavily used to develop such applications due to its power, features and simplicity. There are plenty of open source web applications project written in C# and ASP.NET that can be found in the Codeplex [26] repositories such as Customer Relationship Management (CRM) "e.g ASP.NET CRM" [27], Content Management System (CMS) and e-commerce such as *nopCommerce* which is the best available open source .NET based solution, also written in C# [28].

ASP.NET has received a lot of criticism because it was first released with a
Technology called Web Forms technology which aims to separate HTML code and stuff from application logic, and reduce code by using the data binding capabilities of the server-side .NET control. Web Forms applications are not flexible, but difficult to test and not extendable because they use what is called server-based forms. ASP.NET MVC was released first in 2007 as an alternative to the Web Forms that enables creation MVC-based web applications.

AspectJ/J2EE: *Java 2 Platform Enterprise Edition* (J2EE) technology is not only suited to build dynamic websites but also enterprise applications and server applications. Its simplicity, complete portability on any operating-system and design made it widely popular. J2EE includes/uses APIs such as Java Database Connectivity (JDBC) Database Connectivity which allows applications to interact with databases, Remote Method Invocation (RMI) for building distributed applications and many more. Java security and portability makes J2EE to be the first choice for developing online banking systems and transactional/e-commerce web applications.

AspectJ can be applied/coupled to J2EE in the development of web applications where aspects are used to isolate secondary concerns (such as error-handling and logging), improve the flexibility of your code and reduce the code source size [29]. AspectJ/J2EE applications are clean, easy to maintain and to debug. *SpringSource* dm Server is open source solution which is used to develop enterprise applications that use AspectJ in order to either simplify enterprise application or implement different crosscutting functionalities. AspectJ is heavily used with Spring Framework [30]. AspectJ can be used in Java Servlet [31]

AspectJ/J2EE vs C#/ASP.NET:

J2EE is more mature than ASP.NET [32], thus, it supports AOP using different tools/language such as AspectJ and JBoss AOP framework and it has been adopted by various firms and foundation such as IBM, SAP, Oracle and Apache Software Foundation for developing a wide range of server applications such as *WebSphere*.

Because of Microsoft's products bad reputation in term of security, Java applications became more popular to adopt in developing critical web applications, but it seems that the new .NET infrastructure has been evolved where new security features have been introduced to the .NET framework and CLR environment such as role-based security and authentication.
ASP.NET applications are deployed on Microsoft's Internet Information Services (IIS) and requires the .NET platform to be installed.

**Web applications: AspectJ vs C#**

| Feature | AspectJ | C# |
|---|---|---|
| Dynamic Web Pages | Yes, injection Aspectj code in JSP pages | Yes, as *code behind* ASP.NET web pages. |
| Web Server | Tomcat, JBoss, GlassFish | IIS on Windows OS Mono (Apache, Nginx) |
| Web Framework/Libraries | Using J2EE, Spring or Struts framework. | Using .NET Framework on Windows and Mono ASP.NET Linux and Unix-based OSs. |
| Session management | Yes, i.e using *Spring* which provides *SessionManagementFilter* class and concurrency control [33]. | Yes, provided by ASP.NET State management. |
| Security | Yes, Java web applications run in what is called container, Each application has its container. Thus, the JVM add | Yes (ASP.NET Security Architecture [sr-4]) |
| Model-View-Controller | By adding AspectJ's aspects to J2EE/Spring/Struts applications. | ASP.NET MVC 1.0/2.0 |
| Database interaction | Yes, using the JDBC library | Yes, using ADO.NET library |
| Secure Sockets Layer support | Yes using OpenSSL library. | Yes, SSL certificate can be generated and used to on IIS server. |

---

## 2.1.4  Web services design and composition

AspectJ doesn't provide an API for building web services, but it can be very useful in combination with other framework like spring [30] for building web applications. Web Services, can be seen also as distributed distant programs that communicate together via HTTP.

A distributed system is a collection of independent computers that appears to its users as a single coherent system [35]. One program can be piped to another and so on resulting in kind of a complex program's network. As results, many issues will obviously appear and force programmers to take in consideration. Some of these issues are: communication, fault tolerance, synchronization and security [36].

AOP is the solution to reduce distributed systems' complexity and improve their efficiency by modularizing crosscutting concerns mentioned before. A group of aspects can be added to an existing system with a main task to observe the system's pipeline. For example, "when part of the system fails, a new behaviour emerges which uses certain aspects along with the rest of the system" [36].

.NET Framework provides the *System.Web.Services namespace* which enables programmers to write XML Web Services (WS) that are available over the Web. Windows Communication Foundation (WCF) is an API for building service-oriented applications (SOA) in a unified programming model; it was released with .NET 3.0. WCF is based on the SOA principles which enable distributed applications to communicate together via SOAP messages. Visual Studio IDE is a great tool to create Web Services: the WSDL file is generated and maintained automatically as soon as the code changes, also it provides tools for testing and debugging a WS at run-time [37]. Microsoft BizTalk Server provides services to services that enable different distributed applications/web services to communicate together. BizTalk is based on adapters that support different communication protocol such as HTTP, FTP and SOAP [37].

Unfortunately, the .NET solution is not flexible enough and it is far from benefiting from the AOP approach to modularize crosscutting concerns. There are no mainstream AOP solutions that are suitable for the .NET framework. AspectJ is a cross-platform language that can run on any platform and can be coupled with Java's technology for building either web applications or distributed ones.

*Web Services: AspectJ vs C#*

| Feature | AspectJ | C# |
|---|---|---|
| Web Services including(SOAP, WSDL, UDDI) | Yes. i.e using Apache XML-RPC for Java library [38] or Spring Web Services, JAX-WS 2.1) | Yes. using ASPT.NET web services and .NET's *System.Web.Services namespaces* for Windows-based WS [42] and Mono project's Web Services on other OSs (Linux/Unix, Mac) |
| Web services security | Yes. HTTPS, above libraries support "XML-based standards" such as WS-Policy, WS-Security and WS-Transfer standards) Web Services pipeline watching. | Yes, HTTPS Encryption and signing with SSL[39], WS-Policy, WS-Security standards support) |
| Web Services composition | Using AspectJ and J2EE framework, two or more web services can communicate together by linking them as a network. | Using ASP.NET web services. Also a network of web services is possible by making two or more WS talking together. |

## 2.1.5  OO-based abstraction

AspectJ is an aspect-oriented programming language; it is an extension to OO.
"Aspects are the central unit in AspectJ" [2]. Aspects and classes have many similarities: they share some common elements such as methods and fields, but also aspects introduce the one that is related to AOP: pointcuts and advices. Like classes, an *aspect* can be extended, but only *abstract aspects* can be extended. Unlike classes, the main goal of using aspects is to modularize crosscutting concerns.

C# is a pure object-oriented programming language; in C# everything is an object. Every declared type (object) implicitly or explicitly inherits from the *System.Object* class. It supports all object orientation concepts: abstraction (classes), encapsulation, polymorphism and inheritance. Also, C# supports interfaces, abstracts classes and has introduced partial classes (the class's source code can be split over two or more files) [41].
Multiple inheritance is not supported in C#. C# provides *properties*, also called accessors (Java's getter and setter equivalent) which are class members that provide access either to read or write private fields' values.

C# is suitable to build object-oriented-based applications. C# is designed to be an object-oriented while AspectJ is designed as an aspect-oriented.

### OO-based abstraction, Aspectj vs C#

| Feature | AspectJ | C# |
| --- | --- | --- |
| Central unit | Aspects, abstract Aspects | Classes, interfaces, abstract classes |
| Structural elements | Pointcuts, Advices, Inter-Type declarations for declaring aspect's members (fields, methods, and constructors) | Constructors, Destructors, Methods, Fields (attributes), Delegates, Properties, Indexes, Events, Finalizes, Operators, Nested Classes. |
| Multiple inheritance | No (Aspect Inheritance) | No, can be done by using Interface implementation as work around. |
| Polymorphism | Aspectual polymorphism [44]. Method overloading, method overriding, virtual method. By default AspectJ methods are *virtual* (non-static methods). Using Inter-type declaration. | Yes, through inheritance, based on *Base Type* and *Sub Type* relation. Method overloading, method overriding. C# provides the *virtual* Keyword which is derived from C++. |
| Partial Classes | No, but it provides Inter-type declarations (aka *open classes*) as alternative. | Yes, separation of code (class definition split into two or more source files. |
| Structs support | No, doesn't support Structs declaration. | Yes, derived from C++ with enhanced features. |
| Instantiation | Aspect instantiation: not | Yes, Class instantiation |
| | directly instantiated. You cannot use the *new* keyword to instantiate an aspect. | using the *new* keyword. |
| Extension/Inheritance | Yes, "Aspect extension" Aspects can extend classes and implement interfaces, but Aspects can extend only *abstract* Aspects. | Yes, "class extension". A class may extend another class and may implement one or more interface. Extension methods feature is supported since C# 3.0 but it requires a static class and static method to do so [43]. |
| Accessibility control | Yes, *privileged* for Aspects, *private*, *public* and *protected* for inter-type members. | Yes, C# provides *private*, *public*, *protected*, *internal*, *sealed* and *protected internal* keywords. |

## 2.1.6 Reflection

In AspectJ, join point information is accessed via reflection. There are two types [2]:
- *dynamic* information (dynamic crosscutting): this type of information changes after each call of the same join points. Dynamic information result in the program's dynamic flow.
- *static* information (static crosscutting): is said fixed and doesn't change between multiple execution.

AspectJ provides three objects to be used in advice body:
thisJointPoint, thisjoinPointStaticPart, thisEnclosingJoinPointStaticPart. These objects are similar to *this* keyword in C# or Java which provides access to the current object being executed [2] [45].

Since AspectJ 5, a new reflection API is provided similar to *java.lang.reflect* package. To use this API, code should be compiled by the AspectJ 5 compiler and runs under Java 5 environment [46].

.NET components are pre-compiled modules with a .DLL extension (Dynamic Link Library), they are also known as assemblies which are the smallest unit of execution in the CLR environment. Thanks to the .NET CLR dynamic services, especially the loader manager, these components can be loaded at run-time into the memory and used by different applications directly or through reflection [47].

C# allows reflective programming in different manners. *Reflection* in C# can be used for loading assemblies (DLLs) in order to instantiate some of their types. *Custom Attributes,* which are a special form of customizable code annotation and an extension to the reflection model "similar to Java 1.5 annotation", contain metadata information that can be attached to classes or methods and they are accessed at run-time through reflection.

The .NET framework provides the *System.Reflection.Emit namespace* for code generation at run-time "emitting dynamic methods and assemblies". This namespace provides developers many capabilities such as define assemblies at run-time and saving them to disk, define new types and create instances of these types also at run-time and so on [47].

There are some Java's equivalent libraries available that can be used in AspectJ applications such as ASM *http://asm.ow2.org* and BCEL *http://jakarta.apache.org/bcel*. These libraries provide similar services to the .NET's *Reflection.Emit namespace* for generating byte-code at run-time.

Accessing information dynamically is not suggested in certain applications; reflection has a poor performance and it is unsafe without static type checking [2].

C# provides more features to do reflective programming than AspectJ. AspectJ is not flexible enough to create or modify *aspects* on the fly (at run-time). For code security reason, AspectJ's joint cuts have to be well tested since they are plugged in a way to the program flow which is difficult to do using reflection.

**Reflection, AspectJ vs C#**

| Feature | AspectJ | C# |
|---|---|---|
| Access to program's metadata | Yes, since AspectJ 5.0. Annotation-based development style. | Yes, accessing Attributes through reflection using *System. Reflection* namespace. |
| Generation code at run-time | AspectJ doesn't provide such feature. Can be done using the following librairies. ASM *http://asm.ow2.org* and BCEL *http://jakarta.apache.org/bcel*. | Yes, using .NET's *System.Reflection.Emit* namespace. |
| Dynamic method invocation | Using thisJoinPoint to access the current join point for the advice through reflection. | Yes, method invocation at run-time through reflection [48]. |

## 2.1.7  Aspect-orientation

AspectJ, natively, is an aspect-oriented programming language. AspectJ implements the joint-point model and inter-type declaration implementation. AspectJ has been adopted as a mainstream programming language and there are numerous articles that have been written about it [49]. It is an active project and it is present in many important development projects and domains such as banking systems and web transactional applications where security and logging are implemented in a modular way using aspects [4].

C# doesn't officially provide an implementation of AOP. There are some open source projects [51] but it seems there are no longer active. As it mentioned on the AspectDNG project's website: "the project is not supported any more due to .NET rapid changes" [50]. Due to the .NET platform closed source and its proprietary license.

*Aspect-orientation, AspectJ vs C#*

| Feature | AspectJ | C# |
| --- | --- | --- |
| Aspect-oriented programming | Natively, it is an aspect-oriented programming language | Partially, there are some no-mainstream projects implementing AOP for C#. |
| Modularity | Isolate crosscutting concerns in a modular way. | It is Difficult some times impossible to isolate crosscutting concerns from the program's business logic. |
| Code reusability | Can produce pure uncoupled code | Can't isolate code that crosscut over modules. Partially, can be done using DLLs in order to reduce code coupling degree. |
| Security | Securing UI by detecting Single-Thread UI's rule and web services pipelines failure. | Not applicable since there is no mainstream AO solution for C#. |
| Errors handling/Logging | Cleaner way to handle and log exceptions | Exceptions are handled in *try-catch* blocks and logged within this scope. |

## 2.1.8 Functional programming

C# and AspectJ are both *imperative programming* language which means a program written in one of them consists of sequence of steps/commands that should be executed one by one (step by step), so programs can be seen as a sequence of operations. C# supports generic classes, *generic delegates* which are object-oriented and type-safe functions pointers (similar to function pointers in C++) and anonymous methods [52] which is an advantage over AspectJ (see *Example #1 below*).

AspectJ doesn't support *delegates* but does support generic classes. Also, AspectJ doesn't allow passing a function as parameter to another function. Some functional programming style in AspectJ can be done either by using libraries or by using higher-order functions and lazy functional programming [53]. Interfaces and inner classes can be used as a work around to pass functions as parameters [54]. This practice is not encouraged since this will result in a code much bigger and complex which is not easy to understand by someone who is not familiar with functional programming.

C# has more features for supporting functional programming than AspectJ. Language integrated query (LINQ) has been released with the .NET Framework 3.5 which provides a powerful way to query data in a functional programming style [55].

Recursion is another concept of functional programming that requires functions, also methods, to invoke themselves. C# does support recursion without issues [56] while using it in AspectJ has resulted in major problem: infinite recursion. "if no special precautions are taken, aspects which advise other aspects can easily and unintentionally advise themselves." [57]

***Functional programming features: AspectJ vs C#***

| Feature | AspectJ | C# |
|---------|---------|-----|
| Type inference | No, planned for Java 7 | Yes, functional programming's features have been introduced in C# 3.0 |
| Lambda expression | Using interfaces as a work around. | Yes, since C# 3.0 using LINQ library and *System.Linq.Expression* namespace |
| Anonymous methods | No, use inner classes as work around | Yes, since C# 2.0 where functional programming features has been introduced. |
| Higher-order functions | Using interfaces as work around | Yes using C# delegates. |
| Closure | No, not popular enough an open source project for OpenJDK [58]) | Yes, since C# 3.0 |
| First-class functions(delegates) | No, AspectJ doesn't support delegates, lazy functional programming can be used as work around [53]. | Yes, delegates are C++'s function pointers equivalent. |
| Recursion | Not secure as in C# [57] due to infinite methods loop. | Yes it is fully supported. Can be done using C#'s methods. |

*Example #1: An example showing how to use anonymous functions in C#*

```csharp
using System;

namespace Program
{
    /// <summary>
    /// <autor>Sleiman Rabah</autor>
    /// </summary>
    class Program
    {
        static void Main()
        {
            // Anonymous function to calculate the square of
            // two integer.
            //the first int is the x' type
            //the second int is the return type
            Func<int, int> square = x => x * x;
            Console.WriteLine(square(7)); // Will output 49
        }
    }
}
```

## 2.1.9  Declarative programming

AspectJ and C# enable programmers to specify metadata that can be associated to some elements in the source code and can be retrieved/consulted at run-time through reflection. They both provide a mechanism to define declarative tags that can be added to classes, methods or variables.  This mechanism enables declarative programming style to be used in both languages. It consists in declaring data and rules rather than coding them explicitly.

AspectJ provides what is called Annotations [59] while C# provides Attributes [60]. Programmers can use pre-defined Annotations or Attributes and also they can create their own customized ones.

C# does better support declarative programming than AspectJ since it provides ways to mix imperative and declarative code using LINQ. LINQ can be used to query data from databases (LINQ to SQL), XML (LINQ to XML) and many more [55] (*see Example #2 below*).

Also, C# 4.0 has introduced dynamic programming with the *dynamic* keyword which can be used to work on data type.

Some new technologies have introduced new ways to build UIs based on declarative programming principles such as XAML which was invented by Microsoft for the .NET Framework [61]. Also there are plenty of open source project which are equivalent to XAML and provide the same features as XAML does. Among them we find YAML [62] which is widely used to build AspectJ Swing GUIs. Using these languages, developers can define UI's elements and their properties outside the source code.

**Declarative programming, AspectJ vs C#**

| Feature | AspectJ | C# |
|---|---|---|
| Tags on methods/fields/classes | AspectJ/Java Annotations | C# Attributes |
| LINQ equivalent? | Additional libraries, lamdaJ [63] | Yes, *system.data.linq* library should be used and referenced. |
| Dynamic programming (Dynamic type checking) | No, AspectJ doesn't provide such features. | Yes – since C# 3.0 (*var* keyword) and C# 4.0 (*dynamic* keyword) |
| Declarative programming based on XML [41] | YAML, SwiXML | Using XAML technology which is available since .NET 3.0 and WPF |

*Example #2: An example showing declarative programming using C# and LINQ*

```csharp
using System;
using System.Collections.Generic;

// Importing the LINQ DLL
using System.Linq;
using System.Text;

namespace LinqSample
{
    /// <summary>
    /// <autor>Sleiman Rabah</autor>
    /// </summary>
    class Program
    {
        static void Main(string[] args)
        {
            // Check if a number in a list of integer is odd using LINQ
            // if yes it will result in a list of odd numbers
            List<int> collection = new List<int> { 1, 2, 3, 4, 5, 6, 7 };
            var results = collection.Where(num => num % 2 != 0);
            foreach (var num in results)
            {
                Console.WriteLine("Results:" + num); // Will output 1,3, 5, 7
            }

            // Using "SQL" style to print element from an array
            // which are <= to 7 using LINQ
            int[] numbers = { 3, 2, 13, 15, 5, 7};

            var lowNums =
                from n in numbers
                where n <= 7
                select n;

            Console.WriteLine("Numbers <= 7:");
            foreach (var num in lowNums)
            {
                Console.WriteLine(num);
            }

        }
    }
}
```

## 2.1.10 Batch scripting

Console applications, or command line applications (CLA), are programs that take commands and arguments as input, process them and output result/display messages at the command line. CLA often are executed from the command line and are said text-based applications which use a *TUI* (stands for Text User Interface) to interact with users/other programs using system calls in order to read their commands from the command line and achieve certain tasks like processing files, querying databases, etc. Before commands processing, commands and arguments are checked and validated to be sure that they meet certain constraints. Example of validation can be arguments' data type, commands names whether if they are syntactically correct, arguments count for a given command and so on.

CLA can be done using both AspectJ/Java and C# [64] they both provide this possibility: commands and arguments are received as parameters at the program's entry point which is the main method, usually they are received as an array of strings.

The only difference is that the main method name has to be capitalized due to the C#/.NET naming convention (Pascal case) [65].

The method main signature is as follow:

    **AspectJ:** *public static void main(string[] args)*

    **C#:** *public static void Main(string[] args)*

Building CLA using both languages is a secure solution since they both are type-safe and they provide an exception handling mechanism which can be used when processing commands. While processing a file for example, an exception can be thrown and the program will stop executing.

These type applications have some limitations [66]: they are not flexible enough because they are compiled and usually deployed as executables with extension *.exe* in case of *C#* and a *.jar* in case of Java. If any changes in the code have been done in an existing application/program, it has to be recompiled and re-deployed for the changes to take place which is often a disadvantage since all users have to be notified about changes and required to update their installed copies. There are also other limitations: they are not user-friendly and they usually target a set of trained/experimented users. On the other hands, because they are efficient and they run faster, console applications are suitable for applications that need bigger processing resources especially because there are no graphics rendering/drawing and so on [66].

Also, both languages AspectJ/Java and C# enable to developers to write programs that can invoke/execute external commands and programs. One program can execute another program while it runs, to do so, a program creates a process to run the desired program (as illustrated in *example #3 below* ).

    Pipeline can be used to redirect CLA output to another command or CLA. For example, under Linux operating system, we can run a CLA and use the *grep* command. *e.g*

    [prompt]$  java myProgram | grep [search pattern]

AspectJ can be coupled with Java to build a wide variety of CLA or added to an existing application in order to advantage of AOP in such application. Exception can be logged in files using AspectJ to be later analyzed and debugged.

CLA written in AspectJ or C# can be automated and scheduled to be executed at a given point in time. Script automation is a useful and common way to automate things such as daily reports generation, job alerts and so on. This can be done using Scheduled Tasks tool in Windows or using *crontab* on Linux/Unix system.

AspectJ is more suitable/better to build and deploy portable (cross-platform) CLA since such applications can profit from the AOP's features and AspectJ's portability.

*Batch scripting, AspectJ vs C#*

| Feature | AspectJ | C# |
|---|---|---|
| Run external programs/external commands | Yes, using Java's Process and Runtime classes. | Yes, using Process class in *System.Diagnostics* namespace. |
| Tasks automation | Yes, can be executed as stand alone and executed on a scheduled basis. | Yes, can be executed as stand alone and executed on a scheduled basis. |
| Accept command line arguments | Yes as parameters for the *main* method. i.e *public static void main(string[] args)* | Yes as parameters for the *Main* method. i.e *public static void Main(string[] args)* |
| Need to be recompiled after changing the code source | Yes, code should be recompiled in order to source code changes take effect. | Yes, code should be recompiled in order to source code changes take effect. |

*Example #3 A simple example showing how to execute a program from another program in C#*

```csharp
using System;
using System.Diagnostics;
using System.ComponentModel;

namespace LaunchProcessSample
{
    /// <summary>
    /// <autor>Sleiman Rabah</autor>
    /// </summary>
    class ProcessLauncher
    {
        public static void Main(string[] args)
        {
            // instantiate the System.Diagnostics.Process class
            Process myProcess = new Process();

            try
            {
                // Tell whether the new process will be executed
                // using Shell or not
                myProcess.StartInfo.UseShellExecute = false;
                // Fill out the process name to be started:
                // often it is a program e.g: notepad
                myProcess.StartInfo.FileName = "notepad.exe";
                myProcess.StartInfo.CreateNoWindow = true;
                // Run/launch the process
                myProcess.Start();
            }
            catch (Exception e)
            {
                Console.WriteLine("A problem has occured while
                            starting the process " + e.Message);
            }
        }
    }
}
```

## 2.1.11 UI prototype design

Abstract Window Toolkit (AWT) was the first Java's technology/component set for building graphical user interfaces (GUI). AWT UIs were very slow to run with limited functionalities due to the small number of provided components. Just-in-time compilers (JIT) were introduced in order to improve applications' execution and performance and Java Foundation Classes (JFC) technology was released also to improve productivity and to provide a new rich set of components. JFC provides a group of features and graphic functionalities that can be mixed together for building powerful and extendable desktop applications. Among these features are internationalization which enables to build applications that support different languages and cultural conventions, pluggable look-and-feel which allows to choose a look among wide choices of *look and feel* and *themes* on different platform/operating systems (GTK+, Motif, Windows, Macintosh) and many more. As part of JFC, Java Swing is an API which provides a set of GUI components such as buttons, tables, text box and features for printing, drag and drop, customizing components layout and many more [67].

AspectJ-based Swing enables developers to build either document interface (MDI) using the JdesktopPane container and JinternalFrames or single document interface (SDI) GUIs.

The NetBeans IDE provides a user interface designer (UID) allowing developers to design/prototype GUIs. Also, there are some *plug-in* for the Eclipse IDE which can be used to design interfaces with Swing.

On the other hand, the .NET Framework was first released with the Windows Forms (WinForms) API as a part of it. WinForms is not flexible enough and is said platform-dependent since it is built upon the Windows API (the famous WinAPI or Win32 API) which resulted in a bad design and it doesn't support the model-view-controller architecture. To resolve this problem and push further the UI development, Microsoft has introduced the Windows Presentation Foundation (WPF) a new revolutionary technology for building platform-independent GUIs, [68]. WPF provides much better functionalities than Swing including animations, data templates, effects, templates and many more. Also, WPF is based on the XAML (Extensible Application Markup Language) which is a new declarative language where developers can either manually or using the UI designer describe/declare the properties of the UI's controls and components without the use of the traditional imperative programming [68]. The main goal of XAML is to isolate the graphical content from the code which resulted in a cleaner and better understandable and maintainable code.

WinForms and WPF are not part of .NET's Base Class Library (BCL) which means they are not standardized as of the ECMA/ISO standards. To build applications using the WinForms the *System.Windows.Forms.dll* library should be used and referenced, same for WPF where the *System.Windows.dll* is required. AspectJ-based Swing is built-in Java and no libraries/references are needed to do so (it can be directly imported using the *import* keyword i.e *import javax.swing.\**)[68].

Visual Studio C# comes with a user interface designer (UID) and Toolbox which enable users to add control/components such as buttons, text box, menus, toolbars and so on to the design interface and set their properties and their events handlers.

Swing, WinForms and WPF are popular and used in different development projects, but nowadays, Microsoft's technologies are more popular and suitable to build desktop applications. WinForms or WPF have a high interoperability with the MS SQL Server database system. C# has a lot of advantages over AspectJ-based Swing for prototyping UIs, a WPF code can be simply embedded in a web ASP.NET application without any configuration since WPF can be executed in any browser [69]. There are some discussions and rumors over the      web that AspectJ/Java Swing is dead and it is no longer used in development projects: before Java 1.5, Java Swing left a bad perception behind.

Java Swing library has a single thread-safety rule: it uses a single thread to access all UI's components through the event-dispatching thread [2, chap.11]. This restriction is a problem in complex applications, for example, if a thread is performing database queries and you want to update the UI components. AspectJ resolves this problem: you can write an *aspect* to detect any violations to this rule [2, chap.11].

| Feature | AspectJ | C# |
|---|---|---|
| Graphical user interface | Yes, using Java's built-in Swing and AWT libraries. | Yes, using .NET's Windows Forms and WPF libraries. Built on top of the Base Class Library |
| Built-in? | Yes, part of Java Foundation Classes | You need to import/reference *System.Windows.Forms.dll* for WinForms and *System.Windows.dll* WPF. WPF only on Windows platform [70] |
| Look and feel | Yes such as GTK+, Motif, Windows, Macintosh, etc themes. | Not flexible enough in WinForms. Full customizable UI support in WPF |
| 2D and 3D support | Yes – Java 2D API | Yes, using WPF framework |
| Drag and Drop support | Yes, using *Java.awt.dnd* package. | Partially, only using WPF framework [71] |
| Components' Layout Management | Yes, i.e Border/Grid layout, etc. | Yes, this can be done automatically using Visual Studio's UI designer. |
| Multiple document interface (MDI) | Yes, provided by AWT | Yes, using, *System.Windows.Forms.MdiLayout* class. |
| Model-View-Controller | Yes in components such as JTable and JList, etc. | WinForms no, WPF Model-View-ViewModel [72] |
| Declarative GUI development [41] | Glade XML, SwiXML: not popular yet as much as XAML | Using XAML technology |
| Performance | Relatively Slow | Improved execution on windows Fast |
| IDE/UI designer | NetBeans, Plugins for Eclipse IDE | Visual Studio C# and SharpDevelop for Windows, MonoDevelop . |
| Rich Web UI Development | Yes, using JavaFX framework. | Silverlight framework on windows and Moonlight implementation for Linux and other Unix-based OSs. |
| Deployment | Jar archive (.jar files) | .NET Assemblies: (executables and libraries, .exe and DLLs) |
| Single thread of execution | Yes, Swing library's single-thread rule, thread-safe[73] | Windows Forms (Not by default, but by applying the *STAThread* attribute to the Main method). WPF use by default a single thread of execution. |
| Event handling mechanism | Only Event Listeners, provided by *java.awt.event* package | Yes, Event Handlers and Delegates support |

## 2.2    C++ vs Groovy

## 2.2.1  Source code size

Hello world in C++

```
#include <iostream>

using namespace std;

int main(int argc, char *argv[])
{
cout << "Hello, World!\n";
}
```

Hello world in Groovy

```
println "Hello, world!"
```

## 2.2.2  Default more secure programming practices

First of all, C++'s type strength is strong. C++ language places severe restrictions on the intermixing that is permitted to occur, preventing the compiling or running of source code which uses data in what is considered to be an invalid way. Secondly, C++'s type safety is unsafe. C++ can discourage or prevent some type errors. For example, C++ can prevent result from attempting to perform operations on values that are not of the appropriate data type [74]. Thirdly, C++'s type checking is static. C++ allows many type errors to be caught early in the development cycle. It eliminates the need to repeat type checks every time the program is executed. Program execution may also be made more efficient by omitting runtime type checks and enabling other optimizations. But it has some disadvantage, E.g, C++ will reject some programs that may be well-behaved at run-time, but that cannot be statically determined to be well-typed [74]. Finally, C++ doesn't have garbage collector, in some sense, it will cause dangling pointer bugs, double free bugs and certain kinds of memory leaks.

First of all, Groovy's type strength is also strong which means that Groovy specify one or more restrictions on how operations involving values having been executed. Secondly, Groovy's type safety is also safe. It is just like C++ [74]. Finally, Groovy's type checking is dynamic. Compare to C++, Groovy is more flexible but fewer guarantees, because Groovy accepts and attempts to execute some programs which may be thought as invalid by C++. Groovy may result in runtime type errors. Compare to C++, Groovy make fewer "compile-time" checks on the source code. On the other hand, Groovy' type checking is more sophisticated [74].

## 2.2.3  Web applications development

C++ uses Wt and CppCMS as an open source web application framework to support the development of web applications. It supports Ajax ,template framework ,caching frameworks and security framework, but doesn't support DB migration framework [76]. It also support MVC framework [75]. It provides SQL library and Wt::Dbo for ORM and uses boost. test in testing framework [76] [77].

This is a example of web application in C++:

```
// Displays the current date and time in a Web browser.
#include <iostream>
using std::cout;

#include <ctime> // definitions of time_t, time, localtime and asctime
using std::time_t;
using std::time;
using std::localtime;
using std::asctime;

int main()
{
    time_t currentTime; // variable for storing time

    cout << "Content-Type: text/html\n\n"; // output HTTP header

        // output XML declaration and DOCTYPE
    cout << "<?xml version = \"1.0\"?>"
        << "<!DOCTYPE html PUBLIC \"-//W3C//DTD XHTML 1.1//EN\" "
        << "\"http://www.w3.org/TR/xhtml11/DTD/xhtml11.dtd\">";

time( ¤tTime ); // store time in currentTime
    // output html element and some of its contents
    cout << "<html xmlns = \"http://www.w3.org/1999/xhtml\">"
        << "<head><title>Current date and time</title></head>"
        << "<body><p>" << asctime( localtime( ¤tTime ) )
            << "</p></body></html>";
    return 0;
} // end main
```

Groovy uses Grails as an open source web application framework to support the development of web applications.Grails supports Ajax ,template framework and caching frameworks, but dosenot support form validation framework.It uses active record patten technology for MVC framework.It provides GORM, Hibernate for ORM and uses multiple plugins in DB migration framework [78].

Compare to C++,Grails will take more time to write a program and is more complexity.In some condition,the speed of excution is slow. It is neither high performance nor easy to program.But Grails is more secure and have more feature [79].

## 2.2.4 Web services design and composition

Groovy, as befits a sprightly young scripting language for the Java Virtual Machine, can do XML in a way that is relatively free of bloat and allows the developer to focus on the real problem at hand [80]. Groovy also support WS,SOAP and WSDL [81] [82] [83]. GroovyWS comes with two sets of APIS that are briefly described below using a simple example:Publishing a web-service and Consuming a web service [84].

Compare to C++ in XML, the best part about the Groovy code is not that it is shorter than the equivalent C++ code — only five line. The Groovy code is that it is far more expressive. You can write directly with the XML.The following code show parsing an XML in Groovy,with shorter code:

```
def langs = new XmlParser().parse("languages.xml")
println "type = ${langs.attribute("type")}"
langs.language.each{
  println it.text()
}

//output:
type = current
Java
Groovy
JavaScript
```

C++ is implemented with gSOAP toolkit for SOAP/XML Web services and generic (non-SOAP) C++ XML data bindings. The toolkit analyzes WSDLs and XML schemas (separately or as a combined set) and maps the XML schema types and the SOAP messaging protocols to easy-to-use and efficient C and C++ code 85. It also supports exposing (legacy) C++ applications as SOAP/XML Web services by auto-generating XML serialization code and WSDL specifications. Or you can simply use it to automatically convert XML to/from C++ data. The toolkit supports options to generate pure ANSI C or C++ with or without STL  [86][87].

The gSOAP toolkit is speed, reliability and flexibility, coupled with a proven track record and used by some of the largest technology vendors makes it an ideal platform to develop applications using Web services and XML processing.The gSOAP toolkit offers a comprehensive and transparent C++ XML data binding solution through autocoding techniques [86]. This saves developers substantial time to implement SOAP/XML Web services in C/C++. In addition, the use of XML data bindings significantly simplifies the use of XML in applications by automatically mapping XML to C/C++ data types.

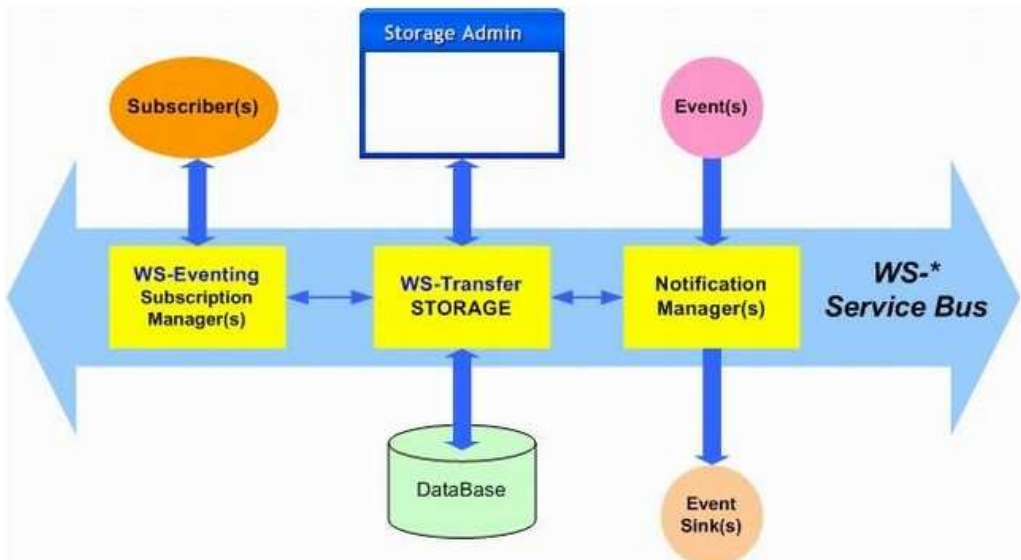

## 2.2.5  OO-based abstraction

Groovy is a pure OO language in which everything is an object. Groovy enables one class to extend another, just as interfaces can. Groovy also support abastract class and interface. In Groovy a class can derive from only one class, but a class can implement multiple interfaces ,in other words, it supports multiple inheritance of types, but only single inheritance of implementation [87] [88].

C++ support OO programming paradigm, it supports inheritance, abstract class and interface. It also supports encapsulation.Full multiple inheritance, including virtual inheritance.Support abstract class,interface and encapsulation [91]. C++ methods can be declared as virtual functions, which means the method to be called is determined by the run-time type of the object [89][90].

## 2.2.6  Reflection

Many programming languages provide built-in reflection mechanism. For example, in Java there is special package java.lang.reflect and in Groovy, org.codehaus.groovy.reflection. But unfortunately C++ doesn't support reflection [92].

Reflection requires some metadata about types to be stored somewhere that can be queried. Since C++ compiles to native machine code and undergoes heavy changes due to optimization, high level view of the application is pretty much lost in the process of compilation, consequently, it won't be possible to query them at run time.

Groovy has the capability of reflection, which enables you to gather information about your script at runtime, including peer into any class or object and obtain detailed information on its properties, methods, interfaces, and constants [84] For example,Groovy can examine the interfaces, public fields and their types, check if a class is a class or an interface. When a class is unknown at compile time,Groovy can use reflection to create objects. Groovy can also use reflection on the reflection classes themselves [93].

## 2.2.7 Aspect-orientation

Bascally, C++does not support aspect-orientation directly.But AspectC++, an aspect-oriented extension of C++ languages,support aspect-orientation. It is based on source-to-source translation, translating AspectC++ source code to C++.It allows modularizing cross-cutting concerns in a single module, an aspect. AspectC++ provides a join point API to provide and access to information about the join point [94][95].

Compare to Groovy, A major different is the modified join point model, which allows to have class, object and control flow join points. The result is a more coherent language design. While it was necessary to make some visible changes to the syntax and grammar of Groovy, we have preserved most language concepts [96]. This should enable users experienced with Groovy and C++ to get familiar with AspectC++ without much effort.However,there are some disavantage on AspectC++.The prototype implementation of an AspectC++ compiler is still in an early stage. The compiler cannot yet deal correctly with all AspectC++ language constructs in all possible contexts [97].

Groovy supports AOP. Adding AOP-like features into Groovy code is as simple as implementing the GroovyInterceptable interface.Any Groovy object that implements this interface will automatically have all of its method calls routed through its invokeMethod()Groovy AOP can provide a shortcut for getter and setter [98].

Tt is easy to implement the powerful features of AOP in Groovy. It's not too difficult to envision how such a capability could be built into an application to handle a lot of the boiler plate code that tends to crowd non AOP applications.

## 2.2.8 Functionnal programming

Groovy is not a functional-only programming language,but it support functional programming. Groovy's functions can be used to define functions which contain no imperative steps [99]. Groovy use lazy evaluation and they typically hide away the hard bits so you don't need to know what magic is happening on your behalf.

FC++ is a library for doing functional programming in C++. The library provides a general framework to support various functional programming aspects such as higher-order polymorphic functions, currying, lazy evaluation, [100] and lambda. In addition to the framework, FC++ also provides a large library of useful functions and data types [101].

## 2.2.9 Declarative programming

Declarative programming is often defined as any style of programming that is not imperative. A number of other common definitions exist that attempt to give the term a definition other than simply contrasting it with imperative programming.Both C++ and Groovy do not support the declarative programming paradigm [102], since they are both support imperative paradigm,which requires an explicitly provided algorithm.In the declarative programming,the program is structured as a collection of properties to find in the expected result, not as a procedure to follow.

## 2.2.10 Batch scripting

Groovy is a scripting language and can be used for command line shell scripting. The simplest way to invoke an external process in Groovy is to use the execute() command on a string [103]. For example, to execute maven from a groovy script run this: "cmd /c mvn".execute().The 'cmd /c' at the start invokes the Windows command shell [105]. The command line is short and clear.Since mvn.bat is a batch script you need this. For Unix you can invoke the system shell [104][84].

C++ is a scripting language and can be used for command line shell scripting.The code use to write command in C++ is more complex and sophosticated and Groovy.Here is the example:

```
string getStdoutFromCommand(string cmd)
{
  // setup
                          string data;
  FILE *stream;
  char buffer[MAX_BUFFER];

  // do it
  stream = popen(cmd.c_str(), "r");
  while ( fgets(buffer, MAX_BUFFER, stream) != NULL )
    data.append(buffer);
  pclose(stream);

  // exit
  return trim(data);
}
```

## 2.2.11 UI prototype design

One of the application development for C++ doing UI is Qt.The Qt toolkit is a C++ class library and a set of tools for building multiplatform GUI programs using a "write once, compile anywhere" approach. Qt lets programmers use a single source tree for applications that will run on Windows 95 to XP.

Compare to Swing/AWT,the program code used in Qt is considerably more intuitive. This is because Swing enforces the use of a Model-View-Controller architecture (MVC) while Qt supports, but does not enforce, such an approach. Qt supports the development of sophisticated user interfaces.Qt has better tool than Swing/AWT. Qt does not enforce particular programming paradigms as Swing/AWT does [106].

Groovy uses Swing/AWT to provide a graphical user interface (GUI) for Groovy programs. The Abstract Window Toolkit (AWT) is Java's original platform-independent windowing, graphics, and user-interface widget toolkit. Swing was developed to provide a more sophisticated set of GUI component than the earlier Abstract Window Toolkit [107].

Compare to Qt,the program code used in Swing/AWT is less intuitive and flexible. Swing/AWT also supports the development of sophisticated user interfaces. Swing/AWT has some problems with runtime- and memory-efficiency which are also problem of Groovy [109]. Swing/AWT may be appropriate for certain projects, especially those without GUIs or with limited GUI functionality. Qt is an overall superior solution, particularly for GUI applications [108].

## 2.3 Haskell vs Java

### 2.3.1 Source code size

Not that the "source code size" criteria is present as one of the official criteria. However, code size may effect the compile performance. While a language are interpreted, the larger code may take more compile time. Moreover, due to the store space and memory size limitation, large source code may also overall running performance. The following code examples show the code size of Haskell and Java.

```haskell
module Main where

main :: IO ()
main = putStrLn "Hello, World!"
```
                    Listing 1: Example Haskell program

```java
public class HelloWorld {

    public static void main(String args[]) {
        System.out.println("Hello world!");
    }
}
```
                         Listing 2: Example Java Program

From these two sample codes, It is clear that source code size of Java (see Listing 2) is longer than that of Haskell (see Listing 1)

### 2.3.2 Default more secure programming practices

I am talking secure programming practices by type system, memory management, and exception handling.

#### 2.1.2.1 Type system

Haskell has a **strong**, **static**, and **automatically inferred** type system based on Hindly-milner type inference.

Firstly, all types in Haskell are strong. Type system of Haskell prevent program from certain kinds of type errors coming from trying to write expressions that don't make sense. Another aspect of Haskell's view of strong typing is that it will not automatically coerce values from one type to another. For example, the following codes are valid in Java:

```java
int a = 10;
double b = (double) a;
```

                         Listing 3: Java explicit casts

But, this is invalid in Haskell, and Haskell compiler will raise a compilation error.

Secondly, in Haskell, type checking is performed during compile-time. The compiler knows the type of every value and expression at compile time, before any code is executed. A Haskell compiler or interpreter will detect type error, and reject our code with an error message running. Haskell's combination of strong and static typing makes it impossible for type errors to occur at runtime.

Finally, a Haskell compiler can automatically deduce the types of almost all expressions in a program. Haskell allows programmer to explicitly declare the type of any value, but the presence of type inference means that this is almost always optional, not something we are required to do.[110]

The Java programming language is also a **strongly** and **statically** typed language. Java require all variables and expressions to have a defined type that is known at compile time. However, other than Haskell, Java supports the use of explicit casts of arithmetic values to other arithmetic types like shown in Codes3. Moreover, Java does not support automatically inferred type system. That means programmers has to declare the types they intend a method or function to use.

### 2.1.2.2 Memory management

Haskell's computation model is very different from that of conventional languages. Data are immutable that means the only way to store every next operation's result is to create new value.[111] Especially, every iteration of a recursive computation creates new values. Therefore, Haskell computations produce a lot of memory garbage - much more than conventional imperative languages. However, GHC is able to efficiently manage garbage collection. The trick is that immutable data never points to younger values. [110] Due to this key property, Haskell uses a simplified garbage collection. At anytime GHC can scan the last values created and free those that are not pointed to from the same set.

Java 2 uses a great automatic memory management in the object lifecycle, thereby keeping the developer from the complexity of explicit memory management. In Java, memory is only allocated to objects, and there is not explicit allocation of memory. At runtime, Java employs a garbage collector that reclaims the resources used by an object once it determines that object is not used in the future. This automatic process makes it safe to throw away unneeded object references because the garbage collector does not collect the object if it is still needed elsewhere. Therefore, in Java the act of letting go of unneeded references never runs the risk of deallocating memory prematurely. [112]

### 2.1.2.3 Exception Handling

Haskell can use monad **Maybe** and **Either** type to achieve exception handling. The Maybe type encapsulates an optional value. A value of type Maybe a either contains a value of type a (represented as Just a), or it is empty (represented as Nothing). Maybe easily handle errors or exceptional cases without resorting to drastic measures such as error [113]. The Either type is similar to the Maybe type, with one key difference: it can carry attached data both for an error and a success ("the Right answer").[114] Appendix 1 show how Maybe and Either handle Haskell exception.

In Java, all exceptions are objects. That is when an exception is thrown, an object is thrown. However, not any object can be thrown -- only those objects whose classes descend from Throwable. Throwable serves as the base class for an entire family of classes, declared in java.lang, and all exception is subclass of Throwable. Throwable has two direct subclasses, Exception and Error. Exceptions are thrown to signal abnormal conditions that can often be handled by some catcher, though it's possible they may not be caught and therefore could result in a dead thread. Errors usually introduce system internal situation, and are thrown for more serious problems, such as OutOfMemoryError, that may not be so easy to handle. In general, Java code only throw exceptions, not errors. Figure 1 show A partial view of the Throwable family.

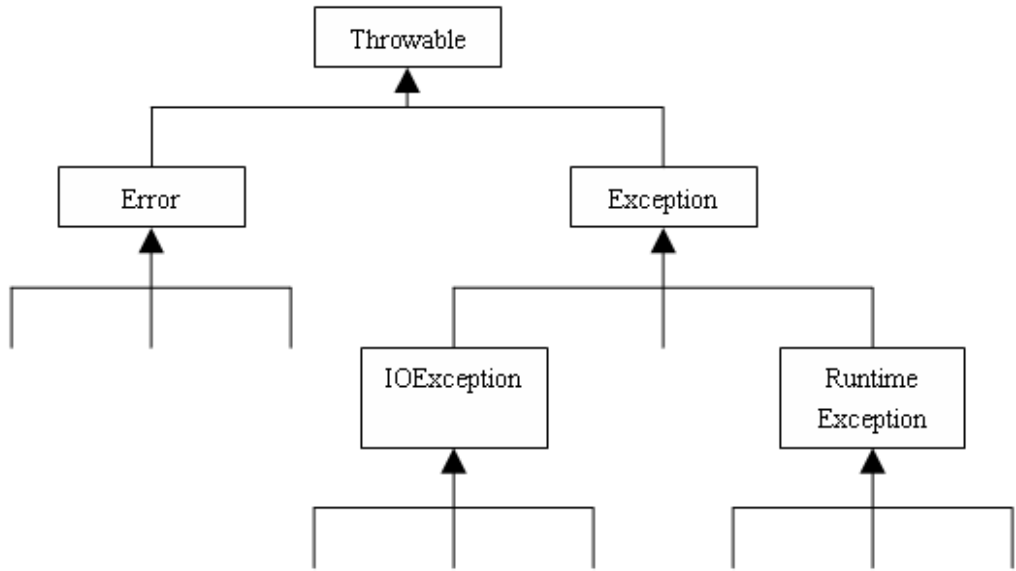

Figure 1: A partial view of the Throwable family.

In addition to throwing objects whose classes are declared in java.lang, Java programmer can also throw objects of own design by declaring new class as a subclass of some member of the Throwable family. To catch an exception, Java uses a try block with one or more catch clauses. Each catch clause specifies one exception type that it is prepared to handle. The try block places a fence around some code that may throw exceptions. If the code delimited by the try block throws an exception, the associated catch clauses will be examined by the Java virtual machine. If the virtual machine finds a catch clause that is prepared to handle the thrown exception, the program continues execution starting with the first statement of that catch clause. Appendix 2 is a sample code to show how Java handle a exception.

### 2.3.3  Web applications development

Web applications are nature distributed applications running on more than one computer and can be access through a network or server. Specifically, web applications are accessed with a web browser and are popular because of the ease of using the browser as a user client. [115]

There are many approaches to Haskell web programming, such as Web Authoring System Haskell (WASH), Haskell Application Server (HAppS), Happstack, and Haskell Server Pages (HSP). WASH is a domain-specific embedded language with type-safe forms handling and threads continuation through client. This gives good back-button and session splitting properties. HApps is a complete system including web server in one program with using of XSLT for output. Happstack is the successor to HAppS. It is a complete system including a web server and database system in one program. It has many template options including HSP, HStringTemplate, Hamlet, XSLT, and more! Haskell Server Pages (HSP) uses preprocessor with dynamic compilation to make XML tags into Haskell expressions. Moreover, another simple, portable, and relatively light-weight approach to Haskell web programming is a CGI library and an XHTML combinator library. Here is a very simple example which just outputs some static HTML using CGI library and XHTML. Appendix 3 shows two simple examples. The first one just outputs some static HTML, and the second one show how to get user input. [116]

Java uses J2EE platform to create web applications. There are too many Java technologies can be used to create web applications, such as Java servlet API, JavaServer Pages, JavaServer Pages Standard Tag Library, JavaServer Faces Technology, Java Message Service API, and so on. In a Java web application, components are either Java servlets, JSP pages, or Web service endpoints. The interaction between a Web client and a Web application is illustrated in Figure 2.

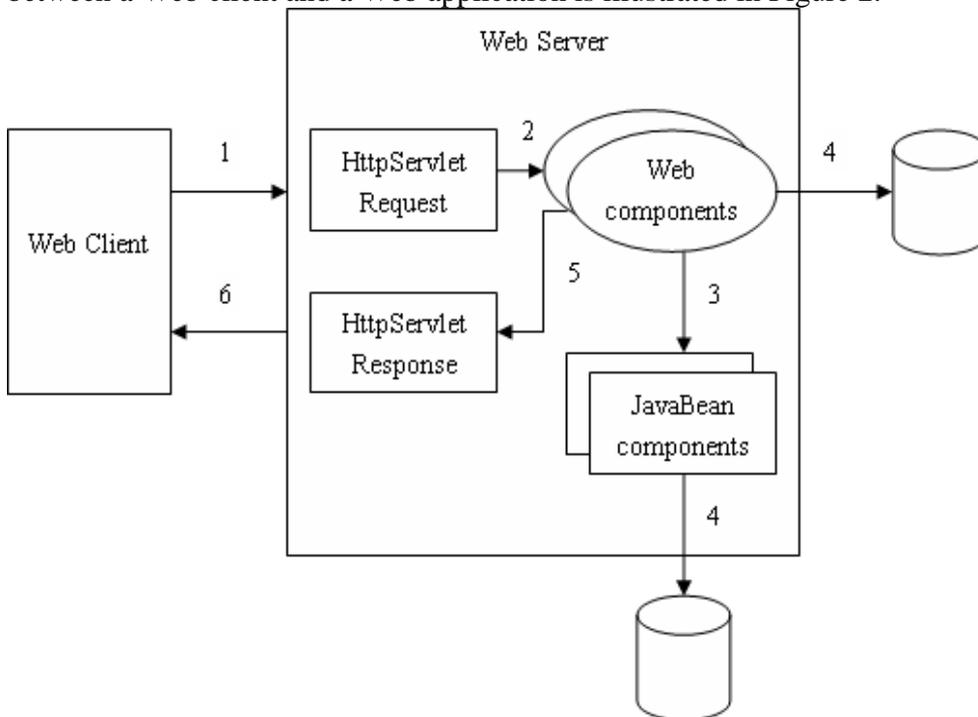

Figure 2, Java Web Application Request Handling [10]

The client sends an HTTP request to the Web server. A Web server that implements Java Servlet and JavaServer Pages technology converts the request into an HTTPServletRequest object and then delivered to a Web component, which can interact with JavaBeans components or a database to get dynamic content. The Web component can then generate an HTTPServletResponse or it can pass the request to another Web component. Finally a Web component generates a HTTPServletResponse object. The Web server converts this object to an HTTP response and returns it to the client. Appendix 4 show JavaServlet handle HttpServlet Request and send HttpServlet Response.

## 2.3.4  Web services design and composition

Since a large Haskell function is composed of a number of smaller functions, it is ideal for composing services. In order to compose service operations, it is necessary to enable the viewing of services as side-effecting functions in Haskell. Many features in Haskell also provide an ideal platform on which various data-processing applications can be elegantly constructed, and this motivates the ability to create atomic services in Haskell. The Haskell Application Interoperation Framework/Architecture (HAIFA) is a library providing the basic components for web-based interoperability, in the form of a service modeling framework. Firstly, the Simple Object Access Protocol (SOAP) can help to set web-service architecture.  Secondly, HaXML, the most well know XML library for Haskell, provides

Haskell2Xml type-class, which performs serialization and deserialization of Haskell types to XML. After created a suitable XML serializer, user can build a client side service interoperation, and then publish the Haskell function as service[117].

Java has been a powerful development platform for Service-Oriented Architecture (SOA) since 2006. Java EE 5, released in May 2006, significantly enhanced the power and usability of the Web Services capabilities on the application server. Java API for XML-based RPC (JAX-RPC) is a technology for building Web services and clients that use remote procedure calls (RPC) and XML. Often used in a distributed client-server model, an RPC mechanism enables clients to execute procedures on other systems. JAX-RPC provides an easy to develop programming model for development of SOAP based Web services [118] In JAX-RPC, a remote procedure call is represented by an XML-based protocol such as SOAP. The SOAP specification defines the envelope structure, encoding rules, and conventions for representing remote procedure calls and responses. These calls and responses are transmitted as SOAP messages (XML files) over HTTP. JAX-RPC can provide a big advantage for both client side and service side -- the platform independence of the Java programming language. Moreover, JAX-RPC is not restrictive. That is a JAX-RPC client can access a Web service not running on the Java platform, and vice versa. [119]

## 2.3.5  OO-based abstraction

Contrast subtype polymorphism of object-oriented languages, Haskell provides type-class-bounded and parametric polymorphism. Ad-hoc polymorphism (user-defined overloading) can be handled by a powerful abstraction mechanism provided by type classes. The basic idea behind type classes is that class declarations allow one to group together related methods (overloaded functions), and instance declarations prove that a type is in the class, by providing appropriate definitions for the methods. Here are some standard Haskell declarations.[120]

```
class Eq a where (==)::a->a->Bool
instance Eq Int where (==) = primIntEq
instance Eq a => Eq [a] where
   (==) [] [] = True
   (==) (x:xs) (y:ys) = (x==y) && (xs==ys)
   (==) _ _ = False

         Listing 4: standard Haskell declarations[120]

```

And then, we can extend the type class hierarchy by introducing new subclasses.

```
class Eq a => Ord a where (<)::a->a->Bool)
instance Ord Int where ...
instance Ord a => Ord [a] where ...

         Listing 5: introducing new subclasses[120]

```

The above class declaration introduces a new subclass Ord which inherits all methods of its superclass Eq. For brevity, we ignore the straightforward instance bodies.

Moreover, the OO features are introduced in Haskell as the O'Haskell library, based on the HList library of extensible polymorphic records with first-class labels and subtyping. Not only O'Haskell provides the conventional OO features, it has also language-engineered several features that are either bleeding-edge or unattainable in mainstream OO languages [121]: for example, first-class classes and class closures; statically type-checked collection classes with bounded polymorphism of implicit collection arguments; multiple inheritance with user-controlled sharing; safe co-variant argument subtyping.[122]The Appendix 5 show an O'Haskell sample code.

Java languages designed mainly for Object-Oriented programming. It support all features of OO programming technique, such as data abstraction, encapsulation, modularity, polymorphism, and inheritance. The Appendix 6 show an Java sample code

## 2.3.6 Reflection

Haskell supports some kinds of reflection such as monads. A monadic-style functional program allows the imperative, behavioral view of a computational effect to be identified with a declarative, databased view in a uniform way. Monadic reflection looks like a formal bridge between the two views. It refines the identification into an observational isomorphism, with explicit reification and reflection functions mediating between views. Such a separation allows the programmer to reason robustly about monadic effects according to the declarative view, while implementing the imperative view much more efficiently in terms of lower-level control and state manipulations. [123] Monadic reflection is essentially a grammar for describing layered monads or monad layering. In Haskell describing also means constructing monads. This is a higher level system so the code looks like functional but the result is monad composition - meaning that without actual monads (which are non-functional) there's nothing real / runnable at the end of the day.

Reflection is a feature in the Java programming language. It allows an executing Java program to observe, examine, and modify internal properties of the program. For example, it's possible for a Java class to obtain the names of all its members and display them. One tangible use of reflection is in JavaBeans, where software components can be manipulated visually via a builder tool. The tool uses reflection to obtain the properties of Java components (classes) as they are dynamically loaded. The following code show how Java reflection working

```
// Without reflection
String stringObj = "This is a string";
system.out.println("Length of string is: "+stringObj.length());

// With reflection
Class cls = Class.forName("java.lang.String");
Method method = cls.getDeclaredMethod("length", new Class[]{});
int length = (Integer)method.invoke(stringObj, new Object[]{});
system.out.println("Length of string is: "+length);
                        Listing 6: Java Reflection
```

## 2.3.7 Aspect-orientation

Type classes can help Haskell to achieve Aspect-Oriented Programming (AOP). Type classes provide some form of type-safe cast to encode AOP in the setting of a strongly typed language by exploiting GHC's overlapping instances. AOP Haskell extends the Haskell syntax by supporting top-level aspect definitions of the form:

N@advice #f1, …, fn# :: ( C => t ) = e

In the definitions, N is the name of the aspect, and each fi is function symbols as joinpoint. C => t is type annotation following the Haskell syntax for types. e is the advice body following Haskell syntax for expressions. The advice will be applied if the type of joinpoint fi is an instance of t such that constrains C are satisfied. And then, by turning advice into type class instance and instrument joinpoints with calls to a "weaving" function – a process that intercept calls to joinpoints and re-direct the control flow to the advice bodies - AOP idioms can be translated to type classes as supported by the Glasgow Haskell Compiler (GHC). Appendix 7 show a AOP Haskell example [120]

Java has a simple and practical extension, AspectJ, which provide aspect-oriented programming (AOP) capabilities for Java. This allows developers to reap the benefits of modularity for concerns that cut across the natural units of modularity. Like any other aspect-oriented compiler, the AspectJ compiler includes a weaving phase that unambiguously resolves the cross-cutting between classes and aspects. AspectJ implements this additional phase by first weaving aspects with regular code and then calling the standard Java compiler.[124] Appendix 8 show a AspectJ example and its running output

## 2.3.8 Functional programming

Haskell is an advanced purely functional programming language. The underlying model of computation is mathematical concept of a function, and programs are executed by evaluating expressions. In Haskell, functions are first-class, which means that they are treated like any other values and can be passed as arguments to other functions or be returned as a result of a function. Being first-class also means that it is possible to define and manipulate functions nested in code blocks. Special attention needs to be given to nested functions, called closures that reference local variables from their scope. If such a function escapes their block after being returned from it, the local variables must be retained in memory, as they might be needed later when the function is called. Usually it is not possible to determine statically the moment of release, which means that an automatic memory management technique has to be used. The following code is a simple Haskell function define.

```
hyp  :: Float → Float → Float
hyp x y = sqrt (x*x + y*y)

             Listing 7: function define in Haskell
```

Basically, Java does not allow composition of functions. In functional programming languages such as Haskell, functions are first-class, while in Java, class is first-class. However, a extension library, Functional Java provide optional functional programming in Java. The library implements several advanced programming concepts that assist in achieving composition-oriented development.[125] The following code is a example of Functional Java

```
final Array<Integer> a = array(3, 2, 1);
final Array<Integer> b = a.map({int i => i + 2});
arrayShow(intShow).println(b); //{5,4,3}
                         Listing 8: Functional Java
```

## 2.3.9 Declarative programming

In computer science, declarative programming is a programming paradigm that expresses the logic of a computation without describing its control flow. [126]That means the language focus on what need to do and not how to do. As a pure functional language, Haskell support declarative programming. The following code shows how to write code in declaration style.

```
filter  :: (a → Bool ) → [a] → [a]
filter p [ ] = [ ]
filter p (x:xs ) | p x = x : rest
                 | otherwise = rest
                 where
                    rest = filter p xs

         Listing 8: filter function with Declaration style
```

From this example, we can see that declaration style define a function by many equations, and each equation uses pattern matching and/or guards to identify the cases it covers.

Java is an imperative programming language which gives a list of instructions to execute in a particular order. However, JDK 1.5 release a powerful language construct, **annotation**. Annotation is a generic mechanism for adding tags with data (metadata) with program elements such as classes, methods, fields, parameters, local variables, and packages.[127] By annotation, declarative programming style can be introduced into Java language. The following codes show how to declaring an annotation and how to use it.

```
// Declaring an Annotation
Package njunit.annotation;
import java.lang.annotation.*;

@Retention(RetentionPolicy.RUNTIME)
@Target({ElementType.METHOD})
public @interface UnitTest {
        String value();
}

//Using an @UnitTest annotation
import njunit.annotation.*;

public class Example {
    @UnitTest(value="Test 1. This test will pass.")
    public void pass() {
        assert 10 > 5;
    }
    @UnitTest("Test 2. This test will fail.")
    public void fail() {
        assert 10 < 5;
    }
}
```

Listing 9: Example of Java Annotation[130]

### 2.3.10  Batch scripting

Both Haskell and Java can create batch scripting because they can process external commands in automated mater. For example, Haskell can use function system and rawSystem to executing an external command. The basic synopsis of system and rawSystem are:

```
system :: String A IO ExitCode
rawSystem :: String A 〖String〗 ->IO ExitCode

           Listing 10: synopisis of system and rawSystem[128]
```

In Linux system we can play them like:

```
system "ls -l"
rawSystem "ls" 〖"-l"〗

             Listing 11: Example of system and rawSystem
```

In Java, Runtime class can be used to executing an external command. The following is an example.

```
import java.io.*;
public class someClass {
       public static void main(String〖〗 args)
       {
             try
             {
                    Process proc = Runtime.getRuntime().exec("ls -l");
                    InputStream p = proc.getInputStream();
                    BufferedReader reader = new BufferedReader(new
InputStreamReader(p));
                    String line;
                    while((line=reader.readLine()) != null){
                          System.out.println(line);
                    }
             }
             catch (Exception e)
             {
                    System.out.println(e.getMessage());
                    e.printStackTrace();
             }
       }
}

                     Listing 12: Example of Java Runtime
```

## 2.3.11 UI prototype design

There are several GUI toolkits available for Haskell. However, there is no standards one and all are more or less incomplete. Generally, there are three level Haskell GUI library, low-level, medium-level, and high-level. Low-level,such as GLFW, GLUT, TclHaskell, Win32, and X11, are going well, but it is doing everything in the IO monad. High-level abstractions, such as FG, FranTk, Fruit,

Fudgets, Grapefruit, GuiTV, Phooey, and wxFruit, are pretty experimental[116]. The medium-level, such as wxHaskell, Gtk2Hs, hoc, and qtHaskell, can support Haskell GUI.[128]

**wxHaskell** is a portable and native GUI library for Haskell, which provides a interface to wxWidgets toolkit[129].

**Gtk2Hs** is a GUI library for Haskell based on Gtk+ which is an extensive and mature multi-platform toolkit for creating graphical user interfaces. [131]

**HOC** (documentation at sourceforge) - provides a Haskell to Objective-C binding which allows users to access to the Cocoa library on MacOS X [132]

**qtHaskell** is is a set of Haskell bindings for the Qt Widget Library from Nokia[133]

Appendix 11 shows how to build GUI using Gtk2Hs.[134]
In Java, there are two components to support GUI design, AWT and Swing. AWT, Abstract Window Toolkit, is Java's original platform-independent windowing, graphics, and user-interface widget toolkit. The AWT is now part of the Java Foundation Classes (JFC) — the standard API for providing a graphical user interface (GUI) for a Java program.
Swing is a built-in GUI component technology of the Java platform. It is successor to AWT. Swing replaces some function of AWT such as using javax.swing.JTextField instead of java.awt.TextField. On the other hand, Swing builds on AWT. For example JtextField is descendant of java.awt.Container. Appendix 10 give a sample code of Java Swing

## 2.4 PHP vs Scala

### 2.4.1 Source code size

Hello world in PHP

```
<?php
  echo "Hello, world!";
?>
```

Hello world in Scala

```
object HelloWorld extends Application {
  println("Hello, world!")
}
```

### 2.4.2 Default more secure programming practices

PHP is a loosing typing and dynamic checking language and the type safety is not that safe. It enforces run-time checking which can provide flexibility to try evaluating compile time error type.
PHP does not require (or support) explicit type definition in variable declaration; a variable's type is determined by the context in which the variable is used [135].

PHP 5.3 introduces garbage collection (GC) to handle the memory allocation and de-allocation problems. In PHP, when there is no variable points to the object, those kinds of objects would be considered as garbage, and it will automated freed in the memory, which avoid the memory leak. When one PHP session is ended, the memory space occupied by those sessions will be destroyed. All the objects in the current applications would also be destroyed in the meantime. Generally, the process of GC in PHP is followed by the start of one session. When session is end, it automatically destroys those files [136].

PHP is popular with its strong extension API after PHP 3.0, which support PHP wrapping with some other existing library functions for certain purpose. This rich API will help PHP integrate with other more completely. As PHP is usually used as a web server module, you should be aware that PHP might be used in threaded an environment which means that global variables might lead to rare conditions [137]. When small PHP program's leak would not be a big problem to OS. If PHP associate with server service(i.e Apache server), it would be a long term running period, then the memory leak would be huge problem to the OS. Even if your program is well-written format to avoid leak issue, it is necessary to have some memory management to avoid them. With the help of extension API, PHP can avoid that leak problem. In extension API (the Zend engine), it provides a set of wrapper function (Like emalloc (), efree () similar to the method malloc (), free () in C language) to handle the memory problems [138].

PHP do support exception handling like java using "try-catch-throw" style to deal with various of types of errors [139].

Static typing, or compile-time type checking, helps you define and verify interface contracts at compile time. Although Scala is static typing languages, unlike some of the other statically typed languages, does not expect you to provide redundant type information. You don't have to specify a type in most cases, and you certainly don't have to repeat it. At the same time, Scala will infer the type (Type inference) and verify proper usage of references at compile time. Scala is considered as a type safety languages. You will gain two benefits from Scala type system [140]:

    1. The compile time checking will give you confidence that the complied code meets certain defense expectation.

    2. Help you express the expectation on your API in a complier-verifiable format

Scala runs on top of JVM which means that your code is compiled into bytecode. That in turn means that JVM uses garbage collection as your Scala program is running - garbage collector (GC) cleans up the memory when objects you allocated in your code become obsolete [141]. Scala also supports a java-like style exception handling, which using "try {…} catch {…}" block to handle error exceptions. But it would not so effectively to handle the run-time exception [142]

In sum up on this criterion, PHP is dynamic type with weak type strength, while Scala is static type with strong type strength and support type inference. Although dynamic type provides flexibility to PHP type system, it also brings unsafe factors. In the memory management part, Scala is only support the garbage collection type, while PHP can gain support from the extension API(zend engine) to have flexibility deal with specified problems(multiple-threading problems). Both languages provide similar syntax exception handling.

### 2.4.3 Web applications development

PHP is associated with HTML to serve as dynamic website development tools.PHP runs on the Apache server to collect user data via HTML form and process them on the server side. After interpreted by web server, it generate website. The PHP web application will collect the user information and store them in the session, thus it is necessary to verify the validity of user information, especially in case of tamping by hacker. Thus, it requires some security issues at the server end. For instance, safety handling the user input in registration validates the user URL etc. This technique would make scripts safer. There are two common forms to for PHP to organize website safety: "one script for serves all" and "one script serve one function". Whatever forms you take, validate every element from outside the server end.

It comes with the following possible benefits for developing web application using PHP [143]:
• It is especially design for the web, generally for the server end.
• It is a robust and proven platform.
• Since this is an open-source, it is constantly upgraded through community development.
• It is also highly customizable, and it is easily adaptable to suit for all uses.
• The web application in PHP reduces the web application development costs further.

Lift is a free leading web application framework that aims to deliver benefits similar to Ruby on Rails. But it is written in Scala, not Ruby. The use of Scala means that any existing Java library and Web container can be used in running Lift applications [144]. Lift applications benefit from native support for advanced web development techniques such as Comet and Ajax. And lift is often used for commercial websites. Now the latest version of Lift 2.0, which provide more industrial optimization and support.

It comes with following possible benefits that Lift-based application [144]:
• Lift is like a breath of fresh air: concise, elegant and robust
• Lift's high performance and scalability
• Lift's built-in support for REST and other web services
• Lift's use of Scala's type-safety so your tests can focus on business logic.
• Built-in security means more time focusing on your application and less time being
  defensive about parameter tampering, SQL injection, Cross Site Scripting and other
  nasty attacks.

## 2.4.4 Web services design and composition

Web services are typically application programming interfaces (API) or Web APIs that are accessed via Hypertext Transfer Protocol (HTTP) and executed on a remote system hosting the requested services. Web services tend to fall into one of two camps: big Web services and RESTful Web services." Big Web services" use Extensible Markup Language (XML) messages that follow the SOAP standard and have been popular with traditional enterprise [145].

In PHP 5, all XML extensions have been rewritten to use the superb libxml2 XML toolkit. PHP 5 comes with a SOAP extension ext/soap, which has even more features than PEAR::SOAP, and is written in C instead of PEAR::SOAP, which is written in PHP. [14] PHP 5 do supports web service design and composition, and support different protocols such as OAuth, SCA, SOAP and XML-RPC [146]. PHPXMLRPC is PHP library for implanting the function of XML-RPC.It designed for ease of use, flexicibility, and completeness. But this library does not consider the speed and memory management issues. With the support for XML-RPC, it enable distributed systems and system interoperability, it allows applications call the methods in other place no matter what machine is, what language written with. Such as C++ function can call the PHP methods. This achieves the purpose of web service. New XML support makes PHP the best language available for processing XML and, coupled with new SOAP support, an ideal platform for creating and using Web Services [147].

Scala do support the Web service building.

Scala's notion of pattern matching naturally extends to the processing of XML data with the help of right-ignoring sequence patterns. In addition, Scala's XM; support is partly from library, with some built-in syntax support. It allows using inline-XML that makes XML in Scala in a convenient form. Since Scala can seamlessly integrate with Java's library, thus Scala can using the XML-RPC in Java, which is a java implementation of XML-RPC protocol for web service purpose. In this context, sequence comprehensions are useful for formulating queries. These features make Scala especially ideal for web services [148].

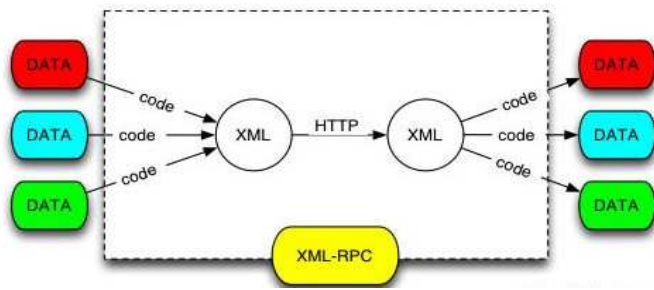

 **Listing 1 How XML-RPC work**

### 2.4.5 OO-based abstraction

Both PHP and Scala do support OO-based abstraction.

PHP 5 support Object Oriented programming (OOP) paradigm, it supports single inheritance, interface (abstract method and class). It also supports encapsulation.

The purpose of inheritance in PHP is to inherit the characteristic of the base class that does make code reusable and structural. Using keyword "new" to instantiate a new object from non-abstract class and keyword "extends" to declare which class inherits from which class. That's somewhat similar way to Java. PHP also support abstract class, which any class extends an abstract class must create an interface (Interface in PHP will define your own common structures and can not be instantiated itself) of the parent abstract methods. Additionally, in order to achieve some level of encapsulation purpose, PHP use some keywords (public, private, protected) to manipulate the visibility of class members (properties or methods). In the source code example A1 [149], it simplify illustrate an abstract class example.

Scala is a pure OO language, everything in Scala is an object. It supports single inheritance and traits, which is like optional implemented interfaces. It is quit similar to the interfaces in Java, traits are used to define objects by specifying the signature of the supported methods. It is allowed for traits to be partially implemented, while there is not allowed to implement the abstract class in PHP. This implies that you can define default implementation for some methods. [151] Compared with the classes, trait may not require constructor parameter. It also provides similar way to encapsulate data members as PHP did. In the source code example A2 [152] it give a trait version that implement abstract class.

To sum up, two languages have similar way in OO-based abstraction. Compared to PHP, Scala have more advanced OOP characteristics (such as Case Classes, Traits, and Companion Objects) which make the data member more flexible in the OO framework.

### 2.4.6 Reflection

Reflection is the process by which a computer program can observe and modify its own structure and behavior. A reflection-oriented program component can monitor the execution of an enclosure of code and can modify itself according to a desired goal related to that enclosure. This is typically accomplished by dynamically assigning program code at runtime [153].

PHP 5 comes with a complete and rich reflection API that adds the ability to *reverse-engineer* classes [154], interfaces, functions, methods and extensions. Additionally, the reflection API offers ways to retrieve doc comments for functions, classes and methods. Using reflection is actually let you know how things actually worked. Reflection itself is handled through various classes, the root of which is the ReflectionClass. There's other class like ReflectionExtension, ReflectionFunction, ReflectionMethod, ReflectionObject, Reflection Parameter [155].

Scala supports the same reflection capabilities that Java and .NET support. [156] The syntax in Scala would be somewhat different depending on the cases. The following code [157] segment shows that some reflection methods in JVM, through java.lang.Object and java.lang.Class:

**Listing2 Reflections on Scala using Traits**

```
trait T[A] {
  val vT: A
  def mT = vT
}
class C extends T[String] {
  val vT = "T"
  val vC = "C"
  def mC = vC
  class C2
  trait T2
}
val c = new C
val clazz = c.getClass              // method from java.lang.Object
val clazz2 = classOf[C]             // Scala method: classOf[C] ~ C.class
val methods = clazz.getMethods      // method from java.lang.Class<T>
```

## 2.4.7 Aspect-orientation

Aspect Oriented Programming is new programming paradigm which enables you to effectively implement crosscutting concerns in new and also existing programming without changing the original code. Unfortunately, it is possible in complied programming languages, but not for scripting language.

Since the PHP is a scripting language, the written code is interpreted, not complied in the server end. In PHP5, it brings to new approach to solve the weaving problems.

To solve this problem of weaving before executing the program, some implementations, of the Aspect-Oriented Programming paradigm in PHP, use another script to read the file that is interpreted firstly, and then all Advices are added on the designated locations in the code. After this, the newly created file is interpreted. This requires another script, time to read the original file, a lot of find and replace operations and finally interpreting a non-optimized new file. This can really slow down the execution time of the PHP code, especially with large programs with a lot of Aspects. Another common used solution is to add some libraries to the PHP interpreter. These libraries then take care of the weaving before executing the file. This solution doesn't remarkably slow down the interpretation and execution process, but needs some extra skills to create the library. Furthermore the PHP installation on the server is modified to enable these libraries. That can result in less stable services. In the source code example A3 [158], it is the second way which provide external library to help achieve the AOP purpose into PHP.

Scala have the ability to interoperate with AspectJ, an extension of Java, which supports Aspect-oriented programming [159]. When using AspectJ in Scala, we have mentioned some detail issues: how to reference Scala execution points and how to invoke Scala code as advice Scala traits is a better choice rather than using AspectJ for the purpose of AOP [159]. However, Scala doesn't have a point cut language, like AspectJ. Using aspect would be more suitable for more "pervasive modification" (e.g. Tracing, police enforcement, security) while using trait is no need to worry about other language (traits is not from the Java or .NET) and fits your entire requirement. [160] In the source code example A4[161], it is an aspect that logs methods calls to the Complex class.

In sum up, Scala provide two ways to achieve AOP purpose: Built in approach (using traits) or Aspect approach (using Aspect), while PHP also provide two ways to handle AOP, but seems rather complicated than Scala did.

## 2.4.8 Functional programming

Since PHP 5.3 support Closure, then PHP can support functional programming with more convenience. Before the Closure come, we need to implement curry function (curry is not built-in function in PHP, need to be defined by developer) to achieve the same goal as Closure, which is a technique to transform functions from multiple arguments to one argument. Assume the following code segment [162]:

```
$add2 = curry ($add, 2); // returns a new function reference, to add (), bound with 2 as the first
                    //argument
$add2 (8);          // return 10
```

**Listing 3 using curry method for functional programming purpose**

Scala support functional programming paradigm.Scala is also a functional language in the sense that every function is treating as value. Scala provides a lightweight syntax for defining anonymous functions, it supports higher-order functions, it allows functions to be nested, and supports currying. Scala's case classes and its built-in support for pattern matching model algebraic types used in many functional programming languages. [163] In the following code segment, it is the code example of high-order function of Scala [165]:

```
def increment(x:Int) = x + 1
def handler( f:Int => Int) = f(2)
println( handler( increment ))      //pass the function as value
```

**Listing 4 High order function**

Scala uses the keyword lazy defers the initialization of a variable until this variable is used. Evaluation of delimited continuations is supported in version 2.8. Tail call optimization is not supported completely, because the JVM lacks tail call support. In simple cases, the Scala compiler can optimize tail calls into loops [166]. In the following code segment, it is code example of lazy evaluation in Scala [167] (Since Scala is not default lazy evaluation, it requires keyword "lazy" to evaluate when in needed):

```
def lazyMultiply(x: => y:) = { lazy val y = x * x }
```

**Listing 5 Lazy evaluation**

## 2.4.9  Declarative programming

Compared to imperative programming, the main point of declarative programming to focus on what to do, and do not concern on how the steps should be.

PHP is a scripting language which can support declarative programming paradigm under the help of annotation. Annotation can be used by tools or libraries which do not affect the semantic structure of program; it adds metadata to the properties (classes, methods) [169]. In addition, PHP can embelled with HTML, which is a declarative user markup language just to tell what the server end should do to support the interface [170],

Since Java can support declarative programming through annotation and reflection [171], and Scala do support both techniques, thus, we consider Scala do also support declarative programming, may have somewhat different in code syntax.

## 2.4.10   Batch Scripting

PHP is a scripting language which more than just web application and can be used for command line shell scripting. PHP supports a CLI SAPI (a layer that PHP interact with other web server (i.e. Apache and M$ IIS)) through as of PHP 4.3.0. The main focus of this SAPI is for developing shell applications with PHP. That is to say we can use CLI as command-line tools same as the standalone server application. This make PHP look more like traditional scripting languages. It is possible to write shell scripts using CGI version of PHP, although CGI have some significant different (In CLI, no header written to the output) with the CLI in web server integration. For the reason of accessibility

(default installed) and consistent, developer prefer using CLI to write PHP shell scripts. [40] In the following, it tells PHP to execute specified files:

```
$ php -f comp6411script.php
```

**command line**

Scala is similar to PHP, which its program can also run as a shell script as a batch command. For instance, The "scalac" command compiles one or multiple Scala source files and generates Java bytecode on the base of JVM, which the Scala complier work similar to Javc(java complier) [172]

```
> scalac HelloWorld.scala
```

**Listing 7 execute the .scala file in command line**

## 2.4.11    UI prototype design

PHP itself does not support UI prototype design□generally, PHP was embelled into HTML file to invoke HTML's features to obtain the graphical interface purpose. Strictly speaking, we should say PHP does not support UI prototype design directly. But after the coming of the PHP-GTK, it reduces the dependence on the UI purpose with HTML. That is to say, we can use PHP create stand-alone GUI application. PHP-GTK [173] is an extension for the PHP programming language that implements language bindings for GTK+. It provides an object-oriented interface to GTK+ classes and functions and greatly simplifies writing client-side cross-platform GUI applications.PHP. We can use this tools create windows, buttons, and text without help by another program (i.e. browser, and text editor).

```
// show an window using gtk library
<?php
$window= new GtkWindow();
$window->connect_object('destroy',array('gtk', 'main_quit'));
$windown->show();
gtk::main();
?>
```

**Listing 8:** A Simple Application in PHP-GTK [174]

Scala's graphical user interface (GUI) development is heavily depending on the Java GUI library. It works well with Java AWT, swing. But there are some swing Scala library to choose like: ScalaGUI [174], scala-swing (Scala API).

```
package example
import scala.gui._    // import from Scala library
 object application extends scala.gui.Application {
 val mainWindow = new container.Window {
 val press = new widget.Button {
     text = "Press me, please"
     subscribe(this)
     toplevel eventloop {
       case this.Click() =>
         field.text = "Wow! Someone pressed me."
     }
   }
 }
```

**Listing 9: A simple segment for create a window and button using ScalaGUI**

But in general case, invoking Java GUI is more convince and with rich library choice. Because there are not much Scala built-in GUI library for use the API.

## 2.5    Scheme vs JavaScript

### 2.5.1    Source code size

Hello world in Scheme

```
(display "Hello World!")
```
Scheme Code sample

Hello world in JavaScript

```
document.write('Hello World!');
```
Javascript Code sample

### 2.5.2    Default more secure programming practices

**Type system:**

Scheme uses strongly but dynamically typed variables. In Scheme, a variable can refer to a value of any type, and the majority of its type checking is performed at run-time. However, at run-time, Scheme rejects any operations which attempt to disregard data types. However, Javascript uses weakly and dynamically typed variables. That is, In Javascript, any kind of variables can be stored into a variable. Programmers just need to simply declare a variable without assigning it a type
Memory Management:

Both Scheme and Javascript uses automatic garbage collection for heap memory management. That is, Scheme memory management system can scan and reclaim all dead objects, i.e. objects that will not be used in future, automatically. This frees the programmer from having to explicitly deallocate memory themselves.

**Exception Handling:**

A **continuation** reifies an *instance* of a computational process at a given point in the process's execution. It contains information such as the process's current stack (including all data whose lifetime is within the process e.g. "local variables"), as well the process' point in the computation [176]. In Scheme, Continuations can form the basis for implementing a simple throw/try/catch-style exception handling mechanism in just a few lines. However, things get rather more complicated if the mechanism is to work in programs that themselves use continuations. Therefore, several Schemes come with built-in exception handling capabilities [177]

Javascript provide an exception handling mechanism allowing preventing application from crashing at run-time. Programmers can use try-catch to handle exception.

### 2.5.3    Web applications development

Many implementation of Scheme can be used to develop a web application, and Schemes with an FFI (foreign function interface) can also support web programming. For example, openScheme and RScheme have build-in CGI, incoming and outgoing HTTP, and HTML generation [178]. PLT Scheme has several web programming packages available, such as complete web server, CGI, MIME, and cookie handling. Gauche also has cgi and html generation bits in its standard library. JScheme can work in a similar way to Java Server Pages. Mod_lisp is an Apache module to easily write web applications in Lisp/Scheme[179]. LAML is a Scheme-based set of libraries for server side web programming. [180]

Originally, Javascript for a web application are created to run on client-side, refer to Client-side Javascript (CSJA). Now, server-side Javascript (SSJA) is available. The first implementation of SSJS was Netscape's LiveWire, which was included in their Enterprise Server 2.0 back in 1996. Since then, a number of other companies have followed suit in offering an alternative to the usual server-side technologies, such as ASP of Microsoft and Jaxer of Aptana. These system support Javascript for accessing database, file system, send email and so on.

### 2.5.4   Web services design and composition

Racket (or PLT Scheme) programming language has a massive set of libraries and tools which are suitable for specific domains such as web applications development, querying databases (*MySQL.plt*) and batch scripting and user interface prototyping [181].

Scheme does support HTTP transactions also it does support web development aside CGI programming. Scheme support also XML documents parsing and writing which is an advantage for supporting web services transactions [182]. Racket provides the *xml* library for parsing and generating XML documents which are represented in a structure type called X-pression [183].

Scheme does support web services [184]. Using Scheme, you can call web services methods through CGI programming facilities. Using Scheme, Amazon Web services (AWS) can be called in order to retrieve data from Amazon's database [185].
There is an implementation of Scheme called Gambit which enables to write a stand alone executable web service. This implementation requires the Gambit-C which is a compiler that generates portables *C* code [186].

JavaScript is a scripting language which runs in web browser as part of them. In web applications, mostly, it is used on the client-side to ensure data validation and interaction with Document Object Model by embedding it within HTML tags.

There are some implementations for JavaScript that run on the server-side. Thus, using Ajax API, a web application can communicate asynchronously with a web server in an interactive mode. This communication enables you to retrieve data form the web server using the XMLHttpRequest object [187].

JavaScript enables to call web services using Ajax (the XMLHttpRequest object) in order to retrieve customized data. *Yahoo!*  and *Amazon w*eb services are a popular example for building XMLHttpRequest-based web services client in JavaScript [188].

### 2.5.5   OO-based abstraction

Scheme support Object-Oriented programming, because it always pass message within the body, just like objects passing. In Scheme, it provides a procedural way to express its OOP features. Instantiate an object instance is represented as a procedure which project operation to methods while a method is somewhat like put the parameter into operation then perform operation on instance [189]. And it also support inheritance and polymorphism (by virtual method).In the listing 1 [190], it shows a class y inherit class x.Some Schemes implementation have their own built-in OO system (i.e. MacScheme, Feel, Oaklisp, XScheme, and PC-Scheme [191]). Most of these are similar to the Common Lisp Object System (CLOS), while others are conceptually closer to C++ and Java. Kawa is fully integrated with the Java object system, i.e. it allows Java classes to be defined and extended in Scheme [192]. The OOP syntax in different implementation of Scheme may vary depends on how it interoperates with the OO concept. In addition, the Scheme provides local encapsulation by closures.

```
(define (y)
 (let ((super (new-part x))
       (self 'nil))
 (let ((y-state 2)
       )
   (define (get-state) y-state)
     (define (set-self! object-part)
      (set! self object-part)
      (send 'set-self! super object-part))
   (define (self message)
     (cond ((eqv? message 'get-state) get-state)
           ((eqv? message 'set-self!) set-self!)
           (else (method-lookup super message))))
   self))) ; end y
```

Listing 1 Class inheritance: Class y inherit from x

JavaScript support OO-based abstraction (inheritance, polymorphism, encapsulation), although it is not a pure OO languages. Unlike Scheme, JavaScript have similar way in syntax with traditional OO language (i.e. java) which would be friendly adapted by java developer. In JavaScript, the straightforward way for constructing OO is the build-in object data type. In JavaScript, objects are implemented as a collection of named properties. Being an interpreted language, JavaScript allows for the creation of any number of properties in an object at any time. Everything is object in JavaScript, except for the primate data types and it classifies 3 types object (native objects, user-defined objects, and host objects (for browser purpose)). In JavaScript, each Object can inherit properties from another object, just like you extend the class for reuse purpose, called its prototype. When evaluating an expression to retrieve a property, JavaScript will check the properties which defined in the object firstly. If it is not found, it then looks at the object's prototype to see if the property is defined there. This continues up the prototype chain until reaching the root prototype [193] . Each object is associated with a prototype which comes from the constructor function from which it is created. The polymorphism achieved by calling correct name properties to involved the appropriate function for corresponding prototype. In the listing 2 [194], it shows a segment of polymorphism include using prototype and methods calls:

```
function A()
{
    this.x = 1;
}
A.prototype.DoIt = function()    // Define Method
{
    this.x += 1;
}
function B()
{
    this.x = 1;
}
B.prototype.DoIt = function()    // Define Method
{
    this.x += 2;
}
a = new A;
b = new B;
a.DoIt();
b.DoIt();
document.write(a.x + ', ' + b.x);
```

Listing 2 Polymorphism in JavaScript

## 2.5.6   Reflection

Scheme dose not support the reflection directly, But since scheme is dynamic type, it provide scheme predicates to determine the object type or the equivalence of two objects (using eq?, eqv?, and equal? as predicates) However, it is somewhat different than we expected, which rule is as follow: Two objects of different types (booleans, the empty list, pairs, numbers, characters, strings, vectors, symbols, and procedures) are distinct. Two objects of the same type with different contents or values are distinct. Thus we need to find a way to fix it. Scheme enables us using marco to reach the function of reflections. Some research paper [195] implements a marco-implemented extension PLT-Scheme for object-oriented programming which can interoperate with Java. That's means we can invoked Java's library within the PLT-scheme environment, so that invoked the reflection API is possible.

Unlike the complicated implementation of Scheme in reflection, JavaScript provide a simple way to do that: do-in statement. Compared to other Scheme, it does not require extra libaries, no namespace, no particular implemented classes, or any other help from other languages. The syntax is just stratiforward:

<div align="center">For (var member in obj){ alert(member); };</div>
<div align="center">Listing 3 for-in statement</div>

The Listing 4 [196] shows a example for reflection which return the object's properties and sort them.
```
function getMembers(obj) {
 var members = new Array();
 var i = 0;
```

```
for (var member in obj) {
  members[i] = member;
  i++;
}

return members.sort();
}
```

<div align="center">Listing 4 reflection in JavaScript</div>

### 2.5.7 Aspect-orientation

PLT Scheme supports AOP and its macro system provides especially powerful support for linguistic extensions. Macro system provides a lightweight implementation of aspects.In Scheme, we can easily define language extensions using its macro system [200]. Scheme macros are effectively functions that rewrite syntax trees; they are more powerful than lexical macros. Scheme macros can define and export the new syntactic forms "around" and "fluid-around" which is frequently used in AOP [199].

The figures below shows an implementation of aspect-scheme for statically -scoped aspects, with base language PLT Scheme.

```
(module aspect-scheme mzscheme
;; previous dynamic aspects part elided
;; statically-scoped aspects
(define-syntax (around stx)
(syntax-case stx ()
[( pc adv body)
(syntax
(w-c-m 'static-aspects
(cons (make-aspect pc adv) (current-static-aspects))
body))]))
(define-syntax (lambda/static stx)
(syntax-case stx ()
[( params body . . . )
(syntax
(let ([aspects (current-static-aspects)])
(_ params
(w-c-m 'static-aspects aspects
(begin body . . . )))))]))
```

JavaScript is not really a place where AOP is used heavily or needed, since the use of callbacks already provides some level of separation of concerns. AOP may have limited utility in JavaScript.But with the JavaScript libraries such as jQuery, Javasrcipt can achieve decent separation of concerns [198]. jQuery also supports the use of custom events, which may be used in a similar manner to some AOP advice types. Indeed, the Ajax Events available for hooking into using jQuery already provide some sort of AOP-like functionality [197].

## 2.5.8   Functional programming

Scheme is a functional programming and support closures with environment.. But scheme is not a pure functional programming which means that there are no variables, no assignment or side effects. Scheme use static scoping; nonlocal references in a function are resolved at the point of function definition. Static scoping is implemented by associating a closure [201].

Scheme characteristics:
· Supports functional programming - but not on an exclusive basis
· Functions are first class data objects
· Uses static binding of free names in procedures and functions
· Types are checked and handled at run time - no static type checking
· Parameters are evaluated before being passed - no lazyness

JavaScript uses functional as a library for functional programming. It defines the standard higher-order functions such as map, reduce, and select [202]. It also defines functions such as curry, rcurry, and partial for partial function application; and compose, guard, and until for function-level programming. And all these functions accept strings, such as 'x -> x+1', 'x+1', or '+1' as synonyms for the more verbose function(x) {return x+1}. All functions in Javascript are objects [203].

Javascript has the ability to write anonymous functions, or functions without a name. It can also pass functions as argument to other functions.Here is an example.

```
var passFunAndApply = function (fn,x,y,z) { return fn(x,y,z); };

var sum = function(x,y,z) {
  return x+y+z;
};

alert( passFunAndApply(sum,3,4,5) ); // 12
```

JavaScript is not one of the languages that use a variety of techniques to optimize function calls.Because of that, invoking a function in javascript is slow. Most current JavaScript implementations are slow with recursion and closures…two cornerstones of functional programming.

### 2.5.9 Declarative programming

Although scheme support funtional programming and the funtional is a subset of declarative language, the scheme language doesn't support declarative programming. Declarative language suits when there is an implicit "behind the scenes" procedure or process that's going to do something uniform with the assertions or statements presented [204].
Scheme focus on what information is desired and what transformations are required. A functional language can help organize the expression of computation [205].

JavaScript is an implementation of the ECMAScript language standard and is typically used to enable programmatic access to computational objects within a host environment. JavaScript supports all the structured programming syntax in C (e.g., if statements, while loops, switch statements, etc.) [205]. Meanwhile, Structured programming can be seen as a subset or subdiscipline of imperative programming, one of the major programming paradigms.
However, imperative programming which requires an explicitly provided algorithm is used in opposition to declarative programming, which expresses what needs to be done, without prescribing how to do it in terms of sequences of actions to be taken. Therefore, We could show Javascript doesn't support declarative programming [205].

### 2.5.10 Batch Scripting

The Racket project provides useful libraries that enable developers to write scripts and automate tasks such as file processing (text or XML files) zip files creation using Racket's *gzip* library and so on. Thus, Racket has what is called *port* which is an Input and Output stream for receiving or redirecting data to a file or a terminal [206]. This is a helpful technique that allows scripts to receive arguments as parameters.

Rackets files could have on of these extensions (*rkt, .rktl, .rktd, .scrbl, .plt, .ss, .scm*) and can run on Windows or on a Unix-based system as executables. This feature enables Racket to be used for batch scripting.

Natively, JavaScript it was designed to run inside a web browser. For security reason, JavaScript doesn't provide support for Batch scripting. *Internet Explorer* (IE) browser provides *ActiveXObject* which enables to execute an external program from a web browser as shown in the code snippet below. To do so, the security settings should be changed in order to enable ActiveX support. This approach is not supported by other web browsers. In addition, JavaScript enables you to automate some of web browser's functionalities. For example opening pop-up windows or bookmarking a web page [205].

```
<script language="javascript" type="text/javascript">
function runNotepad()
{
    var shell = new ActiveXObject("WScript.shell");
    shell.run("notepad.exe", 1, true);
}
</script>
```

### 2.5.11  UI prototype design

There are many GUI and graphics libraries to support scheme for UI prototype design such as Racket Graphics Toolkit.

Javascript is designed to add interactivity to HTML pages. So, it can support DHTML to create user interface and handle user action.

# 3      Consolidated Analysis and Synthesis of the Results

**Default more secure programming practices**

| Criteria 1 | | | | | |
|---|---|---|---|---|---|
| **Features/PL** | **C++** | **AspectJ** | **Haskell** | **PHP** | **Scheme** |
| Memory management | Yes(Manual memory management) | Yes, provided by the JVM | Garbage collection, provided by the GHC | Default using GC(by PHP 5.3) or Manual(by extension API) | Garbage collection |
| Bounds checking | | Yes, for arrays and data structure. | Yes, run time bounds checking | Yes, run time bounds checking | Yes, run time bounds checking |
| Static Type checking | Yes [232] | Yes, at compile time. | Yes check on compile time | No, PHP is dynamic type checking [213] | Yes, check on compile time |
| Dynamic Type checking | No[233] | Partial dynamic checking, i.e down casting. Also, when using reflection. | No, Haskell does not support casting | Yes [214] (no type cascading) | Partial dynamic checking, i.e down casting. |
| Type safety: Strong/weak | N/A – not considered | Yes, Based on Java's access rules. | Yes, , strongly type | Not that safe(running time type checking may not efficient to avoid type error) | Yes, , strongly type |
| Exception handling | N/A - not considered | Yes, exceptions can be thrown in pointcuts. Handlers are used to catch exceptions in AspectJ. Within try-catch/finally block in Java code. | Yes, but more complex than Java | Yes, using try-catch and can throw any kind of exception | Yes, using try-catch with build-in Exception class |
| Compiled/ Interpreted | N/A  - not considered | Compiled, and the Java-byte code interpreted by a JVM compliant. | Compiled or Interpreted by the GHC | Interpreted by web server(i.e. Apache) | Compiled to byteCode, and than interpreted by JVM |
| Conditional compilation | N/A - not considered | Yes (using SCoPE compiler, conditional pointcut evaluator) | Yes, but need to set in GHC options [207] | Yes(but rarely used) | Yes, but need some tricks[1] |
| Assertions | N/A  - not considered | Yes, temporal assertions at run-time[24] | User self-defined | Support(by PHP 4) | Yes, Java support assert |

**Criteria 1**

| Features/PL | C# | Groovy | JavaScript | Java | Scala |
|---|---|---|---|---|---|
| Memory management | Yes, provided by the CLR | Yes (Garbage collection | Garbage collection | Garbage collection, provided by the JVM | Garbage collection by JVM |
| Bounds checking | Yes, also for array bounds (raise *IndexOutOfRange* exception), buffer overflow. It can be disabled in C# [25] | N/A - not considered | Yes, run time bounds checking | Yes, run time bounds checking | Yes, run time bounds checking |
| Static Type checking | Yes at compile time. | Yes [233] | No | Yes, check on compile time | Yes, Scala support static type checking[215] |
| Dynamic Type checking | Yes, since C# 4.0 (.NET System.Dynamic namespace) and d*ynamic* keyword | Yes [233] | a variable can hold an object of any type and cannot be restricted | Partial dynamic checking, i.e down casting. | Strong[215] (with type cascading ) |
| Type safety: Strong/weak | Yes, C# is strongly typed. The *unsafe* keyword can be used for allowing pointers use. | N/A - not considered | No, weakly type | Yes, , strongly type | Safe, it is strongly type system |
| Exception handling | Yes, Within try-catch/finally block. Multiple catch blocks is supported in C#. | Try-catch block | Yes, using try-catch and can throw any kind of exception | Yes, using try-catch with build-in Exception class | Yes, using try-catch, but exceptions are not checked so effectively |
| Compiled/ Interpreted | Compiled, Interpreted by the CLR | N/A - not considered | Interpreted | Compiled to byteCode, and than interpreted by JVM | Complied by JVM |
| Conditional compilation | Yes (using preprocessor directives)[21] | N/A - not considered | Not support | Yes, but need some tricks [208] | Yes(with the help of compiler flag) [217] |
| Assertions | Yes, Managed code assertions  [22] | N/A - not considered | User self-defined | Yes, Java support assert | Support, using **assert, require and assume** |

**Web applications development**

**Criteria 2**

| Features/PL | C++ | AspectJ | Haskell | PHP | Scheme |
|---|---|---|---|---|---|
| Dynamic Web Pages | CSP | Yes, injection Aspectj code in JSP pages | *Haskell* Server Pages (HSP) | Support, the code behind php web page | PLT Scheme can create web page directly |
| Web Server | | Tomcat, JBoss, GlassFish | Haskell Application Server (HAppS) | Apache, Microsoft IIS | PLT Scheme web server |
| Web Framework/Libraries | Using Platinum, Reason, Evocosm, ACF. | Using J2EE, Spring or Struts framework. | Snap, Turbinado *web framework* | Many (i.e. Zend, Symfony, etc…) | SHP framework |
| Session management | N/A - not considered | Yes, i.e using *Spring* which provides *SessionManagementFilter* class and concurrency control [33]. | Yes, Snap have build-in session management. | Yes, using $_SESSION variable | PLT Scheme libsm6 package |
| Security | N/A - not considered | Yes, Java web applications run in what is called container, Each application has its container. Thus, the JVM add | Yes | Not that safe(require extra session management to avoid hacking and hijacking) | Yes, PLT provides a Scheme interface to some of the OpenSSL functionality through its openssl collection |
| Model-View-Controller | Wt/CppCMS/ffead | By adding AspectJ's aspects to J2EE/Spring/Struts applications. | Turbinado MVC framework | Symfony/Mojavi/CakePHP[218] | SHP framework support MVC |
| Database interaction | N/A - not considered | Yes, using the JDBC library | Yes (HaskellDB) | Mysql | Yes (some package in PLaneT) |
| Secure Sockets Layer support | N/A - not considered | Yes using OpenSSL library. | Yes, (hOpenSSL) | Yes(use open SSL function) [219] | Yes, openSSl |
| | N/A - not considered | N/A - not considered | | Yes(store in user session) | |

# Web applications development

**Criteria 2**

| Features/PL | C# | Groovy | JavaScript | Java | Scala Support |
|---|---|---|---|---|---|
| Dynamic Web Pages | Yes, as *code behind* in ASP.NET web pages. | JSP | Using Server-Side JavaScript (SSJS) [jh-4] | JSP | |
| Web Server | IIS on Windows OS Mono (Apache, Nginx) | N/A - not considered | IIS, Apache | Tomcat, JBoss, GlassFish | Tomcat, Glassfish |
| Web Framework/Libraries | Using .NET Framework on Windows and Mono ASP.NET Linux and Unix-based OSs. | Mainly using Grails Framework. | JAXER | J2EE framework | Not much(mainly use Lift) |
| Session management | Yes, provided by ASP.NET State management. | N/A - not considered | Yes, using Jaxer.session | Yes, are represented by an HttpSession object. | Yes, Session Actors |
| Security | Yes (ASP.NET Security Architecture [34]) | N/A - not considered | Yes | Yes | Safe |
| Model-View-Controller | ASP.NET MVC 1.0/2.0 | Grails | Yes, can create MVC structure using Jaxer | J2EE/Spring/Struts | Spring/Pinky |
| Database interaction | Yes, using ADO.NET library | N/A - not considered | Yes, Jaxer.DB | Yes (JDBC) | JDBC |
| Secure Sockets Layer support | Yes, SSL certificate can be generated and used to on IIS server. | N/A - not considered | Yes | Yes, JavaSSL [209] | Yes, can using java ssl support |

**Web services design and composition**

| Features/PL | C++ | AspectJ | Haskell | PHP | Scheme |
|---|---|---|---|---|---|
| Web Services including(SOAP, WSDL, UDDI) | Yes. Supported by gSOAP toolkit. | Yes. i.e using Apache XML-RPC for Java library [38] or Spring Web Services, JAX-WS 2.1) | Yes. i.e using HAIFA | Yes(XML-RPC for PHP [220] ) | Yes, Scheme supports web services but it requires CGI processing utilities [244]. |
| Web services security | Yes.Support HTTPS and WS-Security: authentication,tokens, digital signatures | Yes. HTTPS, above libraries support "XML-based standards" such as WS-Policy, WS-Security and WS-Transfer standards) Web Services pipeline watching. | Yes, hopenSSL help Haskell for HTTPS encrypt. | Partially support(limited by soap interaction with WSDL) [221] | Yes, PLT Scheme provides an *openssl* collection as part of it [255]. |
| Web Services composition | N/A  - not considered | Using AspectJ and J2EE framework, two or more web services can communicate together by linking them as a network. An application written in AspectJ can ensure the WS pipeline | N/A  - not considered | Yes(declared by PHParray or  policy files or inline with WDSL [222]) | Yes, Scheme web services can be built as a web services network where one or more web services can communicate together. |

**Web services design and composition**

| Features/PL | C# | Groovy | JavaScript | Java | Scala |
|---|---|---|---|---|---|
| Web Services including(SOAP, WSDL, UDDI) | Yes. using ASPT.NET web services and .NET's *System.Web.Services namespaces* for Windows-based WS and Mono project's Web Services on other OSs (Linux/Unix, Mac) | Yes. (Supported by Grails) | Yes, it is possible using an AJAX/XMLHttpRequest-based web service client [sr-13]. Gmail and Google services are good examples. | Yes. i.e using Apache XML-RPC for Java library [JL-6] or Spring Web Services, JAX-WS 2.1) | Yes(Apache XML-RPC[223]) |
| Web services security | Yes, HTTPS Encryption and signing with SSL[39], WS-Policy, WS-Security standards support) | Yes, rely on simple keystores or Spring Security for authorization and authentication.(since version 0.5) | Yes, Ajax can be used with any dynamic web programming language that support HTTPS protocol (SSL). | Yes. HTTPS, above libraries support "XML-based standards" such as WS-Policy, WS-Security and WS-Transfer standards) Web Services pipeline watching. | support |
| Web Services composition | Using ASP.NET web services. Also a network of web services is possible by making two or more WS talking together. | N/A  - not considered | No, JavaScript doesn't support web services composition. We can only consume web services using JavaScript. | J2EE WS composition. | Yes(by Apache Axis2) [224] |

| Features/PL | C++ | AspectJ | Haskell | PHP | Scheme |
|---|---|---|---|---|---|
| Central unit | Classes | Aspects, abstract Aspects | Type class | Classes(by PHP 5) | Meta Classes |
| Structural elements | N/A - not considered | Pointcuts, Advices, Inter-Type declarations for declaring aspect's members (fields, methods, and constructors) | Type, Methods, Member | Methods, fields | Meta Object Protocol, simple-object. |
| Multiple inheritance | Yes [235] | No (Aspect Inheritance) | Yes, Haskell support multiple type class inheritance | No(implemented by interface) | Yes, We have multiple inheritance if the instance consults multiple components (not including itself) for behavior. |
| Polymorphism | N/A - not considered | Aspectual polymorphism [44]. Method overloading, method overriding, virtual method. By default AspectJ methods are *virtual* (non-static methods). Using Inter-type declaration. | Yes. The class methods defined by a Haskell class correspond to virtual functions. type class based overloading. | Yes or No, because PHP is weakly type, and does not care about variable type. PHP does not support same function name with different parameter | N/A - not considered |
| Partial Classes | N/A - not considered | No, but it provides Inter-type declarations (aka *open classes*) as alternative. | No. Haskell does not support it | Public/private/protected | N/A - not considered |
| Structs support | N/A - not considered | No, doesn't support Structs declaration. | No, doesn't support Structs declaration. | No, all member function has to write together | N/A - not considered |
| Instantiation | N/A - not considered | Aspect instantiation: not directly instantiated. You cannot use the *new* keyword to instantiate an aspect. | Yes, use *instance* keyword | Support(by Symphony struts) | N/A - not considered |
| Extension/ Inheritance | N/A - not considered | Yes, "Aspect extension" Aspects can extend classes and implement | Yes, Haskell supports a notion of class extension | Instantiate class object using keyword "new" | N/A - not considered |

| Accessibility control | *Public/private* keyword | interfaces, but Aspects can extend only *abstract* Aspects. Yes, *privileged* for Aspects, *private*, *public* and *protected* for inter-type members. | NO, module system must be used to hide or reveal components of a class. | Yes(wrapping with C++ classes) [225] | N/A - not considered |

**OO-based abstraction**

| Features/PL | C# | Groovy | JavaScript | Java | Scala |
|---|---|---|---|---|---|
| Central unit | Classes, interfaces, abstract classes | Classes | Function | Class | Classes |
| Structural elements | Constructors, Destructors, Methods, Fields (attributes), Delegates, Properties, Indexes, Events, Finalizes, Operators, Nested Classes. | N/A - not considered | Variable, function | Constructors, Methods, Members. | Constructors, Methods, fields, traits, case class |
| Multiple inheritance | No, can be done by using Interface implementation as work around. | No, can be done by using Interface implementation. | Yes, it support multiple inheritance | No, but can implement multiple interface | No(can use traits to implement) |
| Polymorphism | Yes, through inheritance, based on *Base Type* and *Sub Type* relation. Method overloading, method overriding. C# provides the *virtual* Keyword which is derived from C++. | N/A - not considered | N/A - not considered | Yes, basic feature for OOP. Method overloading. By default methods are virtual. | Yes, Scala support type polymorphism. Yes, Scala support function overloading |
| Partial Classes | Yes, separation of code (class definition split into two or more source files. | N/A - not considered | N/A - not considered | Yes, can declare interface in a file and implement it in another file. | Public/package/private/ protected |
| Structs support | Yes, derived from C++ with enhanced features. | N/A - not considered | N/A - not considered | No, doesn't support Structs declaration. | Yes(by Scala 2.73 provide partial functions) |
| Instantiation | Yes, Class instantiation using the *new* keyword. | N/A - not considered | N/A - not considered | Yes, Class instantiation using the *new* keyword. | Support(require external build tool for Scala) [226] |
| Extension/ Inheritance | Yes, "class extension". A class may extend another class and may implement one or more Interface. Extension | N/A - not considered | N/A - not considered | Yes, a class can extend a *abstract* class. | Instantiate object using keyword "new" |

| Accessibility control | methods feature is supported since C# 3.0 but it requires a static class and static method to so [43]. Yes, C# provides *private*, *public*, *protected*, *internal*, *sealed* and *protected internal* keywords. | *Public/private* keyword | N/A - not considered | Yes, *private*, *public* and *protected* for inter-type members. | Can extend abstract class |
|---|---|---|---|---|---|

**Criteria 5**

**Reflection**

| Features/PL | C++ | AspectJ | Haskell | PHP | Scheme |
|---|---|---|---|---|---|
| Access to program's metadata | Yes. [234] | Yes, since AspectJ 5.0. Annotation-based development style. | Yes, Monadic reflection | Yes(by using annotations and reflection API supported by PHP 5.3) | Yes, access through library. |
| Generation code at run-time | Yes. [236] | AspectJ doesn't provide such feature. Can be done using the following libraries. ASM *http://asm.ow2.org* and BCEL *http://jakarta.apache. org/bcel* | No, compile-time reflection | Yes(after PHP 5.3, it provide runtime configuration ) | Yes[ |
| Dynamic invocation | N/A  - not considered | Using thisJoinPoint to access the current join point for the advice through reflection. | N/A  - not considered | N/A  - not considered | N/A  - not considered |

**Criteria 5**

<div align="center"><strong>Reflection</strong></div>

| Features/PL | C# | Groovy | JavaScript | Java | Scala |
|---|---|---|---|---|---|
| Access to program's metadata | Yes, accessing Attributes through reflection using *System. Reflection* namespace. | Yes [236]. | Yes, can access directly | Yes, since Java 5.0. Annotation-based development style. | Yes(by reflection API) |
| Generation code at run-time | Yes, using .NET's *System.Reflection.Emit* namespace. | Yes [237] | Yes, compile-time reflection | Yes, such as JUnit. | No, compile time |
| Dynamic invocation | Yes, method invocation at run-time through reflection [48]. | N/A - not considered | N/A - not considered | N/A - not considered | N/A - not considered |

**Aspect orientation**

| Features/PL | C++ | AspectJ | Haskell | PHP | Scheme |
|---|---|---|---|---|---|
| Aspect-oriented programming | Require extra compiler supported | Natively, it is an aspect-oriented programming language | AOP Haskell is an AOP extension for Haskell | Require extra library supported | Yes, as the family of languages with AOP features includes not only academic languages such as Scheme and ML but also industrially popular languages such as Python and Perl, defining aspects in this context takes on immediacy and importance |
| Modularity | isolate crosscutting but need extra compiler support. | Isolate crosscutting concerns in a modular way. | Yes, same as AspectJ. | isolate crosscutting but need extra aop library support. | Yes, object based message passing, Many Schemes provide their own modularization facilities. |
| Code reusability | N/A - not considered | Can produce pure uncoupled code | N/A - not considered | invoked external library to achieve | N/A - not considered |
| Security | Secure(base on java platform) | Securing UI by detecting Single-Thread UI's rule and web services pipelines failure. | N/A - not considered | Require extra library support | N/A - not considered |
| Errors handling/Logging | N/A - not considered | Cleaner way to handle and log exceptions | N/A - not considered | Provide exception API by PHP 5.0 or wrapper functions | N/A - not considered |

**Criteria 6**

**Aspect orientation**

| Features/PL | C# | Groovy | JavaScript | Java | Scala |
|---|---|---|---|---|---|
| Aspect-oriented programming | Partially, there are some no-mainstream projects implementing AOP for C#. | Natively, it is an aspect-oriented programming language | Yes, AspectJS is an open source and free framework for AOP in Javascript | AspectJ is extension of Java for AOP | Natively Support(by AspectJ) |
| Modularity | Difficult some times impossible to isolate crosscutting concerns from the program's business logic. | Same as AspectJ, provide crosscutting for handling on aspect | Yes, we use the main module as the central point through which data is delivered to and from other modules | Same as AspectJ | Same as AspectJ, provide crosscutting for handling on aspect |
| Code reusability | Can't isolate code that crosscut over modules. Partially, can be done using DLLs in order to reduce code coupling degree. | N/A - not considered | N/A - not considered | Same as AspectJ | Less coupled code |
| Security | Not applicable since there is no mainstream AO solution for C#. | Require extra complier support | N/A - not considered | Same as AspectJ | Secured by its access control |
| Errors handling/Logging | Exceptions are handled in *try-catch* blocks and logged within this scope. | N/A - not considered | N/A - not considered | Same as AspectJ | N/A - not considered |

| Features/PL | C++ | AspectJ | Haskell | PHP | Scheme |
|---|---|---|---|---|---|
| **Criteria 7** | | **Functional programming** | | | |
| Type inference | N/A - not considered | No, planned for Java 7 | Yes, Haskell is not necessary to write for each variable | Not support | |
| Lambda expression | Yes [238] [239] | Using interfaces as a work around. | Yes | Yes(since PHP5.3) | Yes [282] |
| Anonymous methods | | No, use inner classes as work around | Yes, using a concise syntax | Yes(since PHP5.3) [227] | |
| Higher-order functions | Yes [239] | Using interfaces as work around | Yes, Haskell can pass function as parameter and return function | Yes, PHP can assign function to variable which can be pass as parameters | Yes [281] [283] |
| Closure | N/A - not considered | No, not popular enough an open source project for OpenJDK [58]) | Yes, | Yes(since PHP5.3) | N/A - not considered |
| First-class functions(delegates) | N/A - not considered | No, AspectJ doesn't support delegates, lazy functional programming can be used as work around [53]. | Yes, function is first class object | Yes[JL - 3] | N/A - not considered |
| Recursion | Yes | Not secure as in C# [57] due to infinite methods loop. | Yes, Haskell using recursion to support loop | Yes | Yes [284] |

# Functional programming

| Features/PL | C# | Groovy | JavaScript | Java | Scala |
|---|---|---|---|---|---|
| Type inference | Yes, functional programming's features have been introduced in C# 3.0 | N/A - not considered | N/A - not considered | Functional Java – a extension of Java, but each variable has to assign a type | Yes |
| Lambda expression | Yes, since C# 3.0 using LINQ library, *System.Linq.Expression* namespace | No [239] [240] | Yes [281] | No, Java does not support it | Yes, defined in a very succinct fashion [227] |
| Anonymous methods | Yes, since C# 2.0 where this feature has been introduced. | N/A - not considered | N/A - not considered | No, Java has anonymous classes, but do not support anonymous functions | Yes, Scala provides a relatively lightweight syntax for defining anonymous functions [277] |
| Higher-order functions | Yes using C# delegates. | Yes [240] | Yes [285] | Java cannot pass function and return function | Yes, can pass function and return function [227] |
| Closure | Yes, since C# 3.0 | N/A - not considered | N/A - not considered | No, but Java simulate some features of closures [211] | Yes |
| First-class functions(delegates) | Yes, delegates are C++'s function pointers equivxalent. | N/A - not considered | N/A - not considered | No, function is not first class object for Java | Yes [278] |
| Recursion | Yes it is fully supported. Can be done using C#'s methods. | Yes | Yes [286] | Yes, Java function can call itself | N/A - not considered |

**Criteria 8**

**Declarative programming**

| Features/PL | C++ | AspectJ | Haskell | PHP | Scheme |
|---|---|---|---|---|---|
| Tags on methods/fields/ classes | N/A | AspectJ/Java Annotations | Annotations | annotations | |
| LINQ equivalent? | N/A | Additional libraries, lamdaJ [63] | Yes, basic feature | Yes (require extra implementation support). | LINQ for R6RS Scheme[287] |
| Dynamic programming (Dynamic type checking) | N/A | No, AspectJ doesn't provide such features. | No, Haskell is static type checking | Yes | Yes, Scheme is dynamic typing |
| Declarative programming based on XML [41] | N/A | YAML, SwiXML | Yes, HAXML | Symphony YAML | XML Schema |

**Criteria 8** — **Declarative programming**

| Features/PL | C# | Groovy | JavaScript | Java | Scala |
|---|---|---|---|---|---|
| Tags on methods/fields/ classes | Using C# Attributes | N/A - not considered | N/A - not considered | Annotations | annotations |
| LINQ equivalent? | Yes, *system.data.linq* library should be used and referenced. | N/A - not considered | N/A - not considered | Additional libraries, lamdaJ | Yes |
| Dynamic programming (Dynamic type checking) | Yes – since C# 3.0 (*var* keyword) and C# 4.0 (*dynamic* keyword) | N/A - not considered | N/A - not considered | No | No(static type checking) |
| Declarative programming based on XML [41] | Using XAML technology which is available since .NET 3.0 and WPF | N/A - not considered | N/A - not considered | YAML, SwiXML | Bad support on XML [228] |

**Criteria 9**

<div align="center"><b>Batch scripting</b></div>

| Features/PL | C++ | AspectJ | Haskell | PHP | Scheme |
|---|---|---|---|---|---|
| Run external programs/external commands | Yes [241] | Yes, using Java's Process and Runtime classes. | Yes, using system or rawsystem function | Yes, such as exec, system | Yes, using scheme/system [279] library which allows running external commands and programs. |
| Tasks automation | Yes(by COM objects) | Yes, can be executed as stand alone and executed on a scheduled basis. | Yes | Yes(use cur command) | Yes, Racket's files can be executed as standalone programs on Windows or on any Unix-based OSs. |
| Accept command line arguments | Yes [242] | Yes as parameters for the *main* method. i.e *public static void main(string[] args)* | Yes | Yes(by PHP CLI script) | Yes, on Unix-based terminal or MS DOS using scheme/cmdline from Racket libraries [279]. |
| Need to be recompiled after changing the code source | Not compared | Yes, code should be recompiled in order to source code changes take effect. | Yes | Yes [230] | Yes, need to recompile Racket scripts. |

**Criteria 9**

| Features/PL | C# | Groovy | JavaScript | Java | Scala |
|---|---|---|---|---|---|
| Run external programs/external commands | Yes, using Process class in *System.Diagnostics* namespace. | Yes [243] | Only with Internet Explorer using ActiveXObject. | Yes, using Runtime class | Yes |
| Tasks automation | Yes, can be executed as stand alone and executed on a scheduled basis. | Yes(by Groovy Monkey) | Yes, nearly in every web browsers with support for JavaScript. | Yes, | Yes |
| Accept command line arguments | Yes as parameters for the *Main* method. i.e *public static void Main(string[] args)* | Yes(using groovy script and listen mode) | No, JavaScript run with a browser and not from the command line. | Yes | Yes |
| Has to recompile after changing the code source | Yes, code should be recompiled in order to source code changes take effect. | N/A  - not considered | No, no need to recompile JavaScript after changing the source code | Yes | Yes |

| Features/PL | C++ | AspectJ | Haskell | PHP | Scheme |
|---|---|---|---|---|---|
| Graphical user interface | Yes, using Qt, a C++ class library. | Yes, using Java's built-in Swing and AWT libraries. | WxHaskell, Gtk2Hs, HOC, qtHaskell and so on | Yes(PHP-GTK or embelled with HTML) | Yes, Racket Graphics Toolkit, PLT MrEd Graphical Toolbox |
| Built-in? | N/A - not considered | Yes, part of Java Foundation Classes | No, all are extension of Haskell | Yes(using PHP-GTK+ by PHP 4.0) | No, |
| Look and feel | N/A - not considered | Yes such as GTK+, Motif, Windows, Macintosh, etc themes. | More complex than Java | | Yes |
| 2D and 3D support | Yes using OpenGL | Yes – Java 2D API | Yes, FRAN [212] | | Yes – sgl library to support 3D |
| Drag and Drop support | N/A - not considered | Yes, using *Java.awt.dnd* package. | Yes, Gtk2Hs support D&D | Yes, PHP-GTK 2 | Yes, pasteboard in MrEd [1] |
| Components' Layout Management | N/A - not considered | Yes, i.e Border/Grid layout, etc. | Yes, using layout container | N/A - not considered | Yes, specifies the layout of a window by assigning each GUI element to a parent containerrs [2] |
| Multiple document interface (MDI) | Yes using QT | Yes, provided by AWT | Yes, wxHaskell has basic MDI support | Yes | Yes, using frame% |
| Model-View-Controller | Support but not enforce. | Yes in components such as JTable and JList, etc. | Yes | | Yes, Racket GUI Application Framework |
| Declarative GUI development [41] | N/A - not considered | Glade XML, SwiXML: not popular yet as much as XAML | Yes, FranTk | GTK+ | Using XAML |
| Performance | N/A - not considered | Relatively Slow | Working well on both Linux and windows | Relatively Slow | Provide good memory performance |
| IDE/UI designer | N/A - not considered | NetBeans, Plugins for Eclipse IDE | Glade | Zend, Dreamware | WinScheme Editor |
| Rich Web UI Development | N/A - not considered | Yes, using JavaFX framework. | HsWTK | | Yes |
| Deployment | N/A - not considered | Jar archive (.jar files) | .hs file | | Yes, Event Listeners |
| Single thread of execution | N/A - not considered | Yes, Swing library's single-thread rule, thread-safe[73] | Yes. | Yes, PHP does not support multiple thread | N/A - not considered |
| Event handling mechanism | N/A - not considered | Only Event Listeners, provided by *java.awt.event* package | Yes, Event Handlers | N/A | N/A |

**Criteria 10**

<div align="center">

**UI prototype design**

</div>

| Features/PL | C# | Groovy | JavaScript | Java | Scala |
|---|---|---|---|---|---|
| Graphical user interface | Yes, using .NET's Windows Forms and WPF libraries. | Yes, using Java's built-in Swing and AWT libraries. | Support DHTML to implement GUI | Yes, using Java's built-in Swing and AWT libraries. | Yes(using java ... ScalaGUI) |
| Built-in? | Built on top of the Base Class Library You need to import/reference *System.Windows.Forms.dll* for WinForms and *System.Windows.dll* WPF. WPF only on Windows platform [70] | N/A - not considered | Javascript has not built in GUI library. | Yes, part of Java Foundation Classes | Yes(using ScalaGUI, ... library as Java GUI) |
| Look and feel | Not flexible enough in WinForms. Full customizable UI support in WPF | N/A - not considered | Very flexible, and easy to implement powerful UI on webpage, | Yes such as GTK+, Motif, Windows, Macintosh, etc themes. | N/A - not considered |
| 2D and 3D support | Yes, using WPF framework | N/A - not considered | Javascript can support Html to show 2D graphic | Yes – Java 2D API and Java 3D API | N/A - not considered |
| Drag and Drop support | Partially, only using WPF framework [71] | N/A - not considered | No, does not support | Yes, java.awt.dnd | N/A - not considered |
| Components' Layout Management | Yes, this can be done automatically using Visual Studio's UI designer. | N/A - not considered | Yes, base on DHTML and Css | Yes, Several AWT and Swing classes provide layout managers | N/A - not considered |
| Multiple document interface (MDI) | Yes, using, *System.Windows.Forms.MdiLayout* class. | Yes, using Java Swing library. | No | Yes, using JDesktopPane | N/A - not considered |
| Model-View-Controller | WinForms no, WPF Model-View-ViewModel [72] | Yes,enforce by Swing | No | Yes in components such as JTable and JList, etc. | N/A - not considered |
| Declarative GUI development [41] | Using XAML technology | N/A - not considered | No, | using XAML | XAML |
| Performance | Improved execution on windows Fast | | Javascript is a slow language | Relatively Slow | Slow as java graphic |
| IDE/UI designer | Visual Studio C# and SharpDevelop for Windows, MonoDevelop . | N/A - not considered | No | NetBeans, Plugins for Eclipse IDE | Scala for eclipse, Net b... |
| Rich Web UI Development | Silverlight framework on windows and Moonlight implementation for Linux and other Unix-based OSs. | N/A - not considered | Yes, Javascript can only run in single thread | Yes, using JavaFX framework. | Silverlight framewo... windows and M... implementation for L... other Unix-based OSs. |
| Deployment | .NET Assemblies: (executables and libraries, .exe and DLLs) | N/A - not considered | | Jar archive (.jar files) | N/A - not considered |
| Single thread of execution | Windows Forms (Not by default, but by applying the **STAThread attribute to the Main method**). WPF use by default a single thread of execution. | N/A - not considered | Yes, Javascript can only run in single thread | Yes, Swing library's single-thread rule, thread-safe[46] | N/A - not considered |
| Event handling mechanism | Yes, Event Handlers and Delegates support | N/A - not considered | Yes, can handle event coming from webpage | Only Event Listeners | N/A - not considered |

## 3.1    Criteria 1: Default more secure programming practices

The comparison tables show that the compared languages have different type system. Some of them are strongly typed and others are weakly typed. PHP, Groovy and JavaScript are dynamic typed which is not secure enough because type checking is performed at run-time. In C# 4.0, it is possible to use dynamic typing using the *dynamic* keyword. C#, Java, Haskell and  Scalar are said strongly typed.

Among these languages, only C++ doesn't provide automatic memory management mechanism such as garbage collection and memory layout, etc which is not safe. In C++ dangling pointers, memory leak and double pointers free are well known issues where programmers are called to pay more attention to prevent them.

All languages provide an exception handling mechanism for detecting errors and prevent applications from crashing. C# and Java have the best exception handling implementation by providing a set of customized exception classes.

Haskell support exception handling, but it is a complex task to achieve.

## 3.2    Criteria 2: Web applications development

We have found that all 10 languages support web applications development. However, we have found that ASP.NET/C#, PHP, Java/J2EE are particularly suited for easy web design. It is difficult to say which one is better than others. These three technologies provide powerful feature and have been adopted in large web applications development project. Thus, during the last decade they also gained popularity.

PHP is an easiest language to learn, but it seems that it is not safe enough and doesn't provide a strong type safety since it is dynamically typed.

We have found out also that the remaining languages can also be used in web applications as CGI scripts, but are quite inconvenient and error-prone to program in them, maintain, and deploy. Hence, they require external libraries which are hard to configure, maintain and deploy.

## 3.3    Criteria 3: Web services design and composition

PHP, Java, C#, Groovy and Scala enable developers to write web services. As shown in the comparison tables, we have found that ASP.NET/C#, PHP, Java/J2EE are particularly suited for developing WS. It is seems that Java/J2EE is better than others languages. Java is cross-platform and can be coupled with AspectJ to implement a secure, stable composed web services.

ASP.NET/C# is an easiest language to learn. Visual Studio is a powerful IDE which provides helpful features for building and debugging web services applications.

Also, we have found that is possible to build such applications using the remaining languages. But it is painful since they are natively not designed/suited for kind of applications and some external libraries are required to do so.

## 3.4    Criteria 4: OO-based abstraction

According to our comparison table, Java, C#, Groovy, Scala are pure object-oriented programming languages. C++, PHP 5.0 also support OO development but they are not pure OOP and don't provide all OO features found in Java and C#.
 C# and Java are particularly more suited for OO development. It is difficult to say which one is better than others. But C# has introduced more OO features such as Properties, Indexers, delegates, etc.
Java and C# have a similar syntax and they are easy to learn, and both are strongly typed.

Other partially support OOP, but in some of them OOP has been introduced lately or as extension libraries. However, they are not suited to be adopted in a large OO development project.

## 3.5    Criteria 5: Reflection

Our comparison tables let us conclude that all compared languages support reflection and provide different mechanism to implement it. Some of them don't require external libraries especially C# which provides the *System.Reflection.Emit*  namespace that allow to generate code at run-time. AspectJ and Java indirectly provide this feature using external libraries such as ASM and BCEL.

For others languages, reflection is supported but not mainly used in development.

## 3.6    Criteria 6: Aspect-orientation

Based on our research, AspectJ was the first aspect-oriented language. Since it extends Java, it gained a wide popularity and has become mainstream language for AOP. Other languages may support AO if there are extensions available. For example, Groovy, Haskell and Scala have already mainstream extension for AOP. Unfortunately, there are no mainstream AOP extension for C#.
To learn AspectJ, programmers have to be familiar with Java language since it inherits syntax and features from Java.
  C++ is not suited to AOP. Dealing with pointers in AO it is not safe enough and can result in memory issues.

## 3.7    Criteria 7: Functional programming

Obviously, Haskell and Scheme are natively functional programming languages. They provide all functional programming features, especially Haskell which is pure functional language. Basically Java, Groovy, AspectJ and Scala don't support functional programming. Type inference is planned to be released with Java 7.

C# 3.0 has been introduced with the release of LINQ which provides functional programming capabilities such as lambda expression, anonymous methods, and type inference and high-order functions.

## 3.8    Criteria 8: Declarative programming

Our tables prove that Haskell is the best language to support declarative programming. Haskell is pure functional programming language with minimized side effect.
Other languages provide what is called annotation mechanism which is considered as declarative approach since programmers provide information used by the program at run-time.
A new declarative methodology has been introduced as extension to OOP based on XML technology. XAML was the first one introduced by Microsoft which enables developers to describe structured values and objects before coding.
C# is the best language which is suited for mixing OOP and declarative style using LINQ services especially for querying database and XML files.

## 3.9    Criteria 9: Batch scripting

According to our comparison studies, batch scripting is possible in all languages except JavaScript which is possible using ActiveXObject and Internet Explorer browser.

PHP, C#, Scala and Java are the best. A program written in one of them can be automated (scheduled using *Scheduled tasks* in Windows and *Crontab* in a Unix-based system). PHP and C# are used to automate report generation such as sales report, job alerts, etc.

PHP is more suited one since is not compiled, so if some changes has occurred on the source code, there is no need to recompile an already deployed script.

## 3.10    Criteria 10: UI prototype design

After comparing more than ten features, we have found that is possible to create UI in any one of these 10 languages. C#, Java, Scala, Groovy and AspcetJ are particularly suited for GUI design.
Java, Scala, Groovy and AspectJ GUIs are based on Java AWT/Swing libraries.
There are a lot of extension libraries to support GUI design in Haskell but these libraries are not standardized and it is hard to make them working together.
We conclude that PHP and JavaScript are more suitable for web pages UI design since they can easily mixed/embedded in HTML code.

C# is the most suitable for designing and developing GUI. Visual Studio IDE provides a UI designer with a big set of pre-built component. Thus, C# can use WPF technology to create powerful highly customizable UIs.

# 4 Conclusion

We have performed a thorough and detailed comparison of our selected programming languages to study within the specified criteria. We have found that, by their native design, each language is suited for a specific field. Some of them can be fully applied to the compared criteria without extensions.

C# and Java have demonstrated that they are the more suitable for both web and desktop applications development. They are strongly typed which enable to write more secure programs. On the other hands, AspectJ as an extension of Java language is the best mainstream aspect-oriented language. PHP and JavaScript are particularly suitable for web development. Haskell and Scheme are better for functional programming. C++ is more suitable for system development and desktop applications. Scala and Groovy have extended Java.

## 4.1 Future work

We'd like to refine our analysis of our languages within the stated criteria further as we become more familiar with them over time. We also plan on expanding our analysis onto other criteria and languages and provide more programming snippets as proof-of-concept illustrations.

## 4.2 Acknowledgments

We would like to acknowledge the following people and entities who made this work possible:
- Faculty of Engineering and Computer Science, Concordia University, Montreal, Canada.
- Concordia University Libraries for access to the invaluable digital libraries of ACM, IEEE, Springer and others to do our research.
- Wikipedia contributors with the wealth of information.
- Our poor families, wives, husbands, children, parents, and pets to help us to get through the suering and sleepless nights and oer all their help and understanding while we were away from them while doing this project.
 Our POD, Yi Ji, for the introductions into AspectJ and Java reflection.

# Acronym and abbreviation

**IDE:**  Integrated Development Environment
**CLR**:  Common Language Runtime
**JVM**:  Java Virtual Machine
**DOS**:  Disk Operating System
**GUI**:  Graphical User Interface
**WPF**:  Windows Presentation Foundation
**API**:  Application Programming Interface
**WCF**:  Windows Communication Foundation
**RAD**:  Rapid Application Development
**API**:  Application Programming Interface
**JDBC**:  Java Database Connectivity
**SOA**:  Service Oriented Architecture
**J2EE:**  Java 2 Enterprise Edition
**TUI:**  Text User Interface
**CLA**:  Command Line Application
**MDI:**  Multiple Document Interface
**JFC:**  Java Foundation Classes
**LINQ**:  Language Integrated Query
**UI**:  User Interface
**BCL**:  Basic Class Library

# References

[1] The C++ Resources Network,  http://www.cplusplus.com/
[2]  AspectJ in Action, *Practical Aspect-Oriented Programming,* Ramnivas Laddad.
    ISBN: 1930110936
[3] Programming paradigm, Wikipedia, http://en.wikipedia.org/wiki/Programming_paradigm
[4] AspectJ projecrt, Eclipse foundation, http://www.eclipse.org/aspectj/
[5] Haskell project, http://www.haskell.org/
[6] Scheme implementation choices, http://web.mit.edu/~axch/www/scheme/choices.html
[7] Gambit project, http://dynamo.iro.umontreal.ca/~gambit/wiki/index.php/Main_Page
[8] PLT Scheme project, http://www.plt-scheme.org/
[9] The Scheme programming language, http://www.scheme.com/tspl3/
[10] Groovy (programming language), Wikipedia,
    http://en.wikipedia.org/wiki/Groovy_(programming_language)
[11] Java programming language,
    http://en.wikipedia.org/wiki/Java_%28programming_language%29#Practices
[12] [17] Introducing the Scala, "The Scala Programming languages".
    Retrieve from: http://www.scala-lang.org/node/25
[13] Visual C# Developer Center, http://msdn.microsoft.com/en-ca/vcsharp/default.aspx
[14] C# programming language , wikiepdia,
    http://en.wikipedia.org/wiki/C_Sharp_%28programming_language%29
[15] ECMA standard, http://www.ecma-international.org/default.htm

[16] Unsafe Code Tutorial,
     http://msdn.microsoft.com/en-us/library/aa288474%28VS.71%29.aspx
[17] Garbage collection (computer science), Wikipedia,
    http://en.wikipedia.org/wiki/Garbage_collection_%28computer_science%29
[18] Data Types (C# Programming Guide),
    http://msdn.microsoft.com/en-us/library/ms173104%28VS.80%29.aspx
[19] Bounds checking, Wikipedia, http://en.wikipedia.org/wiki/Bounds_checking
[20] COMP 6411 lecture notes, Joey Paquet and Serguei A.  Mokhov.
[21]Comaparison of Java and C#,
    http://en.wikipedia.org/wiki/Comparison_of_Java_and_C_Sharp#Conditional_compilation
[22]Assertions in Managed Code,
     http://msdn.microsoft.com/en-us/library/ttcc4x86%28v=VS.71%29.aspx
 [23]Static Conditional Pointcut Evaluator for AspectJ,
    http://www.graco.c.u-tokyo.ac.jp/ppp/index.php?Projects%2Fscope
[24]  J-LO, the Java Logical Observer, A tool for runtime-checking temporal assertions,
        http://www.sable.mcgill.ca/~ebodde/rv//JLO/
[25] MSDN's blog, Array Bounds Check Elimination in the CLR,
     http://blogs.msdn.com/b/clrcodegeneration/archive/2009/08/13/array-bounds-check-elimination-
in-the-clr.aspx
[26] Project Hosting for Open Source Software, http://www.codeplex.com/
[27] Asp.net CRM, http://crm.codeplex.com/
[28] nopCommerce. Open Source online shop e-commerce solution,
        http://nopcommerce.codeplex.com/
[29] Lesiecki, N. "Applyinq AspectJ to J2EE application development," *Software, IEEE* , vol.23, no.1,
     pp.24-32, Jan.-Feb. 2006 doi: 10.1109/MS.2006.1
[30]  SpringSourse Tool Suite, http://www.springsource.org/
[31]  Frequently Asked Questions, Eclipse network resource
        http://www.eclipse.org/aspectj/doc/released/faq.php#q:aspectjandj2ee
[53-] [32] J2EE vs. Microsoft.NET,
        http://media.techtarget.com/tss/static/articles/pdf/J2EE-vs-DotNET.pdf
[33] SpringSource, Spring project,

http://static.springsource.org/spring-security/site/docs/3.0.x/reference/session-mgmt.html
[34]ASP.NET Security Architecture, http://msdn.microsoft.com/en-us/library/yedba920.aspx
[35] Tanenbaum, A. S. and van Steen, M. "Distributed Systems: Principles and Paradigms"
[36] Sasa Subotic and Judith Bishop :"Emergent behaviour of aspects in high performance and
     distributed computing" ,Year of Publication: 2005,ISBN:1-59593-258-5
[37] Conceptual Overview, http://msdn.microsoft.com/en-us/library/ms731190.aspx
[38] Apache XML-RPC project, http://ws.apache.org/xmlrpc/
[39] HTTP Security and ASP.NET Web Services,
     http://msdn.microsoft.com/en-us/library/ms996415.aspx
[41] Object-Oriented Programming (C# and Visual Basic), http://msdn.microsoft.com/en-
us/library/dd460654.aspx
[42] System.Web.Services Namespace, http://msdn.microsoft.com/en-us/library/9xe4bs0s.aspx
[43] Extension Methods (C# Programming Guide),
     http://msdn.microsoft.com/en-us/library/bb383977.aspx

[44] Erik Ernst and David H. Lorenz: "Aspects and polymorphism in AspectJ", Year of Publication:
       2003, ISBN:1-58113-660-9
[45] New Reflection Interfaces,
     http://www.eclipse.org/aspectj/doc/next/adk15notebook/reflection.html
[46] AspectJ, Eclipse network resource, http://www.eclipse.org/aspectj/doc/released/faq.php
[47] Reflection Overview,
     http://msdn.microsoft.com/en-us/library/f7ykdhsy%28v=VS.80%29.aspx
 [48] MSDN, Reflection the C# programming guide,
     http://msdn.microsoft.com/en-us/library/ms173183%28VS.80%29.aspx
 [49] ASpectJ publications,
     http://dev.eclipse.org/viewcvs/indextech.cgi/aspectj-home/publications.html
[50] Tigris.org Open Source Software Engineering Tools, http://aspectdng.tigris.org/
[51] Operating system + middleware, http://www.dcl.hpi.uni-potsdam.de/research/loom/
[52]  "Delegates and functional programming in C#", David R. Naugler Southeast Missouri State
     University, Cape Girardeau, MO, 2004
[53] "Lazy functional programming in Java ", Anthony H. Dekker Defence Science and Technology
     Organisation, Canberra ACT SSN:0362-1340
[54] "Functional programming in Java", David R. Naugler, Southeast Missouri State University Cape
     Girardeau MO, ISSN:1937-4771
[55] Functional Programming for Everyday .NET Development,
     http://msdn.microsoft.com/en-us/magazine/ee309512.aspx
[56] C# Recursion, http://www.meshplex.org/wiki/C_Sharp/Recursion
[57] Avoiding Infinite Recursion with Stratified Aspects,
     http://www.sable.mcgill.ca/~ebodde/meta/
[58] Project Lambda, http://openjdk.java.net/projects/lambda/
[59] An Annotation Based Development Style,
     http://www.eclipse.org/aspectj/doc/next/adk15notebook/ataspectj.html
[60] Introduction to Attributes, http://msdn.microsoft.com/en-us/library/Aa288059
[61] XAML Overview (WPF), http://msdn.microsoft.com/en-us/library/ms752059.aspx
[62] YAML, Wikipedia, http://en.wikipedia.org/wiki/YAML
[63] Lambdaj, http://code.google.com/p/lambdaj/
[41] Louridas, P.; , "Declarative GUI Programming in Microsoft Windows," *Software, IEEE* , vol.24,
     no.4, pp.16-19, July-Aug. 2007 doi: 10.1109/MS.2007.105
[38] [64] Basic Console Application (C#), http://msdn.microsoft.com/en-us/library/bb251798.aspx
[65] Capitalization Styles, http://msdn.microsoft.com/en-us/library/x2dbyw72%28v=VS.71%29.aspx
[66] Text-based (computing), Wikipedia, http://en.wikipedia.org/wiki/Text-based_%28computing%29
[67] Java SE Desktop Overview, http://java.sun.com/javase/technologies/desktop/
[68] Windows Forms, Wikipedia, http://en.wikipedia.org/wiki/Windows_Forms
[69] Windows Presentation Foundation on the Web: Web Browser Applications,
http://msdn.microsoft.com/en-us/library/aa480223.aspx#wpfandwbas_topic1

[70] Mono project, WPF, http://www.mono-project.com/WPF
[71] MSDN, .Net framework 4, Drag and Drop Overview, http://msdn.microsoft.com/en-us/library/ms742859.aspx
[72] WPF Apps With The Model-View-ViewModel Design Pattern, http://msdn.microsoft.com/en-us/magazine/dd419663.aspx

[73] Threads and Swing, http://java.sun.com/products/jfc/tsc/articles/threads/threads1.html

[74] Comparison of programming languages, Wikipedia, http://en.wikipedia.org/wiki/Comparison_of_programming_languages
[75] Web Applications, http://msdn.microsoft.com/en-us/library/ms235434(VS.80).aspx
[76] WT, http://www.webtoolkit.eu/wt
[77] C++ Applications, http://www2.research.att.com/~bs/applications.html
[78] Web Development in Groovy using Groovlets, http://www.javabeat.net/articles/58-web-development-in-groovy-using-groovlets-1.html
[79] Groovy and Grails – A Getting Started Guide, http://www.indicthreads.com/1481/groovy-and-grails-a-getting-started-guide/
[80] Groovying XML, http://www.techbookreport.com/tutorials/groovy_xml_01.html
[81] GroovyWS,The Groovy Resources Network, http://groovy.codehaus.org/GroovyWS
[82] Groovy SOAP, The Groovy Resources Network, http://groovy.codehaus.org/Groovy+SOAP
[83] Practically Groovy: Building, parsing, and slurping XML, http://www.ibm.com/developerworks/java/library/j-pg05199/index.html
[84] *Savage, W. J. (W. John). Groovy programming : an introduction for Java developers. Amsterdam ; Boston : Morgan Kaufmann Publishers.*

[85] WS-Eventing for WCF , http://www.codeproject.com/KB/WCF/WSEventing.aspx
[86] The gSOAP Toolkit for SOAP Web Services and XML-Based Applications , http://gsoap2.sourceforge.net/
[87] Walkthrough: Creating an XML Web Service Using C++ and the CLR, http://msdn.microsoft.com/en-us/library/a86z84tw(VS.80).aspx

[88] JN3025-Inheritance, The Groovy Resources Network, http://groovy.codehaus.org/JN3025-Inheritance
[89] Abstract Classes (C++), http://msdn.microsoft.com/en-us/library/c8whxhf1.aspx
[90] Abstract classes(IBM) , http://publib.boulder.ibm.com/infocenter/comphelp/v8v101/index.jsp?topic=/com.ibm.xlcpp8a.doc/language/ref/cplr142.htm
[91] Liang, Y. Daniel.  Introduction to programming with C++. Upper Saddle River, NJ ; Montreal : Prentice Hall.
[92] Q&A for professional and enthusiast programmers , http://stackoverflow.com/questions/359237/why-does-c-not-have-reflection
[93] JN3535-Reflection, The Groovy Resources Network, http://groovy.codehaus.org/JN3535-Reflection
[94] AspectC++, Wikipedia, http://en.wikipedia.org/wiki/AspectC%2B%2B
[95] The Home of AspectC++, http://www.aspectc.org/Home.1.0.html
[96] Groovy AOP, http://chanwit.blogspot.com/2007/12/groovy-aop-part-4-getter-and-setter.html
[97] Easy AOP with GroovyInterceptable, http://www.justinspradlin.com/programming/easy-aop-with-groovyinterceptable/
[98] *Suman Roychoudhury, Jeff Gray, Jing Zhang, Purushotham Bangalore, and Anthony Skjellum.  A Program Transformation Technique to Support AOP within C++ Templates. Dept. of Computer and Information Sciences, University of Alabama at Birmingham, Alabama, USA*
[99] Functional Programming with Groovy, The Groovy Resources Network, http://groovy.codehaus.org/Functional+Programming+with+Groovy
[100] Functional Programming in C++, http://www.cc.gatech.edu/~yannis/fc++/

[101] Boost.FC++, http://www.cc.gatech.edu/~yannis/fc++/boostpaper/fcpp.html
[102] Imperative programming, Wikipedia, http://en.wikipedia.org/wiki/Imperative_programming
[103] How to make System command calls in Java/Groovy,
http://stackoverflow.com/questions/2701547/how-to-make-system-command-calls-in-java-groovy
[104] Windows Services, The Groovy Resources Network,
http://groovy.codehaus.org/Windows+Services
[105] robertbody C++ & *.BAT, http://www.robertbody.com/prog/cpp-bat.html
[106] GUI Programming with Groovy, The Groovy Resources Network,
http://groovy.codehaus.org/GUI+Programming+with+Groovy
[107] The Groovy Resources Network, http://groovy.codehaus.org/api/index.html
*[108] Abdul-Jawad, Bashar. Groovy and Grails recipes. Berkeley, Calif. : Apress ; New York, N.Y. :
Distributed to the book trade worldwide by Springer-Verlag New York.*
*[109] König, Dierk. Groovy in action. Greenwich, [Conn.] : Manning.*

[110] Bryan O'Sullivan, Don Stewart, and John Goerzen(2008). Real World Haskell, (Chapter
        17) O'Reilly Media
[111] Haskell project, memory management,
        http://www.haskell.org/haskellwiki/GHC/Memory_Management
[112] Sun Microsystems (April 2006), Memory Management in the Java HotSopt Virtual
        Machine
[113] Haskell project, documentation,
        http://www.haskell.org/ghc/docs/6.12.2/html/libraries/base-4.2.0.1/Data-Maybe.html
[114] Bryan O'Sullivan, Don Stewart, and John Goerzen(2008). Real World Haskell, (Chapter
        19) O'Reilly Media
[115] Wikipedia, Web application, http://en.wikipedia.org/wiki/Web_application
[116] Practical web programming in Haskell, Haskell wiki
        http://www.haskell.org/haskellwiki/Practical_web_programming_in_Haskell

[117] Eric Armstrong , Jennifer Ball, Stephanie Bodoff, Debbie Bode, Carson, Ian Evans, Dale Green,
Kim Haase, Eric Jendrock , J2EE 1.4 Tutalial, 2004
[118] Simon Foster, University of Sheffield, HAIFA : An XML Based Interoperability
        Solution for Haskell
[119] Java API for XML-Based RPC (JAX-RPC),
        http://java.sun.com/webservices/jaxrpc/overview.html
[120] Martin Sulzmann, Meng Wang, Aspect-Oriented Programming with Type Classes
[121] Haskell's overlooked object system, http://homepages.cwi.nl/~ralf/OOHaskell/
[122] Oleg Kiselyov, Ralf lammel, Haskell's overlooked object system, 10 Sep, 2005[123] Andrzej
Filinski, Monadic Reflection in Haskell, Datalogisk institut,
        Københavns Universitet, 2006
[124] Aspect-Oriented Programming in Java,
        http://www.voelter.de/data/articles/aop/aop.html
[125] Functional Java project, http://functionaljava.org/
[126] Declarative programming, Wikipedia, http://en.wikipedia.org/wiki/Declarative_programming
[127] Declarative    Programming    in    Java    using    Annotations    and    Reflection,
http://www.riedquat.de/articles/javaDecl
[128]    System.Cmd,    http://www.haskell.org/ghc/docs/6.12.1/html/libraries/process-1.0.1.2/System-
Cmd.html
[129]        Applications        and        libraries/GUI        libraries,
http://www.haskell.org/haskellwiki/Applications_and_libraries/GUI_libraries
[130] Declarative Programming in Java, http://onjava.com/pub/a/onjava/2004/04/21/declarative.html
[131] Haskell, http://www.haskell.org/haskellwiki/WxHaskell
[132] Gtk2Hs, http://haskell.org/gtk2hs/
[133] Haskell/GUI, Wikepedia, http://en.wikibooks.org/wiki/Haskell/GUI
[134] QtHaskell, http://qthaskell.berlios.de/
[135] Garbage collection, Retrieve from:

http://phparch.cn/index.php/php/62-articles-and-reviews/299-PHP-%E4%B9%8B%E5%9E%83%E5%9C%BE%E5%9B%9E%E6%94%B6

[136] Basic memory management, PHP Manual,
Retrieve from: http://www.php.net/manual/en/internals2.memory.management.php

[137] [13] Web service, Wikipedia, Retrieve: http://en.wikipedia.org/wiki/Web_service#cite_note-0

[138] A.Gutsmans, S.S Bakken and D.Rethans,Chpater 15.2 Memory Management, PHP 5 Power programming, Indianapolis: Prentice Hall

[139] PHP exception handling,
Retrieve from: http://www.microshell.com/programming/php/php-exception-handling/

[140] D.Wampler and A. Payne (2009), Chapter 5 sensible typing ,Programming in Scala,O'REILLY

[141] M.Huniewicz, Scala-performance-garbage collection analysis,
Retrieve from:http://blog.m1key.me/2010/04/scala-performance-garbage-collection.html

[142] J.EICHAR, Exception handling,
Retrieve from:http://daily-scala.blogspot.com/2009/09/exception-handling.html

[143] PHP Web application, Retrieve from: http://en.wikipedia.org/wiki/PHP

[144] D.Polak, Lift 1.0 released, Scala Blog
Retrieve from: http://www.scala-blogs.org/2009/02/lift-10-released.html

[145] A.Gutsmans, S.S Bakken and D.Rethans, chapter 1.3.1 XML and Web Services, chapter 8.7 PHP's SOAP Extension, PHP 5 Power programming, Indianapolis: Prentice Hall

[146] Web service, PHP Manual, Retrieve from: http://www.php.net/manual/en/refs.webservice.php

[147] A.Gutsmans, S.S Bakken and D.Rethans, Forword, PHP 5 Power programming(pp.xii), Indianapolis: Prentice Hall

[148] http://www.scala-lang.org/node/25

[149] XML-RPC working graph, Retrieve from: http://www.xmlrpc.com/

[150] K.Waterson, Abstract Classes, Retrieve from:
http://phpro.org/tutorials/Object-Oriented-Programming-with-PHP.html

[151] Tours of Traits, Retrieve from: http://www.scala-lang.org/node/126

[152] Tours of Traits, Retrieve from: http://www.scala-lang.org/node/126

[153] Reflection (Computer Science), Wikipedia,
Retrieve from: http://en.wikipedia.org/wiki/Reflection_%28computer_science%29#PHP

[154] Reflection, Retrieve from: http://www.tuxradar.com/practicalphp/16/4/0

[155] The Reflection class, PHP Manual.
Retrieve from:http://www.php.net/manual/en/class.reflection.php

[156] Chapter 12 Scala type system, programming Scala

[157] Sample code example for Scala reflection,
Retrieve from: http://gpiancastelli.altervista.org/scala-it/esempi/cap-12/jvm-script.scala

[158] D.Sheiko, Aspect-Oritented Programming in PHP,
Retrieved from: http://www.weberdev.com/ViewArticle/Aspect-Oriented-Programming-and-PHP

[159] D.Wampler and A. Payne (2009), chapter 14: Scala tools, libraries, and IDE support: Java library interpretability, Programming in Scala (pp.369-381), O'REILLY

[160] D. Wamper, Trait vs. Aspects in Scala, 2008,
Retrieve from: http://blog.objectmentor.com/articles/2008/09/27/traits-vs-aspects-in-scala

[161] D.Wampler and A. Payne (2009), chapter 14: Scala tools, libraries, and IDE support: Java library interpretability, Programming in Scala (pp.378-379), O'REILLY
Available on: http://gpiancastelli.altervista.org/scala-it/esempi/cap-14/aspectj/complex.scala
            http://gpiancastelli.altervista.org/scala-it/esempi/cap-14/aspectj/complex-main.scala
        http://gpiancastelli.altervista.org/scala-it/esempi/cap-14/aspectj/LogComplex.aj

[162] T.K Nieleson, The state of functional programming in PHP,
Retrieve:http://www.sitepoint.com/blogs/2007/12/15/the-state-of-functional-programming-in-php/

[163][164][165]E.Begoli, Scala vs. F#: Comparing with functional programming features, 2010,
Retrieve:http://www.developer.com/article.php/3883051/Scala-vs-F-Comparing-Functional-Programming-Features.htm

[166] Scala (Programming language), Wikipedia
Retrieve: http://en.wikipedia.org/wiki/Scala_(programming_language)#Functional_programming

[168] F.Kleine and S.Schmidt, Declarative Development using Annotations in PHP [pdf document], international PHP 2007 conference-spring edition
[169] Declarative Programming, Wikipedia
Retrieve from: http://en.wikipedia.org/wiki/Declarative_programming
[170] Declarative Programming in Java using Annotation and reflection,
Retrieve from:http://www.riedquat.de/articles/javaDecl
[171] A.Gutsmans, S.S Bakken and D.Rethans, Chapter 16.2 PHP shell scripts,PHP 5 Power programming, Indianapolis: Prentice Hall
[172] A.Gutsmans, S.S Bakken and D.Rethans, Chapter 14 Scala Tools, Libraries and IDE Support: CommandLineTool, PHP 5 Power programming, Indianapolis: Prentice Hall
[173] S.Mattocks (2010), introduction, Pro PHP-GTK, Apress
[174] S.Mattocks (2010), Chapter1 introduction php-gtk, Pro PHP-GTK (p11), Apress
[175] ScalaGUI, ScalaWiki, Retrieve from: http://scala.sygneca.com/code/scalagui

[176] Continuation, Wikipedia, http://en.wikipedia.org/wiki/Continuation
[177] Scheme-faq-programming, http://community.schemewiki.org/?scheme-faq-programming
[178] Rscheme, Wikipedia, http://en.wikipedia.org/wiki/RScheme
[179] Mod_lisp,
http://www.fractalconcept.com/asp/npl5/sdataQG97Qx90SRn$DM==/sdataQuvY9x3g$ecX
[180] The LAML, http://www.cs.aau.dk/~normark/laml/
[181] The Racket project, http://racket-lang.org/
[182] http://classes.eclab.byu.edu/330/wiki/index.cgi?XMLandScheme
[183] http://docs.racket-lang.org/xml/index.html
[184] http://okmij.org/ftp/Scheme/search-mslib.scm
[185] http://classes.eclab.byu.edu/330/wiki/index.cgi?SchemeAndAmazon
[186] http://www.devx.com/opensource/Article/42778/1763/page/5
[187] Aptana project, Wikipedia, http://en.wikipedia.org/wiki/Aptana
[188] Make Yahoo! Web Service REST Calls with JavaScript and XMLHttpRequest,
        http://developer.yahoo.com/javascript/howto-ajax.html
[189] N.Adams and J.Rees, Object-oriented programming in scheme
[190] A class y that inherits from x, http://www.cs.aau.dk/~normark/prog3-03/html/notes/oop-scheme-self-demo-note-program-2.html
[191] How do I do object-oriented programming in Scheme, http://www.faqs.org/faqs/scheme-faq/part1/section-6.html
[192] Object-Oriented programming systems in Scheme, http://www.cs.indiana.edu/scheme-repository/code.oop.html
[193] JavaScript Object-Oriented Programming, http://articles.sitepoint.com/article/oriented-programming-2/4
[194] Object Oriented Programming in JavaScript, http://mckoss.com/jscript/object.htm
[195] K.E Gary and Mathew Flatt, Compiling Java to PLT Scheme
[196] Reflection in Javascript, http://lpetr.org/blog/archives/reflection-in-javascript
[197] Scheme (programming language), wikipedia,
http://en.wikipedia.org/wiki/Scheme_(programming_language)
[198] Lambda calculus, Wikipedia, http://en.wikipedia.org/wiki/Lambda_calculus
[199] Functional programming, Wikipedia, http://en.wikipedia.org/wiki/Functional_programming
[200] Frequently Asked Questions for comp.lang.functional,
http://www.cs.nott.ac.uk/~gmh/faq.html
[201] Functional Javascript , http://www.hunlock.com/blogs/Functional_Javascript
[202] Category:Functional languages, http://en.wikipedia.org/wiki/Category:Functional_languages
[203] The Relation Reflection Scheme, http://onlinelibrary.wiley.com/doi/10.1002/malq.200710035/pdf
[204] JavaScript, Wikipedia, http://en.wikipedia.org/wiki/JavaScript
[205] Structured programming, Wikipedia, http://en.wikipedia.org/wiki/Structured_programming
[205] http://www.tildemark.com/programming/javascript/adding-bookmark-this-to-your-website-using-javascr.html

[206]  The Racket project,  http://racket-lang.org/

[207] Adding Bookmark This link to your website using javascript,
http://www.haskell.org/ghc/docs/6.10.2/html/users_guide/options-phases.html

[208] Comparison of Java and C Sharp, Wikipedia,
http://www.javapractices.com/topic/TopicAction.do?Id=64

[209] Bpsinfo.com, http://www.bpsinfo.com/javassl/

[210] Haskell/Polymorphism, Wikibooks, http://en.wikibooks.org/wiki/Haskell/Polymorphism

[211] Closure (computer science), Wikipedia,
http://en.wikipedia.org/wiki/Closure_%28computer_science%29#Java

[212] Fran version 1.16, http://conal.net/Fran/

[213][214][215][216] Comparison of programming languages, Wikipedia,
http://en.wikipedia.org/wiki/Comparison_of_programming_languages#Type_systems

[217] Meta-Programming with Scala: Conditional Compilation and Loop Unrolling,
http://michid.wordpress.com/2008/10/29/meta-programming-with-scala-conditional-
compilation-and-loop-unrolling/

[218] Top 10 PHP MVC frameworks, http://www.mustap.com/phpzone_post_73_top-10-php-
mvc-frameworks

[219] PHP introduciton, http://www.php.net/manual/en/intro.openssl.php

[220] http://phpxmlrpc.sourceforge.net/

[221] S.Abeysinghe, PHP Web Services: Getting Started, retrieve: http://wso2.org/library/3032

[222] The apache software foundation, http://ws.apache.org/axis2/

[223] Using Scala to update LiveJournal tags, http://rafaelnaufal.com/blog/2009/05/23/using-
scala-to-update-livejournal-tags-part-i/

[224] WS-SecurityPolicy With PHP, http://www.dimuthu.org/blog/2008/11/19/ws-
securitypolicy-with-php/

[225] http://devzone.zend.com/article/4486

[226] scala-on-struts, http://github.com/leonm/scala-on-struts

[227] Scala vs. F#: Comparing Functional Programming Features,
http://www.developer.com/article.php/3883051/Scala-vs-F-Comparing-Functional-
Programming-Features.htm

[228] Working with Scala's XML Support, http://www.codecommit.com/blog/scala/working-
with-scalas-xml-support

[230] http://www.web-tech-india.com/articles/php/compiling_php_apache/#why

[231] http://days2010.scala-lang.org/node/92

[232] Comparison of programming languages, Wikipedia,
http://en.wikipedia.org/wiki/Comparison_of_programming_languages

[233] Type system, Wikipedia, http://en.wikipedia.org/wiki/Type_system
[234] Reflection (computer science), Wikipedia,
http://en.wikipedia.org/wiki/Reflection_(computer_science)
[235] Multiple Inheritance and virtual Base Classes ,
http://www.deitel.com/articles/cplusplus_tutorials/20060225/virtualBaseClass/
[236] Reflection in C++, http://msdn.microsoft.com/en-us/library/y0114hz2(VS.80).aspx
[237] Introduction, http://www.garret.ru/cppreflection/docs/reflect.html
[238] Functional Programming in C++ , http://www.cc.gatech.edu/~yannis/fc++/
[239] Functional programming, Wikipedia,
         http://en.wikipedia.org/wiki/Functional_programming
[240] Functional Programming with Groovy ,
         http://groovy.codehaus.org/Functional+Programming+with+Groovy
[241] Calling external program in C++ ,
         http://www.velocityreviews.com/forums/t287554-calling-external-program-in-c.html
[242] shell commands and c++ , http://forums.macrumors.com/archive/index.php/t-92081.html
[243] Executing external processes in Groovy ,
http://startbigthinksmall.wordpress.com/2010/04/23/antexec-executing-external-processes-in-
groovy/
[244] http://classes.eclab.byu.edu/330/wiki/index.cgi?SchemeAndAmazon
[255] http://www.plt-scheme.org/software/openssl/
[266] http://javascriptsoapclient.codeplex.com/
[277] http://www.scala-lang.org/node/133
[278] http://en.wikipedia.org/wiki/First-class_function
[279] http://docs.racket-lang.org/scheme/index.html?q=arguments
[280] The Relation Reflection Scheme,
         http://onlinelibrary.wiley.com/doi/10.1002/malq.200710035/pdf
[281] Scheme (programming language), wikipedia,
         http://en.wikipedia.org/wiki/Scheme_(programming_language)
[282] Lambda calculus, http://en.wikipedia.org/wiki/Lambda_calculus
[283] Functional programming, http://en.wikipedia.org/wiki/Functional_programming
[284] Frequently Asked Questions for comp.lang.functional,
         http://www.cs.nott.ac.uk/~gmh/faq.html
[285] Functional Javascript , http://www.hunlock.com/blogs/Functional_Javascript
[286] Category:Functional languages,
         http://en.wikipedia.org/wiki/Category:Functional_languages
[287] Wisdom and Wonder,
         http://www.wisdomandwonder.com/link/1027/linq-for-r6rs-scheme

# Appendix

# A Source code Examples

**C# Console application**:

Simple calculator written in C# using Windows Forms:

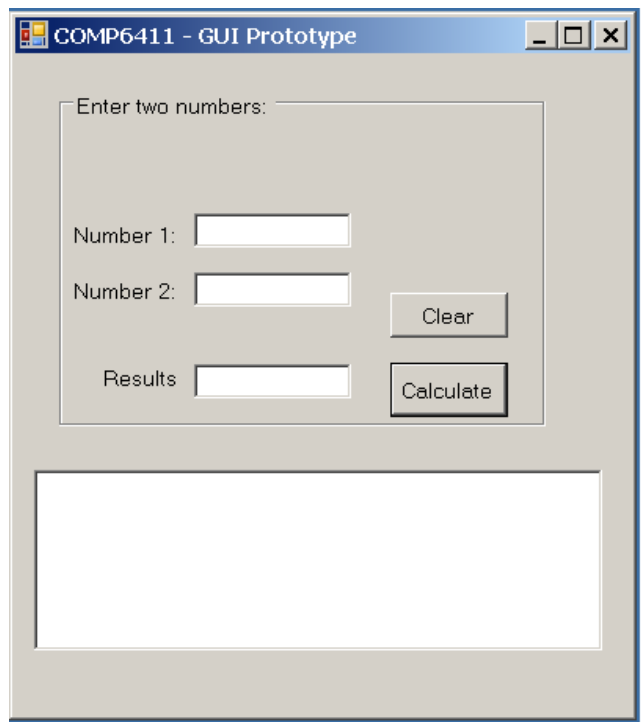

```csharp
namespace COMP6411.GuiPrototype
{
    partial class Form1
    {
        /// <summary>
        /// Required designer variable.
        ///
        /// <autor>Sleiman Rabah</autor>
        /// </summary>
        private System.ComponentModel.IContainer components =
null;

        /// <summary>
        /// Clean up any resources being used.
        /// </summary>
        /// <param name="disposing">true if managed resources
should be disposed; otherwise, false.</param>
        protected override void Dispose(bool disposing)
        {
            if (disposing && (components != null))
            {
                components.Dispose();
            }
            base.Dispose(disposing);
```

```csharp
#region Windows Form Designer generated code

        /// <summary>
        /// Required method for Designer support - do not modify
        /// the contents of this method with the code editor.
        /// </summary>
        private void InitializeComponent()
        {
            this.btnCalculate = new System.Windows.Forms.Button();
            this.label1 = new System.Windows.Forms.Label();
            this.txtNumber1 = new System.Windows.Forms.TextBox();
            this.label2 = new System.Windows.Forms.Label();
            this.txtNumber2 = new System.Windows.Forms.TextBox();
            this.txtResult = new System.Windows.Forms.TextBox();
            this.label3 = new System.Windows.Forms.Label();
            this.groupBox1 = new System.Windows.Forms.GroupBox();
            this.lstErrors = new System.Windows.Forms.ListBox();
            this.btnClear = new System.Windows.Forms.Button();
            this.groupBox1.SuspendLayout();
            this.SuspendLayout();
            this.btnCalculate.Location =
                    new System.Drawing.Point(211, 174);
            this.btnCalculate.Name = "btnCalculate";
            this.btnCalculate.Size = new System.Drawing.Size(75, 35);
            this.btnCalculate.TabIndex = 0;
            this.btnCalculate.Text = "Calculate";
            this.btnCalculate.UseVisualStyleBackColor = true;
            this.btnCalculate.Click += new
                        System.EventHandler(this.btnCalculate_Click);
            //
            // label1
            //
            this.label1.AutoSize = true;
            this.label1.Location = new System.Drawing.Point(6, 83);
            this.label1.Name = "label1";
            this.label1.Size = new System.Drawing.Size(74, 17);
            this.label1.TabIndex = 1;
            this.label1.Text = "Number 1:";
            //
            // txtNumber1
            //
            this.txtNumber1.Location = new System.Drawing.Point(86,
78);
            this.txtNumber1.Name = "txtNumber1";
            this.txtNumber1.Size = new System.Drawing.Size(100, 22);
            this.txtNumber1.TabIndex = 2;
            //
            // label2
            //
            this.label2.AutoSize = true;
            this.label2.Location = new System.Drawing.Point(6, 119);
            this.label2.Name = "label2";
            this.label2.Size = new System.Drawing.Size(74, 17);
            this.label2.TabIndex = 3;
```

```csharp
            this.label2.Text = "Number 2:";
            //
            // txtNumber2
            //
            this.txtNumber2.Location =
                        new System.Drawing.Point(86, 116);
            this.txtNumber2.Name = "txtNumber2";
            this.txtNumber2.Size = new System.Drawing.Size(100, 22);
            this.txtNumber2.TabIndex = 4;
            //
            // txtResult
            //
            this.txtResult.Location =
             new System.Drawing.Point(86, 175);
            this.txtResult.Name = "txtResult";
            this.txtResult.Size = new System.Drawing.Size(100, 22);
            this.txtResult.TabIndex = 5;
            //
            // label3
            //
            this.label3.AutoSize = true;
            this.label3.Location = new System.Drawing.Point(25, 175);
            this.label3.Name = "label3";
            this.label3.Size = new System.Drawing.Size(55, 17);
            this.label3.TabIndex = 6;
            this.label3.Text = "Results";
            //
            // groupBox1
            //
            this.groupBox1.Controls.Add(this.btnClear);
            this.groupBox1.Controls.Add(this.txtResult);
            this.groupBox1.Controls.Add(this.label3);
            this.groupBox1.Controls.Add(this.btnCalculate);
            this.groupBox1.Controls.Add(this.txtNumber1);
            this.groupBox1.Controls.Add(this.label1);
            this.groupBox1.Controls.Add(this.label2);
            this.groupBox1.Controls.Add(this.txtNumber2);
            this.groupBox1.Location = new System.Drawing.Point(28, 24);
            this.groupBox1.Name = "groupBox1";
            this.groupBox1.Size = new System.Drawing.Size(310, 215);
            this.groupBox1.TabIndex = 7;
            this.groupBox1.TabStop = false;
            this.groupBox1.Text = "Enter two numbers:";
            //
            // lstErrors
            //
            this.lstErrors.FormattingEnabled = true;
            this.lstErrors.ItemHeight = 16;
            this.lstErrors.Location =
                    new System.Drawing.Point(12, 267);
            this.lstErrors.Name = "lstErrors";
            this.lstErrors.Size = new System.Drawing.Size(362, 116);
            this.lstErrors.TabIndex = 8;
            //
            // btnClear
            //
```

```csharp
            this.btnClear.Location =
                new System.Drawing.Point(211, 129);
            this.btnClear.Name = "btnClear";
            this.btnClear.Size = new System.Drawing.Size(75, 29);
            this.btnClear.TabIndex = 7;
            this.btnClear.Text = "Clear";
            this.btnClear.UseVisualStyleBackColor = true;
            this.btnClear.Click += new
                System.EventHandler(this.btnClear_Click);
            //
            // Form1
            //
            this.AutoScaleDimensions =
                new System.Drawing.SizeF(8F, 16F);
            this.AutoScaleMode =
                System.Windows.Forms.AutoScaleMode.Font;
            this.ClientSize = new System.Drawing.Size(392, 424);
            this.Controls.Add(this.lstErrors);
            this.Controls.Add(this.groupBox1);
            this.Name = "Form1";
            this.Text = "COMP6411 - GUI Prototype";
            this.groupBox1.ResumeLayout(false);
            this.groupBox1.PerformLayout();
            this.ResumeLayout(false);

        }

        #endregion

        private System.Windows.Forms.Button btnCalculate;
        private System.Windows.Forms.Label label1;
        private System.Windows.Forms.TextBox txtNumber1;
        private System.Windows.Forms.Label label2;
        private System.Windows.Forms.TextBox txtNumber2;
        private System.Windows.Forms.TextBox txtResult;
        private System.Windows.Forms.Label label3;
        private System.Windows.Forms.GroupBox groupBox1;
        private System.Windows.Forms.ListBox lstErrors;
        private System.Windows.Forms.Button btnClear;
    }
}
```

```csharp
using System;
using System.Collections.Generic;
using System.Linq;
using System.Windows.Forms;

namespace COMP6411.GuiPrototype
{
    /// <summary>
    /// <autor>Sleiman Rabah</autor>
    /// </summary>
    static class Program
    {
        /// <summary>
        /// The main entry point for the application.
        /// </summary>
        [STAThread]
        static void Main()
        {
            Application.EnableVisualStyles();
            Application.SetCompatibleTextRenderingDefault(false);
            Application.Run(new Form1());
        }
    }
}
```

```csharp
using System.ComponentModel;
using System.Data;
using System.Drawing;
using System.Linq;
using System.Text;
using System.Windows.Forms;

namespace COMP6411.GuiPrototype
{
    /// <summary>
    ///
    /// A UI from which calculates two integer.
    ///
    /// <author> Sleiman Rabah</author>
    /// </summary>
    public partial class Form1 : Form
    {
        public Form1()
        {
            InitializeComponent();
        }

        /// <summary>
        ///  Calculate the results based on the user entred values.
        /// </summary>
        /// <param name="sender"></param>
        /// <param name="e"></param>
        private void btnCalculate_Click(object sender, EventArgs e)
        {
            int result = 0;

            try
            {
            if (this.txtNumber1.Text != ""
               || this.txtNumber2.Text != "")
                {
                  result = Int32.Parse(txtNumber1.Text) +
                    Int32.Parse(this.txtNumber2.Text);
                    this.txtResult.Text = result.ToString();
                }
            }
            catch (Exception ex)
            {
                this.lstErrors.Items.Add(ex.Message);
            }
        }

        /// <summary>
        /// Clear the UI fields.
        /// </summary>
        /// <param name="sender"></param>
        /// <param name="e"></param>
        private void btnClear_Click(object sender, EventArgs e)
        {
            this.txtNumber2.Text = String.Empty;
            this.txtNumber1.Text = String.Empty;
            this.txtResult.Text = String.Empty;
            this.lstErrors.Items.Clear();
        }
    }
}
```

C# Console application:
     Launching processes/executing a program from a C# code.

```csharp
using System;
using System.Diagnostics;
using System.ComponentModel;

namespace COMP6411.ProcessLauncher
{
    /// <summary>
    /// <autor>Sleman Rabah</autor>
    /// </summary>
    class ProcessLauncher
    {
        // Notepad command.
        public static string szNotepade = "-note";
        // Calculator command.
        public static string szCalculator = "-calc";
        public static string szNone = "-err";

        public static void Main(string[] args)
        {
            try
            {
                if (IsValid(args))
                {
                    // instantiate the System.Diagnostics.Process class
                    Process myProcess = new Process();

                    // Tell whether the new process will be executed
                    // using Shell or not
                    myProcess.StartInfo.UseShellExecute = false;
                    // Fill out the process name to be started:
                    // often it is a program e.g: notepad
                    if (args[0].Equals(ProcessLauncher.szCalculator))
                    {
                        myProcess.StartInfo.FileName = "calc.exe";
                    }
                    else if
                        (args[0].Equals(ProcessLauncher.szNotepade))
                    {
                        myProcess.StartInfo.FileName = "notepad.exe";
                    }
                    else
                    {
                        myProcess.StartInfo.FileName = "no_one.exe";
                    }
                    myProcess.StartInfo.CreateNoWindow = true;
                    // Run/launch the processs
                    myProcess.Start();
                }
                else
                {
                    // If arguments are invalid, display usage.
```

```csharp
                            Console.WriteLine("Usage: ProcessLauncher.exe
                                              [arg1]");
                            Console.WriteLine("\t -calc to run
                                                  Calculator");
                            Console.WriteLine("\t -note to run Notepad");
                            Console.WriteLine("\t -err to raise an
                                              error");
                        }
                    }
                    catch (Exception e)
                    {
                        Console.WriteLine("A problem has occured while
                                          starting the process " + e.Message);
                    }
                }

            /// <summary>
            ///  Validates user commands.
            /// </summary>
            /// <param name="args">an array containing user's
                commands</param>
            /// <returns>boolean indicating whether commands are valid
                or not</returns>
            public static bool IsValid(string[] args)
            {
                if (args.Length == 1 )
                {
                    if ((args[0].Equals(ProcessLauncher.szCalculator)
                      ||args[0].Equals(ProcessLauncher.szNotepade)) ||
                      args[0].Equals(ProcessLauncher.szNone))
                    {
                        return true;
                    }
                }

                return false;
            }
        }
}
```

UI application in AspectJ and Java Swing:

Same application as shown above  (using C# and Windows Forms)

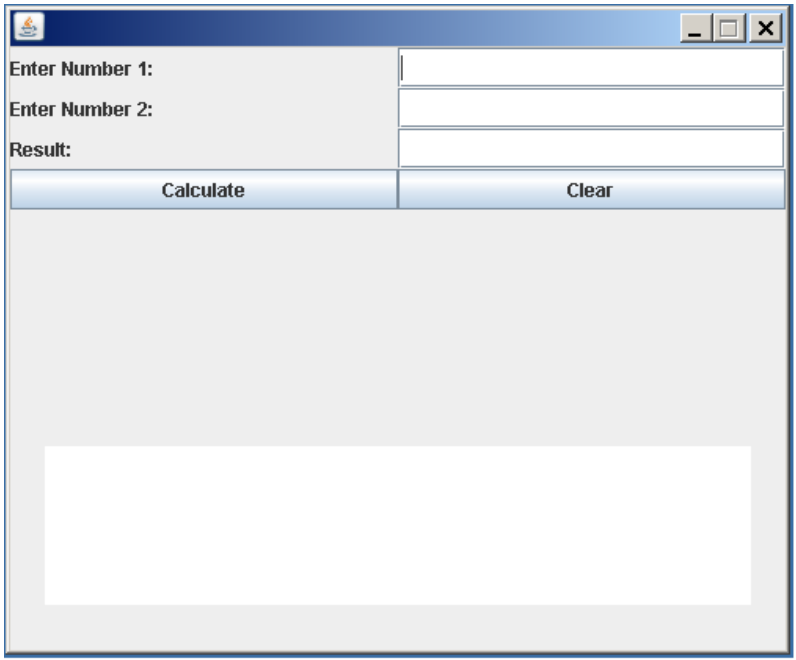

package com.comp6411.aspect.gui;

import java.awt.BorderLayout;
import java.awt.FlowLayout;
import java.awt.GridLayout;
import java.awt.event.ActionEvent;
import java.awt.event.ActionListener;
import javax.swing.JButton;
import javax.swing.JLabel;
import javax.swing.JPanel;
import javax.swing.JTextArea;
import javax.swing.JTextField;
import javax.swing.WindowConstants;

/**
 *
 * A class for building a Swing-based GUI.
 * It calculates two integers.
 *
 * @author Sleiman Rabah
 */
public class CalculatorSample extends javax.swing.JFrame {
        private JPanel jPanel1;
        private JButton btnCalculate;
        private JTextArea txtErrors;
        private JTextField txtNumber2;
        private JTextField txtNumber1;
        private JTextField txtResult;
        private JLabel lblNumber2;
        private JLabel lblNumber1;
        private JLabel lblResult;
        private JPanel jPanel2;

```java
private JButton btnClear;

/**
 * Auto-generated main method to display this JFrame
 */
public static void main(String[] args) {

        CalculatorSample inst = new CalculatorSample();
        inst.setLocationRelativeTo(null);
        inst.setVisible(true);
}

public CalculatorSample() {
        super();
        initGUI();
}

private void initGUI() {
        try {
                setDefaultCloseOperation(WindowConstants.DISPOSE_ON_CLOSE);
                {
                        jPanel1 = new JPanel();
                        jPanel1.setLayout(new GridLayout(0, 2));
                        getContentPane().add(jPanel1, BorderLayout.NORTH);

                        // --
                        lblResult = new JLabel();
                        lblResult.setText("Result: ");
                        // --
                        lblNumber1 = new JLabel();
                        lblNumber1.setText("Enter Number 1:");
                        // --
                        txtNumber2 = new JTextField();
                        // --
                        lblNumber2 = new JLabel();
                        lblNumber2.setText("Enter Number 2:");
                        // --
                        txtNumber1 = new JTextField();
                        //--
                        txtResult= new JTextField();
                        // --
                        btnCalculate = new JButton();
                        btnCalculate.setText("Calculate");
                        btnCalculate.addActionListener(new ActionListener() {
                                public void actionPerformed(ActionEvent e) {
                                        calculateResult();
                                }
                        });
                        // --
                        btnClear = new JButton();
                        btnClear.setText("Clear");
                        btnClear.addActionListener(new ActionListener() {
                                public void actionPerformed(ActionEvent e) {
                                        clearFields();
                                }
                        });
                        // --
                        jPanel1.add(lblNumber1);
                        jPanel1.add(txtNumber1);
                        jPanel1.add(lblNumber2);
                        jPanel1.add(txtNumber2);
```

```
                    jPanel1.add(lblResult);
                    jPanel1.add(txtResult);
                    jPanel1.add(btnCalculate);
                    jPanel1.add(btnClear);

                    jPanel2 = new JPanel();
                    FlowLayout jPanel2Layout = new FlowLayout();
                    jPanel2Layout.setAlignOnBaseline(true);
                    getContentPane().add(jPanel2, BorderLayout.SOUTH);
                    jPanel2.setLayout(jPanel2Layout);
                    jPanel2.setPreferredSize(new java.awt.Dimension(492, 136));
                    txtErrors = new JTextArea();
                    jPanel2.add(txtErrors);
                    txtErrors.setPreferredSize(new java.awt.Dimension(450, 102));
            }
            pack();
            this.setSize(500, 417);
            this.setResizable(false);
        } catch (Exception e) {
                    // add your error handling code here
                    e.printStackTrace();
        }
}

/**
 * Clear text fields.
 */
public void clearFields() {

            this.txtNumber1.setText("");
            this.txtNumber2.setText("");
            this.txtErrors.setText("");
            this.txtResult.setText("");

}

/**
 * Calculate the result and update the UI fields.
 */
public void calculateResult() {

            try {
                    int result = 0;

                    result = Integer.parseInt(this.txtNumber1.getText()) +
Integer.parseInt(this.txtNumber2.getText());
                    this.txtResult.setText(""+ result);

            } catch (Exception e) {
                    this.txtErrors.setText("An error has occured, only number are allowed: "+
e.getMessage());
            }
        }
}

package com.comp6411.aspect.gui;

import java.util.logging.Level;
import java.util.logging.Logger;
import org.aspectj.lang.Signature;
```

```
/**
 *
 * An aspect that intercepts the GUI methods and handles its exceptions.
 *
 * @author Sleiman Rabah
 *
 */
public aspect AspectGuiSample {

        Logger logger = Logger.getLogger("Log MethodEntries");

        /**
         * Pointcut on public methods.
         */
        pointcut all_publics():
                call(public * CalculatorSample.*(..));

        /**
         * Pointcut to handle exception.
         */
        pointcut exceptionHandler() :
                call(* *.*(..)) && !within(AspectGuiSample);

        /**
         * Catch and log any exception.
         *
         * @param ex the exception thrown in ProcessLauncher.java
         */
        after() throwing(Throwable ex) : exceptionHandler(){

                logger.setLevel(Level.WARNING);
        Signature methodSignature = thisJoinPoint.getSignature();
        System.err.println("AspectJ has caught an exception in method: "+
methodSignature.getDeclaringTypeName()
                        + "." + methodSignature.getName());
        System.err.println("AspectJ-Exception Trace:" + ex.getMessage());

        //--
     logger.info(methodSignature.getDeclaringType().getName());
        }

        /**
         * Advice to log methods entering: All called Swing methods will be displayed.
         * NOTICE: extra stuff can be done if desired before entring a method.
         */
    before(): all_publics() {

    Signature methodSignature = thisJoinPoint.getSignature();
    System.out.println("Entring method:" + methodSignature.getName()+ "()");
    }

    /**
```

```
    * Advice to log methods' execution ending: also, All called Swing methods will be displayed.
    *
    *  NOTICE: extra stuff can be done if desired after returning from a method.
    */
   after() returning() : all_publics() {
      Signature methodSignature = thisJoinPoint.getSignature();
      System.out.println("Leaving method: "+ methodSignature.getName()+ "()");
   }

}
```

## Console application in AspectJ:

```
package com.comp6411.aspect.batchscripting;

import java.io.IOException;
import java.io.InputStream;
import java.io.OutputStream;

/**
 * A class which creates a process and runs a program in it.
 *
 * @author Sleiman Rabah
 */
public class ProcessLauncher {

        /**
         * Notepad command.
         */
        public static String szNotepade = "-note";
        /**
         * Calculator command.
         */
        public static String szCalculator = "-calc";
        public static String szNone = "-err";

        public static void main(String[] args) throws IOException {

                if (ProcessLauncher.isValid(args)) {

                        if (args[0].equals(ProcessLauncher.szCalculator)) {

                                new ProcessLauncher().launchProcess("calc.exe");

                        }
                        else if (args[0].equals(ProcessLauncher.szNotepade)) {
                                new ProcessLauncher().launchProcess("notepad.exe");
                        } else {
                                new ProcessLauncher().launchProcess("no_one.exe");
                        }

                } else {
                        // If arguments are invalid, display usage.
```

```java
                System.out.println("Usage: ProcessLauncher [arg1]");
                System.out.println("\t -calc to run Calculator");
                System.out.println("\t -note to run Notepad");
                System.out.println("\t -err to raise an error");
        }
    }

    /**
     *
     * @param iPorgramToRun The name of the program to be executed.
     * @throws IOException Throws an exception if failed to execute the process.
     */
    public void launchProcess(String iPorgramToRun) throws IOException {
            Runtime myRunTime = Runtime.getRuntime();
            Process myProcess = myRunTime.exec(iPorgramToRun);
    }

    /**
     * A method which validates entered command (user's commands)
     *
     * @param args An array containing user's commands
     * @return Boolean indicating whether commands are valid or not
     */
    public static boolean isValid(String[] args) {

            if (args.length == 1) {
                    if ( args[0].equals(ProcessLauncher.szCalculator) ||
                                    args[0].equals(ProcessLauncher.szNotepade) ||
                                    args[0].equals(ProcessLauncher.szNone)
                                    ) {
                            return true;
                    }
            }

            return false;
    }
}
```

```
package com.comp6411.aspect.batchscripting;

import java.util.logging.Level;
import java.util.logging.Logger;
import org.aspectj.lang.Signature;

/**
 * An aspect which handles ProcessLaunchers's exceptions.
 *
 * @author Sleiman Rabah
 */
public aspect AspectProcessLauncher {

        Logger logger = Logger.getLogger("Log MethodEntries");

        /**
         * Pointcut on public methods.
         */
        pointcut all_publics():
                call(public * ProcessLauncher.*(..));

        /**
         * Pointcut to handle exception.
         */
        pointcut exceptionHandler() :
                call(* *.*(..)) && !within(AspectProcessLauncher);

        /**
         * Catch and log any exception.
         *
         * @param ex the exception thrown in ProcessLauncher.java
         */
        after() throwing(Throwable ex) : exceptionHandler(){

                logger.setLevel(Level.WARNING);
        Signature methodSignature = thisJoinPoint.getSignature();
        System.err.println("AspectJ has caught an exception: "+ methodSignature.getName());
        System.err.println("AspectJ-Exception Trace:" + ex.getMessage());

        //--
    logger.info(methodSignature.getDeclaringType().getName());
        }

        /**
         * Advice to log methods entering.
         */
    before(): all_publics() {

        Signature methodSignature = thisJoinPoint.getSignature();
        System.out.println("Entring method:" + methodSignature.getName()+ "()");
    }

    /**
     * Advice to log methods' execution ending.
     */
```

```
after() returning () : all_publics() {
    Signature methodSignature = thisJoinPoint.getSignature();
    System.out.println("Leaving method: "+ methodSignature.getName()+ "()");
}

}
```

## A1.PHP Criteria 4

The following program was compiled and run using the Zend Development Environment available at http://www.zend.com/en/downloads/ with the following options. To execute it run the following: File-NEWFILE, then write code in Editor, Menu-Debug-Go.

```php
//Inheritance, Abstract Class, interface, encapsulation sample in PHP
<?php
abstract class mathematics{
/*** child class must define these methods ***/
abstract protected function getMessage();
abstract protected function addTwo($num1);
/**
* method common to both classes
**/
public function showMessage() {
  echo $this->getMessage();
}
} /*** end of class ***/
class myMath extends mathematics{
/**
* Prefix to the answer
* @return string
**/
protected function getMessage(){
  return "The anwser is: ";
}
/**
* add two to a number
* @access public
* @param $num1 A number to be added to
* @return int
**/
public function addTwo($num1) {
  return $num1+2;
}
} /*** end of class ***/
/*** a new instance of myMath ***/
$myMath = new myMath;
/*** show the message ***/
$myMath->showMessage();
/*** do the math ***/
echo $myMath->addTwo(4);
```

?>

## A2. Scala Criteria4

The following program was compiled and run using the Scala-ide for eclipse (Eclipse 3.51 with JVM) which is available at http://www.scala-ide.org/ and http://www.eclipse.org/downloads/packages/release/galileo/sr2 with the following options. After installed the Scala PDT, build a scala project, and create file in the default package, put code in the scala file, then run them as Scala application.

```scala
//Inheritance, interface, encapsulation in Scala
package com.example

trait Similarity {
  def isSimilar(x: Any): Boolean
  def isNotSimilar(x: Any): Boolean = !isSimilar(x)
}

class Point(xc: Int, yc: Int) extends Similarity {
  var x: Int = xc
  var y: Int = yc
  def isSimilar(obj: Any) =
    obj.isInstanceOf[Point] &&
    obj.asInstanceOf[Point].x == x
}
object TraitsTest extends Application {
  val p1 = new Point(2, 3)
  val p2 = new Point(2, 4)
  val p3 = new Point(3, 3)
  println(p1.isNotSimilar(p2))
  println(p1.isNotSimilar(p3))
  println(p1.isNotSimilar(2))
}
```

## A3. PHP Criteria 6

The following program was compiled and run using the Zend Development Environment available at http://www.zend.com/en/downloads/ with the following options. To execute it run the following: Menu->File->NEWFILE, then write code in Editor, Menu->Debug->Go.

```
<?PHP
include("aop.lib.php");   // aop.lib.php is an external supported AOP library
                // , provide in the .zipfile with the submitted report
$aspect1 = new Aspect();
$pc1 = $aspect1->pointcut("call Sample::Sample or call Sample::Sample2");
$pc1->_before("print 'PreProcess<br />';");
$pc1->_after("print 'PostProcess<br />';");
$pc1->destroy();

Class Sample {
var $aspect;
function Sample($aspect1) {
$this->aspect = &$aspect1;
Advice::_before($this->aspect);
print 'Some business logic of Sample<br />';
Advice::_after($this->aspect);
}
function Sample2() {
Advice::_before($this->aspect);
print 'Some business logic of Sample2<br />';
Advice::_after($this->aspect);
}

}

$Sample = new Sample(&$aspect1);
$Sample->Sample2();
?>
```

**A4. Scala Criteria 6**

The following program was compiled and run using the Scala-ide for eclipse (Eclipse 3.51 with JVM and AJDT plug-in) which is available at http://www.scala-ide.org/ , http://www.eclipse.org/downloads/packages/release/galileo/sr2 , and http://www.eclipse.org/ajdt/downloads with the following options. After installed the Scala pdt and AJDT plug-in, create the Aspect project, then create a package , then put two .scala file into package. Then create an aspect in the package named LogComplex.aj. Finally, then run as Aspect program.

// code-examples/ToolsLibs/aspectj/complex.scala

package example.aspectj

```
case class Complex(real: Double, imaginary: Double) {
  def +(that: Complex) =
    new Complex(real + that.real, imaginary + that.imaginary)
  def -(that: Complex) =
    new Complex(real - that.real, imaginary - that.imaginary)
}
```

// code-examples/ToolsLibs/aspectj/complex-main.scala

package example.aspectj

```
object ComplexMain {
  def main(args: Array[String]) {
    val c1 = Complex(1.0, 2.0)
    val c2 = Complex(3.0, 4.0)
    val c12 = c1 + c2
    println(c12)
  }
}

// code-examples/ToolsLibs/aspectj/LogComplex.aj
// define point cut, advice for scala file.
package example.aspectj;
public aspect LogComplex {
  public pointcut newInstances(double real, double imag):
    execution(Complex.new(..)) && args(real, imag);

  public pointcut plusInvocations(Complex self, Complex other):
    execution(Complex Complex.$plus(Complex)) && this(self) && args(other);

  before(double real, double imag): newInstances(real, imag) {
    System.out.println("new Complex(" + real + "," + imag + ") called.");
  }

  before(Complex self, Complex other): plusInvocations(self, other) {
    System.out.println("Calling " + self + ".+(" + other + ")");
  }

  after(Complex self, Complex other) returning(Complex c):
    plusInvocations(self, other) {
    System.out.println("Complex.+ returned " + c);
  }
}
```

C++ vs Grouvy:

## A.1  C++ Criteria 4

The following program was compiled and run using the GCC compiler in VC2008 which is available at http://www.microsoft.com/downloads/details.aspx?FamilyID=9b2da534-3e03-4391-8a4d-074b9f2bc1bf&displaylang=en. To execute it run the following code in the cpp file..

```
#ifndef  FRUIT_H
#define  FRUIT_H
#include<string>
Using namespace std;

Class Fruit
{
    Public:
    Virtual void identify(){cout<<"Fruit"<<endl;}
}

Class Apple:public Fruit
{
    Public:
    Void identify(){cout<<"Apple"<<endl;}
}
Class Orange:public Fruit
{
    Public:
    Void identify(){cout<<"Orange"<<endl;}
}

#include<iostream>
Using namespace std;
#include<iomanip>
#include "Fruit.h"

Int main()
{
    Fruit f;
    Apple a;
    Orange o;
    f.identify();
    a.identify();
    o.identify();
    return 0;
}
```

## A.2 Groovy Criteria 4

The following program was compiled and run using the Groovy for eclipse (Eclipse 3.51 with JVM) which is available at http://groovy.codehaus.org/ and http://www.eclipse.org/downloads/packages/release/galileo/sr2 with the following options. After installed the Groovy, build a Groovy project, and create a Groovy class in the default package, put code in the class, then run them as Groovy application.

```
abstract class A{
  public int prev //field
  int signature //property

  abstract String sayFly(int k)//abastract method
}
class B extends A{
  String sayBirds(int n){  "There are $n birds!" }
  String sayFly(int k){"There are $k flys!"}
}
def b= new B()
//def a= new A()
assert b.sayBirds(17) == 'There are 17 birds!'

assert b.sayFly(10) == 'There are 10 flys!'

b.signature= 19
assert b.signature == 19 //property 'signature' from A acts as
part of B
assert b.getSignature() == 19
```

## A.3 C++ Criteria 6

The following program was compiled and run using the GPL under VC2008 which is available at http://groovy.codehaus.org/ and http://www.aspectc.org/Download.2.0.html with the following options. After installed the AspectC++, build a project, and create a class in the default package, put code in the class, then run them as AspectC++ application.

```
aspect Logging
{
    ostream * _out ; // ordinary attributes
    public:
    void bind_stream (ostream *o) { _out = o; } // member
function
    pointcut virtual logged_classes () = 0; // pure virtual
pointcut
    // some advice
    advice execution(" % ...::%(...) ") && within(
logged_classes ()) :
    before () {
    *_out << "executing " << JoinPoint :: signature () << endl;
    }
};
```

A.4  Groovy Criteria 6
The following program was compiled and run using the Groovy for eclipse (Eclipse 3.51 with JVM) which is available at http://groovy.codehaus.org/ and http://www.eclipse.org/downloads/packages/release/galileo/sr2 with the following options. After installed the Groovy, build a Groovy project, and create a Groovy class in the default package, put code in the class, then run them as Groovy application.

```
class SimplePOGO implements GroovyInterceptable {
    void simpleMethod1(){
        System.out.println("simpleMethod1() called")
    }

    void simpleMethod2(String param1, Integer param2){
        System.out.println("simpleMethod2(${param1},${param2})
called")
        System.out.println("sleeping...")
        Timer.sleep(2000)
    }

    def invokeMethod(String name, args){
        System.out.println("time before ${name} called: ${new Date()}")

        //Get the method that was originally called.
        def calledMethod = SimplePOGO.metaClass.getMetaMethod(name,
args)

        //The "?" operator first checks to see that the "calledMethod" is not
            //null (i.e. it exists).
        calledMethod?.invoke(this, args)

        System.out.println("time after ${name} called: ${new Date()}\n")
    }
}

simplePogo = new SimplePOGO()
simplePogo.simpleMethod1()
simplePogo.simpleMethod2("stringParam", 24)
```

## A.1 Haskell, Criteria: Default secure programming practices, Exception handling part

The following program show **Maybe** handle exception

```
divBy :: Integral a => a -> [a] -> Maybe [a]
divBy _ [] = Just []
divBy _ (0:_) = Nothing
divBy numerator (denom:xs) =
   case divBy numerator xs of
     Nothing -> Nothing
     Just results -> Just ((numerator `div` denom) : results)
                 Use of Maybe [6]
```

The following program show **Either** handle exception

```
divBy :: Integral a => a -> [a] -> Either String [a]
divBy _ [] = Right []
divBy _ (0:_) = Left "divBy: division by 0"
divBy numerator (denom:xs) =
   case divBy numerator xs of
     Left x -> Left x
     Right results -> Right ((numerator `div` denom) : results)
                 Use of Either [6]
```

## A.2 Java, Criteria: Default secure programming practices, Exception Handling part

The following program show Java exception handling.

```
public class sample{
        public int divide( int a, int b){
                if ( b == 0) throw new DivisionByZeroException( );
                return a/b;
        }
        public static void main (String arg[]){
                try{
                        int result = divide( 5, 0);
                }
                catch(DivisionByZeroException e){
                        system.out.println(e.getMessage());
                }
        }
}
class DivisionByZeroException extends Exception{
        DivisionByZeroException( ){
                super("Division by 0");
        }
        DivisionByZeroException(String msg){
                Super( msg );
        }
}
```
**Java exception handling**

## *A.3 Haskell, Criteria: Web application*

```
import Network.CGI
import Text.XHtml

page :: Html
page = body << h1 << "Hello World!"

cgiMain :: CGI CGIResult
cgiMain = output $ renderHtml page

main :: IO ()
main = runCGI $ handleErrors cgiMain
```
**Haskell web application 1 – out put text [7]**

```
import Network.CGI
import Text.XHtml

inputForm = form << [paragraph << ("My name is " +++ textfield "name"),
            submit "" "Submit"]
greet n = paragraph << ("Hello " ++ n ++ "!")
page t b = header << thetitle << t +++ body << b
cgiMain = do mn <- getInput "name"
        let x = maybe inputForm greet mn
        output $ renderHtml $ page "Input example" x
main = runCGI $ handleErrors cgiMain
-- Get the value of an input variable, for example from a form.
-- If the variable has multiple values, the first one is returned.
getInput :: String -> CGI (Maybe String)
```
**Haskell web application 2 — Get user input [7]**

## A.4 Java, Criteria: Web application

```
import java.io.IOException;
import java.io.PrintWriter;
import javax.servlet.ServletException;
import javax.servlet.http.HttpServlet;
import javax.servlet.http.HttpServletRequest;
import javax.servlet.http.HttpServletResponse;

public final class Hello extends HttpServlet {
    public void doGet(    HttpServletRequest request, HttpServletResponse
                            response)throws IOException, ServletException {
        response.setContentType("text/html");
        PrintWriter writer = response.getWriter();
        writer.println("<html>");
        writer.println("<head>");
        writer.println("<title>A Sample Application</title>");
        writer.println("</head>");
        writer.println("<body>");
        writer.println("Hello world");
        writer.println("</body>");
        writer.println("</html>");
    }
}
```
**Java web application — Output a text [7]**

### A.5 Haskell, Criteria: OO-based abstraction

```
shape newx newy self
    = do
        -- Create references for private state
        x <- newIORef newx
        y <- newIORef newy
        -- Return object as record of methods
        returnIO $ getX          .=. readIORef x
                .*. getY          .=. readIORef y
                .*. setX          .=. writeIORef x
                .*. setY          .=. writeIORef y
                .*. moveTo        .=. (\newx newy -> do
                                    (self # setX) newx
                                    (self # setY) newy )
                .*. rMoveTo .=. (\deltax deltay ->
                                    do
                                        x <- self # getX
                    y <- self # getY
                (self # moveTo) (x + deltax) (y + deltay) )
          .*. emptyRecord
```

**Object generator for shapes [13]**

```
rectangle newx newy width height self
  = do
      -- Invoke object generator of superclass
      super <- shape newx newy self
      -- Create references for extended state
      w <- newIORef width
      h <- newIORef height
      -- Return object as   record of methods
      returnIO $
              getWidth .=.    readIORef  w
        .*.     getHeight .=.    readIORef  h
        .*.     setWidth .=.    (\neww ->  writeIORef w neww)
        .*.     setHeight .=.    (\newh ->  writeIORef h newh)
        .*.     draw .=.
                do -- Implementation of the abstract draw method
                  putStr "Drawing a Rectangle at:(" <<
                self # getX << ls "," << self # getY <<
        ls "), width " << self # getWidth <<
        ls ", height " << self # getHeight << ls "\n"
      -- Rectangle records start from shape records
    .*. super
```

**Object generator for rectangles [13]**

```
circle newx newy newradius self
= do
    super <- shape newx newy self
    ...
    returnIO ... .*. super
```

**Object generator for circles [13]**

```
myOOP = do
    -- Construct objects
    s1 <- mfix (rectangle (10::Int) (20::Int) 5 6)
    s2 <- mfix (circle (15::Int) 25 8)
            -- Create a homogeneous list of different shapes
            let scribble = consLub s1 (consLub s2 nilLub)
            -- Loop over list with normal monadic map
            mapM_ (\shape -> do
            shape # draw
            (shape # rMoveTo) 100 100
            shape # draw)
        scribble
```

**Object construction and invocation as a monadic sequence [13]**

## A.6 Java, Criteria: OO-based abstraction

```
public class Shape{
        protected int x;
        protected int y;
        public Shape(int x, int y){
                this.x = x;
                this.y = y;
    }
        public int getX(){ return x; }
        public int getY(){ return y; }
        public void setX(int x){ this.x = x; }
        public void setY(int y){ this.y = y; }
        public void moveTo(int x, int y){
        this.x = x;
        this.y = y;
    }
    public void rMoveTo(int deltaX, int deltaY){
        this.x += deltaX;
        this.y += deltaY;
    }
}
```
                    **Object generator for shapes**

```
public class Rectangle extends Shape{
        private int w;
        private int h;
        public Rectangle (int x, int y, int w, int h){
                super(x, y);
                this.w = w;
                this.h = h;
    }
        public int getW(){ return w; }
        public int getH(){ return h; }
        public void setW(int w){ this.w = w; }
        public void setH(int h){ this.h = h; }
    public void draw(){
        system.out.println("Drawing a Rectangle at: " + x +", " + y + "Width: " +
    w + "Height: " + h);
    }
}
```
                    **Object generator for rectangle**

```
public class Circle extends Shape{
        // similar to class Rectangle
}
                        Object generator for circle
```

```
public class Main{
        public static void main(String arg[]){
        Shape s1 = new Rectangle(10, 20, 5, 6);
        Shape s2 = new Cirle(15, 25, 8);
        s1.draw();
        s2.draw();
        s1.moveTo(20, 30);
        s1.rMoveTo(10, 20);
    }
}
                        Main class
```

## A.7 Haskell, Criteria: Aspect-oriented

```
import List(sort)
insert x [] = [x]
insert x (y:ys)
| x <= y = x:y:ys
| otherwise = y : insert x ys
insertionSort [] = []
insertionSort xs =
insert (head xs) (insertionSort (tail xs))
-- sortedness aspect
N1@advice #insert# :: Ord a => a -> [a] -> [a] =
\x -> \ys ->
let zs = proceed x ys
in if (isSorted ys) && (isSorted zs)
then zs else error "Bug"
where
isSorted xs = (sort xs) == xs
-- efficiency aspect
N2@advice #insert# :: Int -> [Int] -> [Int] =
\x -> \ys ->
if x == 0 then x:ys
else proceed x ys
                        AOP Haskell Example[12]
```

### A.8 Java, Criteria: Aspect-oriented

```
package com.reg.dev.aspects;

public class HelloWorld {
    public void print() {
        System.out.println("Hello World");
    }
    public static void main(String [] args) {
        HelloWorld hw = new HelloWorld();
        hw.print();
    }
}
//the aspect pointcuts are dynamically weaved to create the following output:
Entering print
Hello World
Exiting print
```
**AspectJ Example and output**

## A.9 Haskell, Criteria: UI Prototype design

```
import Graphics.UI.Gtk
    main :: IO ()
    main = do
      initGUI
      window  <- windowNew
      hbox    <- hBoxNew True 10
      button1 <- buttonNewWithLabel "Button 1"
      button2 <- buttonNewWithLabel "Button 2"
      set window [windowDefaultWidth := 200, windowDefaultHeight := 200,
            containerBorderWidth := 10, containerChild := hbox]
      boxPackStart hbox button1 PackGrow 0
      boxPackStart hbox button2 PackGrow 0
      onDestroy window mainQuit
      widgetShowAll window
      mainGUI
```

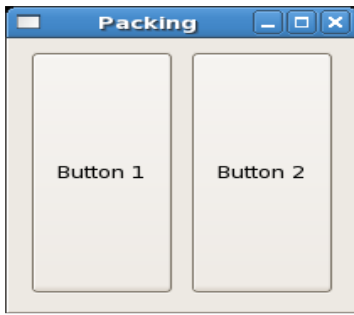

**Gtk2Hs sample and output**

A.10 Java, Criteria: *UI Prototype design*

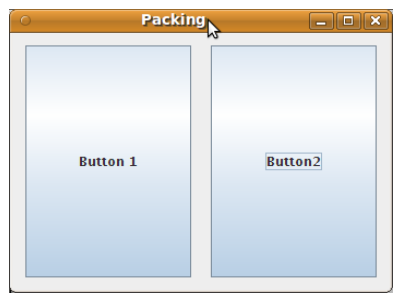

`Sample code output`

```java
public class Example extends javax.swing.JFrame {
    public Example(){ initComponents();}
    private void initComponents() {
        mainFrame = new javax.swing.JFrame();
        button1 = new javax.swing.JButton();
        button2 = new javax.swing.JButton();
        javax.swing.GroupLayout mainFrameLayout = new
        javax.swing.GroupLayout(mainFrame.getContentPane());
        mainFrame.getContentPane().setLayout(mainFrameLayout);
        mainFrameLayout.setHorizontalGroup(
          mainFrameLayout.createParallelGroup(javax.swing.    GroupLayout.Alignment.LEADING)
          .addGap(0, 400, Short.MAX_VALUE));
        mainFrameLayout.setVerticalGroup(
          mainFrameLayout.createParallelGroup(javax.swing.    GroupLayout.Alignment.LEADING)
          .addGap(0, 300, Short.MAX_VALUE));
      setDefaultCloseOperation(javax.swing.WindowConstants.EXIT_ON_CLOSE);
        setTitle("Packing");
        button1.setText("Button 1");
        button2.setText("Button2");
        javax.swing.GroupLayout layout = new          javax.swing.GroupLayout(getContentPane());
        getContentPane().setLayout(layout);
        layout.setHorizontalGroup(
          layout.createParallelGroup(javax.swing.GroupLayo        ut.Alignment.LEADING)
          .addGroup(layout.createSequentialGroup()
          .addContainerGap()
          .addComponent(button1,javax.swing.GroupLayout.PREFERRED_SIZE, 152,
javax.swing.GroupLayout.PREFERRED_SIZE)
          .addGap(18, 18, 18)
          .addComponent(button2, javax.swing.GroupLayout.PREFERRED_SIZE, 152,
javax.swing.GroupLayout.PREFERRED_SIZE)
          .addContainerGap(javax.swing.GroupLayout.DEFAULT_SIZE, Short.MAX_VALUE)));
    layout.setVerticalGroup(
       layout.createParallelGroup(javax.swing.GroupLayout.Alignment.LEADING)
          .addGroup(layout.createSequentialGroup()
          .addContainerGap()
          .addGroup(layout.createParallelGroup(javax.swing.GroupLayout.Alignment.BASELINE)
          .addComponent(button1, javax.swing.GroupLayout.PREFERRED_SIZE, 223,
javax.swing.GroupLayout.PREFERRED_SIZE)
             .addComponent(button2, javax.swing.GroupLayout.PREFERRED_SIZE, 223,
javax.swing.GroupLayout.PREFERRED_SIZE))
  .addContainerGap(javax.swing.GroupLayout.DEFAULT_SIZE, Short.MAX_VALUE)));
        pack();
    }
    public static void main(String args[]) {
        java.awt.EventQueue.invokeLater(new Runnable() {
                public void run() { new Example().setVisible(true); }});
  }
    private javax.swing.JButton button1, button2;
    private javax.swing.JFrame mainFrame;
}
```

**Java Swing sample**

# Java Swing Sample codes